\newcommand{\M}{{\mathcal M}} 
\newcommand{\R}{I\!\!R}   
\newcommand{\N}{I\!\!N}   
\newcommand{\gr}{g_{\scriptscriptstyle{\rm R}}}   
\newcommand{\iip}[2]{\big\langle #1,#2\big\rangle}  
\newcommand{\rip}[2]{\big\langle #1,#2\big\rangle_{\scriptscriptstyle{\text{(R)}}}}                                       
\newcommand{\nablar}{\nabla^{\!\scriptscriptstyle{\text{(R)}}}}

\newcommand{\bpg}{{\mathcal B}_{p,\gamma}^{\scriptscriptstyle{(2)}}(k)}
\newcommand{\bbpg}{{B}_{p,\gamma}(k)}

\newcommand{\B}{{\mathcal B}^{\scriptscriptstyle{(2)}}(k)}
\newcommand{\Bp}{{\mathcal B}_p^{\scriptscriptstyle{(2)}}(k)}
\newcommand{\Ba}[1]{{\mathcal B}^{\scriptscriptstyle{(#1)}}(k)}
\newcommand{\Bpa}[1]{{\mathcal B}_p^{\scriptscriptstyle{(#1)}}(k)}
\newcommand{\opg}{\Omega_{p,\gamma}^{\scriptscriptstyle{(2)}}}

\newcommand{\Opga}[1]{\Omega_{p,\gamma}^{\scriptscriptstyle{(#1)}}}

\newcommand{\Ogpa}[1]{\Omega_{\gamma,p}^{\scriptscriptstyle{(#1)}}}
\newcommand{\Cpg}{{\mathcal C}^0_{p,\gamma}}

\newcommand{\Opa}[1]{\Omega_{p}^{\scriptscriptstyle{(#1)}}}
\newcommand{\nablak}{\nabla^{\{k\}}}
\newcommand{\Rk}{R^{\{k\}}}
\newcommand{\Ik}{I^{\{k\}}}
\newcommand{\opga}[1]{\Omega_{p,\gamma}^{\scriptscriptstyle{(#1)}}}
\newcommand{\ogpa}[1]{\Omega_{\gamma,p}^{\scriptscriptstyle{(#1)}}}
\newcommand{\bpga}[1]{{\mathcal B}_{p,\gamma}^{\scriptscriptstyle{(#1)}}(k)}

\newcommand{\Tt}[1]{{\mathbf T}_{#1}}
\newcommand{\hip}[3]{{\langle\! \langle}#2,#3{\rangle \!\rangle}_{#1}}
\newcommand{\Vvert}{\vert\!\vert\!\vert}
\newcommand{\Tsigma}{{\mathcal T}_\sigma} 
\newcommand{\tsig}{{\mathtt r}_\sigma}    

\newcommand{\Dds}{{\frac{\rm D}{{\rm d}s}}}
\newcommand{\Ddt}{{\frac{\rm D}{{\rm d}t}}}
\newcommand{\dds}{{\frac{\rm d}{{\rm d}s}}}
\newcommand{\ddso}{{\frac{\rm d}{{\rm d}s}}\Big\vert_{s=0}}
\newcommand{\Ddso}{{\frac{\rm D}{{\rm d}s}}\Big\vert_{s=0}}
\newcommand{\ddt}{{\frac{\rm d}{{\rm d}t}}}
\newcommand{\ddto}{{\frac{\rm d}{{\rm d}t}}\Big\vert_{t=0}}
\newcommand{\Ddto}{{\frac{\rm D}{{\rm d}t}}\Big\vert_{t=0}}

\newfont{\bolditalic}{cmbxti10 scaled1440} 

\documentclass[twoside,letterpaper,draft,10pt]{amsart}

\usepackage{amsmath}               
\usepackage{amsthm}

\numberwithin{equation}{section}   

\title[Morse Theory for Relativistic Brachistochrones]%
{Morse Theory for the Travel Time Brachistochrones
in Stationary Spacetimes}
\author[F.\ Giannoni]{Fabio Giannoni}
\address{Dipartimento di Matematica e Fisica, Universit\'a di Camerino, Italy}
\email{giannoni@campus.unicam.it}
\author[P.\ Piccione]{Paolo Piccione}
\address{Departamento de Matem\'atica, Universidade de S\~ao Paulo, Brazil}
\email{piccione@ime.usp.br}
\urladdr{http://www.ime.usp.br/\~{}piccione}
\author[D.\ V.\ Tausk]{Daniel V.\ Tausk}
\address{Departamento de Matem\'atica, Universidade de S\~ao Paulo, Brazil}
\email{tausk@ime.usp.br}

\theoremstyle{plain}\newtheorem{regolaritabpg}{Proposition}[section]
\theoremstyle{remark}\newtheorem{remregolarita}[regolaritabpg]{Remark}
\theoremstyle{plain}\newtheorem{corTsigma}[regolaritabpg]{Corollary}
\theoremstyle{definition}\newtheorem{defbrach}[regolaritabpg]{Definition}
\theoremstyle{plain}\newtheorem{carbrach}[regolaritabpg]{Corollary}
\theoremstyle{remark}\newtheorem{comm1}[regolaritabpg]{Remark}
\theoremstyle{remark}\newtheorem{comm2}[regolaritabpg]{Remark}
\theoremstyle{plain}\newtheorem{regolaritabpg1}[regolaritabpg]{Proposition}
\theoremstyle{definition}\newtheorem{defpticrit1}[regolaritabpg]{Definition}

\theoremstyle{plain}\newtheorem{formalizzazione}{Proposition}[section]
\theoremstyle{plain}\newtheorem{calcvar1}[formalizzazione]{Lemma}
\theoremstyle{plain}\newtheorem{calcvar2}[formalizzazione]{Lemma}
\theoremstyle{plain}\newtheorem{calcvar3}[formalizzazione]{Lemma}
\theoremstyle{plain}\newtheorem{regolaritaL2}[formalizzazione]{Lemma}
\theoremstyle{plain}\newtheorem{diffeqbrach}[formalizzazione]{Proposition}
\theoremstyle{plain}\newtheorem{pticritici12}[formalizzazione]{Proposition}

\theoremstyle{plain}\newtheorem{improve}{Proposition}[section]
\theoremstyle{definition}\newtheorem{deflocmin}[improve]{Definition}
\theoremstyle{plain}\newtheorem{stessipunti}[improve]{Lemma}
\theoremstyle{plain}\newtheorem{smoothnessD}[improve]{Proposition}
\theoremstyle{plain}\newtheorem{first}[improve]{Proposition} 

\theoremstyle{plain}\newtheorem{abstract1}{Lemma}[section]
\theoremstyle{plain}\newtheorem{princsec}[abstract1]{Corollary}

\theoremstyle{remark}\newtheorem{commentoinu}{Remark}[section]
\theoremstyle{remark}\newtheorem{remfocali}[commentoinu]{Remark} 
\theoremstyle{remark}\newtheorem{remarkKillJac}[commentoinu]{Remark} 
\theoremstyle{plain}\newtheorem{cargammaJacobi}[commentoinu]{Lemma}
\theoremstyle{definition}\newtheorem{defmorseindex}[commentoinu]{Definition} 
\theoremstyle{remark}\newtheorem{commentoindici}[commentoinu]{Remark}
\theoremstyle{plain}\newtheorem{hessianoE}[commentoinu]{Proposition}
\theoremstyle{plain}\newtheorem{MorseRiem}[commentoinu]{Theorem} 
\theoremstyle{plain}\newtheorem{secmorseindexth}[commentoinu]{Theorem}

\theoremstyle{definition}\newtheorem{defbJacobi}{Definition}[section]
\theoremstyle{plain}\newtheorem{diffeqbKacobi}[defbJacobi]{Proposition}
\theoremstyle{plain}\newtheorem{converse}[defbJacobi]{Proposition}
\theoremstyle{plain}\newtheorem{corA}[defbJacobi]{Corollary}
\theoremstyle{plain}\newtheorem{corB}[defbJacobi]{Corollary}
\theoremstyle{plain}\newtheorem{carat}[defbJacobi]{Proposition}
\theoremstyle{definition}\newtheorem{defbfoc}[defbJacobi]{Definition}
\theoremstyle{plain}\newtheorem{propC}[defbJacobi]{Proposition}
\theoremstyle{plain}\newtheorem{propD}[defbJacobi]{Proposition}
\theoremstyle{plain}\newtheorem{immagine}[defbJacobi]{Proposition}
\theoremstyle{plain}\newtheorem{corE}[defbJacobi]{Corollary}
\theoremstyle{plain}\newtheorem{finaleMorse}[defbJacobi]{Theorem}
\theoremstyle{plain}\newtheorem{maimax}[defbJacobi]{Corollary}
\theoremstyle{plain}\newtheorem{minimilocali}[defbJacobi]{Corollary}

\theoremstyle{plain}\newtheorem{thm8.1}{Theorem}[section]
\theoremstyle{remark}\newtheorem{rem8.2}[thm8.1]{Remark}
\theoremstyle{plain}\newtheorem{prop8.3}[thm8.1]{Proposition}
\theoremstyle{plain}\newtheorem{thm8.4}[thm8.1]{Theorem}

\theoremstyle{plain}\newtheorem{abstract2}{Lemma}[section]
\theoremstyle{definition}\newtheorem{defvariazione}[abstract2]{Definition} 
\theoremstyle{plain}\newtheorem{hessianoF}[abstract2]{Proposition}
\theoremstyle{plain}\newtheorem{rembordo}[abstract2]{Remark}

\begin{document}

\begin{abstract}
The travel time brachistochrone curves in a general relativistic framework are
timelike curves, satisfying a suitable conservation law with respect
to a an observer field, that are stationary points of the travel time
functional.
In this paper we develop a global variational theory for
brachistochrones joining an event $p$ and 
the worldline of an observer $\gamma$ in a stationary spacetime $\M$.
More specifically, using the method of Lagrange multipliers, we compute
the first and the second variation of the travel time functional, obtaining
two variational principles relating the geometry of the brachistochrones
with the geometry of geodesics in a suitable Riemannian structure.
We present an extension of the classical Morse Theory for Riemannian
geodesics to the case of travel time brachistochrones, and we prove
a Morse Index Theorem for brachistochrones. Finally, using techniques
from Global Analysis, we prove the Morse relations for the travel time functional
and we establish some existence and multiplicity results for brachistochrones.
\end{abstract}

\maketitle

\begin{section}{Introduction: the General Relativistic Brachistochrone Problem}
\label{sec:intro}

\setcounter{page}{1}            %

The classical brachistochrone problem dates back to the end of the
seventeenth century, when Johann Bernoulli challenged his contemporaries
to solve the following problem. 
\begin{quote}
If in a vertical plane two points $A$ and $B$ are given, then it is required to specify
the orbit $AMB$ of the movable point $M$, along which it, starting from
$A$, and under the influence of its own weight, arrives at $B$ in the shortest possible
time.
\emph{(Acta Eruditorum, June 1696)}
\end{quote}
This problem attracted the attention of many important mathematicians of
the time, including Newton, Leibniz, L'H\^opital, and Johann's brother, Jackob
Bernoulli. 
The papers written on the subject may be considered the fundaments of a new
field in mathematics, the {\em Calculus of Variations}. A beautiful historical
exposition of the brachistochrone problem may be found in Reference~\cite{SW},
where the authors' thesis is that the brachistochrone problem also {\em marks
the birth of Optimal Control}.

Still now the classical brachistochrone problem is very popular, and its importance
is witnessed by the fact that there is hardly any book on Calculus of Variations
that does not use this problem as a takeoff point. The well known solution
to the brachistochrone problem is a cycloid, which is the curve described
by a point $P$ in a circle that rolls without slipping.

This problem has several generalizations, e.g.,
the homogeneous gravitational field could be replaced with an arbitrary
Newtonian potential, and instead of releasing the particle from rest one
could prescribe an arbitrary value for the initial speed, leaving the initial
direction of the velocity undetermined. 

In modern terminology, the Newtonian brachistochrone problem can be stated as
follows. Given a manifold $\M_0$ endowed with a Riemannian metric $g_0$,
to be interpreted as the state space, and a smooth function $V:\M_0\longmapsto\R$,
representing the gravitational potential, a brachistochrone of energy $E>0$
between two points $x_0$ and $x_1$ of $\M$ is a  
curve $x:[0,T_x]\longmapsto\M$ joining $x_0$ and $x_1$ that extremizes the travel time
$T_x$ in the space of all  curves $y$ joining $x_0$ and $x_1$ and
satisfying the conservation of energy law:
\begin{equation}\label{eq:consnewton}
\frac12\,g(\dot y,\dot y)+V(y)\equiv E 
\end{equation}
(throughout this paper we will consider the motion of particles with unit mass).
A well known variational principle states that a curve $x$ joining $x_0$ and $x_1$
is a brachistochrone of fixed energy $E$ if and only if $x$ is a geodesic with respect
to the conformal Riemannian metric $\phi_E\cdot g_0$, with conformal factor
$\phi_E=(E-V)^{-1}$.

The brachistochrone problem
can also be formulated in the context of general relativity. 
We want to emphasize here that the original solution to the brachistochrone
problem offered by Johann Bernoulli, which lacked mathematical rigor,
can be made absolutely rigorous in a general relativistic context. 
Namely, the trajectory of a freely falling
massive object, which is represented by a timelike geodesic in a 
Lorentzian manifold, is characterized by extremizing its arrival time
measured by means of a smooth parameterization of the receiving observer.
This is the so called general relativistic timelike Fermat Principle,
suggested in \cite{Ko} and rigorously proven in \cite{GMP}.

The first relativistic versions of the brachistochrone problem appear in
\cite{GB} and \cite{K}. 
V.\ Perlick (see \cite{Pe2}) has determined the brachistochrone equation in a 
 stationary Lorentz\-ian manifold of splitting type, 
 and two of the authors, together with
J.\ Verderesi, in \cite{GPV} have generalized Perlick's result to the case of an
arbitrary stationary manifold by reformulating the brachistochrone problem
in the context of sub-Riemannian geometry.
We recall that a stationary metric that satisfies the Einstein's equations
describes a time-independent gravitational field in General Relativity. 

The variational principle proven in \cite{GPV} was then used in
\cite{GP} to prove some results concerning the existence and the multiplicity of
relativistic brachistochrones with a given value of energy between a fixed event
and a fixed observer of a stationary spacetime.

We formulate the general relativistic brachistochrone problem for
the travel time as follows.

Let $(\M,g)$ be a 4-dimensional Lorentzian manifold, i.e., an arbitrary
spacetime in the sense of general relativity and fix a timelike
smooth vector field $Y$ on $\M$. For simplicity, we
assume that $Y$ is complete, i.e., its integral lines
are defined over the entire real line. 
 The integral curves of $Y$ can be
interpreted as the worldlines of {\it
observers}. Please note that we do not require $Y$ to be normalized, i.e., in
general the
worldlines of our observers are not parameterized by proper time. The reason is
that in the stationary case, i.e., if $(\M,g)$ admits a timelike Killing vector
field, it is convenient to choose this Killing vector field for $Y$ and not a
renormalized version of it.

To formulate the brachistochrone problem with respect to our
arbitrarily chosen observer field $Y$, we fix a point $p$ in $\M$, a
(maximal) integral curve $\gamma:\R\longmapsto\M$ of $Y$ and a real number $k>0$. 
The {\em trial paths\/} for our
variational problem are all timelike smooth curves 
$\sigma : [0,1] \longmapsto \M$ which are nowhere tangent to $Y$ and
satisfy the following conditions:

\begin{eqnarray}\label{eq:ini}
&&\sigma (0) = p;
\\ \label{eq:fin}
&&\sigma (1) \in \gamma(\R);
\\
\label{eq:energy}
&&g({\dot \sigma}(0), Y({\sigma (0)})) = - k \: \big( -g({\dot \sigma},{\dot
\sigma}) \big) ^{1/2};
\\
\label{eq:nba}
&&g(\nabla_{\dot \sigma} {\dot \sigma} , {\dot \sigma}) = 0;
\\
\label{eq:nbb}
&&g(\nabla_{\dot \sigma}{\dot \sigma} , Y ) = 0.
\end{eqnarray}

\noindent Here $\nabla$ denotes the Levi-Civita connection of the 
Lorentzian metric
$g$.
We denote by ${\mathcal B}_{p,\gamma}(k)$ the set of trial paths;
in the rest of the paper we will be working with suitable completions
of this space.

If we interpret each integral curve of $Y$ as a ``point in space'',
(\ref{eq:ini}) and (\ref{eq:fin}) mean that all trial paths connect
the same two points in space, where the starting time is fixed whereas the
arrival time is not. Condition (\ref{eq:energy}) says that all trial
paths start with the same speed with respect to the observer field $Y$. 

Observe that, in order to simplify the mathematics, 
we have chosen to parameterize our trial curves on the interval $[0,1]$, 
rather than using a proper time parameterization over intervals
varying with the curves. 
By condition (\ref{eq:nba}), the quantity ${\Tsigma}$ defined by
$-{\Tsigma}^2 = g({\dot \sigma}, {\dot \sigma})$ is a constant for each trial path
$\sigma$ (but takes different values for different trial paths). This implies that
the curve parameter $t$ along $\sigma$ is related to proper time $\tau$ by an
affine transformation, $\tau = {\Tsigma} t + const$. As a consequence, the
4-velocity along each trial path is given by ${\Tsigma}^{-1}{\dot \sigma}$, whereas
the 4-acceleration is given by ${\Tsigma}^{-2}\nabla_{\dot \sigma} {\dot \sigma}$.
Hence, conditions (\ref{eq:nba}) and (\ref{eq:nbb}) require the 4-acceleration to
be perpendicular to the plane spanned by ${\dot \sigma}$ and $Y$. In other words,
with respect to the observer field $Y$ there are only forces perpendicular to
the direction of motion. Such forces can be interpreted as constraint forces
supplied by a
frictionless slide which is at rest with respect to the observer field $Y$.

The brachistochrone problem can now be formulated in the following way. 

Among
all trial paths that satisfy the above-mentioned conditions, we want to find
those curves for which the travel time is minimal or, more generally,
stationary.  

A different general relativistic brachistochrone
problem can be formulated, by requiring that
the solutions be stationary points for the {\em arrival time\/}
functional,  given by $AT(\sigma)=\gamma^{-1}(\sigma(1))$.
In other words, $AT(\sigma)$
is the value of the proper time of the receiver at the arrival event.
In physical terms, the two brachistochrone problems differ by the 
way of measuring time: in the first case the time is measured by a 
watch traveling along the  trajectory of the mass, in the second case 
the time is measured by the observer that receives the mass at the
end of its trajectory. The two variational problems are essentially
different; in this paper we stick to the first problem, while the
"arrival time brachistochrones" are the subject of a followup paper.

\smallskip
If $(\M,g)$ is a stationary spacetime and $Y$ is a {\em Killing\/} vector field,
i.e., the flow of $Y$ preserves the metric $g$, then the condition (\ref{eq:nbb})
means that the product $g(\dot\sigma,Y)$ is constant along $\sigma$. The value
of this constant can be easily computed using condition (\ref{eq:energy}), 
that gives $g(\dot\sigma,Y)\equiv-k {\Tsigma}$. Hence, in the stationary case,
the conditions (\ref{eq:energy}) and (\ref{eq:nbb}) can be resumed in the
condition:
\begin{equation}\label{eq:nbb2}
g(\dot\sigma,Y)=-k {\Tsigma}.
\end{equation}
Again, observe that the value of the travel time ${\Tsigma}$ appears in
formula (\ref{eq:nbb2}) because of our choice of the parameterization
on the interval $[0,1]$ of our trial curves.
The condition (\ref{eq:nbb2}) is the relativistic counterpart of the  energy
conservation law (\ref{eq:consnewton}) in the Newtonian case.
Although physically meaningful, the mathematical approach to the
general relativistic brachistochrone problem in the non stationary
case presents difficulties of higher order than in the stationary case.
For instance, it is not even clear whether the non stationary brachistochrones
are solutions to a second order differential equation; in Reference~\cite{PP},
the authors used a Lagrange multiplier technique to derive a system of
differential equations for the brachistochrones and for the Lagrangian multipliers.
Unfortunately, it does not seem to be possible to eliminate the Lagrangian
multipliers from the system without introducing integrals, unless in the
stationary case. Thus, it looks as
if the brachistochrones in the non-stationary case are not determined by a
second-order differential equation, but rather by an integro-differential
equation.

For these technical reasons, in this paper we will only study the case of
a manifold $\M$ with metric $g$ which is stationary with respect to the
observer field $Y$.\smallskip

%

The purpose of this article is to present a complete variational theory
for travel time brachistochrones in a stationary Lorentzian manifold and,
in particular,  it will be developed a full-fledged infinite
dimensional Morse theory for the critical points of the travel time. 

We present below a list of the main results proven in this paper:
\begin{itemize}
\item the general-relativistic brachistochrone problem in a stationary
Lorentzian manifold is presented in a context of Global Analysis 
on infinite dimensional Hilbertian manifolds (Section~\ref{sec:setup});
\smallskip

\item the travel time brachistochrones are smooth curves; they can be
characterized as the  only solutions of a second order differential
equation (formula~(\ref{eq:diffeq}) and Proposition~\ref{thm:improve});
\smallskip

\item the brachistochrones can also be characterized as
local minimizers for the travel time, and, equivalently, as curves
whose {\em spatial\/} part is a geodesic with respect to a suitable
Riemannian structure on $\M$ (Proposition~\ref{thm:first});
\smallskip

\item it is computed a second order variation formula
for the travel time functional, which is characterized by a Morse
Index Theorem (Theorem~\ref{thm:finaleMorse}). In analogy with the
Riemannian geodesic problem, this theorem
relates the nature of a stationary point for the travel time with
some metrical properties of $\M$ and with the {\em convexity\/}
of the timelike curve $\gamma$ representing the observer
and measured by the second fundamental form of $\gamma$;
\smallskip

\item under suitable completeness hypotheses for $\M$, we prove the
global Morse relations for the travel time functional in a completion
of the space ${\mathcal B}_{p,\gamma}(k)$ (Section~\ref{sec:global});
thanks to this relations one obtains estimates on the number of
brachistochrones of fixed energy $k$ between $p$ and $\gamma$,
according to the topology and the metric of $\M$.
\end{itemize}

From a strictly mathematical point of view, the paper presents some
technicalities that is worth discussing. The main difficulties in our
variational problem are due to the presence of the double constraint
given by (\ref{eq:nba}) and (\ref{eq:nbb}) (or (\ref{eq:nbb2})), which are,
respectively, quadratic and linear in the first derivative.

Due to this kind of constraint, in order to put a differentiable structure
on the set ${\mathcal B}_{p,\gamma}(k)$ of trial paths, one
needs to consider a Hilbert space completion of ${\mathcal B}_{p,\gamma}(k)$
made in a Sobolev space of curves having at least the $C^1$-regularity,
and thus one is forced to consider curves of class $H^2$ (see 
formula~(\ref{eq:defbpg}), Proposition~\ref{thm:regolaritabpg} 
and Remark~\ref{thm:remregolarita}).

However, the $H^2$-approach has the disadvantage of introducing
new difficulties, especially for the following reasons:
\begin{itemize}
\item the Riesz duality in the Hilbert spaces $H^i$ involves products
of functions and also their derivatives, resulting in lengthy and complicated
calculations when using the Lagrange multipliers method;
\smallskip
\item the arrival time functional {\em does not\/} satisfy good compactness
properties in the space of $H^2$-curves, like the Palais--Smale condition
(see Appendix~\ref{sec:example}), which is
an essential tool  for developing an  infinite dimensional Morse Theory.

\end{itemize}
The problem of duality in Hilbert spaces of curves with {\em high\/} 
regularity is faced through the introduction of a suitable formalism
based on the theory of {\em distribution\/} and {\em generalized functions},
whose technical details are worked out at the beginning of Section~\ref{sec:LM}. 
Unavoidably, the results needed are stated and proven in a formal way,
and this part of the paper turns out to be rather technical nature. 
Even though these
results are essential from a formal point of view, the reader should not be
intimidated by Proposition~\ref{thm:formalizzazione} and the few subsequent
Lemmas, and should keep his/her attention to the main issue of the paper.

As to the problem of lack of compactness for the travel time functional, the
crucial observation here is that, if one is only interested in a local
differentiable structure, then around each smooth ($C^2$) curve $\sigma$  in
${\mathcal B}_{p,\gamma}(k)$ it can be defined a differentiable chart on the set
of $H^1$-curves that are uniformly close to $\sigma$
(see Proposition~\ref{thm:regolaritabpg1}).
Since the solution to our variational problem are proven to be
curves of class $C^2$, then one can relax the requirement of convergence for
the Palais--Smale sequences, which allows to prove the global Morse relations
for the arrival time functional (Section~\ref{sec:global}).

\medskip

The paper is organized according to the following outline.

In Section~\ref{sec:setup} we discuss the variational setup, where we define
our main function spaces and functionals, proving their differentiability
in the setting of infinite dimensional Hilbertian manifolds.

In Section~\ref{sec:LM} we present a Lagrange multiplier approach to
the brachistochrone problem, and we derive some conditions on the
curves that are extrema for the travel time functional and their
corresponding multipliers. Moreover, we obtain a differential equation that
is satisfied by the brachistochrones.

Section~\ref{sec:diffeq} is devoted to the proof of the variational principle
for brachistochrones, that extends the principle proven in \cite{GPV} for
local minimizers of the travel time. We also prove that the differential
equation determined in Section~\ref{sec:LM} is the equation obtained
by the above variational principle.

In Section~\ref{sec:second} we study the second variation of the travel time $T$
at a given brachistochrone. We prove a second order variational principle
for brachistochrones,
that relates the Hessian $H^T$ to the Hessian of the energy functional of 
a suitable Riemannian metric on $\M$.

In Section~\ref{sec:index} we recall some known facts about the Morse Index Theorem
for orthogonal geodesics between submanifolds in Riemannian geometry, and
we prove a slightly different version of the theorem for the case of
a manifold admitting a Killing vector field. This result
(Theorem~\ref{thm:secmorseindexth}), which has
some interest on its own and for this reason it is stated in a general
form, is then used in the next section to prove a brachistochrone
version of the Morse Index Theorem.

In Section~\ref{sec:morse}, in analogy with the classical Morse theory for
Riemannian geodesics, we define the
notions of Jacobi fields and focal points along a brachistochrone, 
and we prove a version of
the Morse Index Theorem for brachistochrones. 
Some immediate consequence of the theory
concerning the local nature of the critical points of $T$ are derived.

Section~\ref{sec:global} is dedicated to the proof of the global
Morse relations, from which we obtain some results on the multiplicity
of brachistochrones with a given value of the energy between an event
and an observer.

Finally, the paper has two short appendices containing some
side results. In Appendix~\ref{sec:explicit} we show the explicit calculation
of the second variation of the travel time functional at a given
brachistochrone. In Appendix~\ref{sec:example} we discuss a simple
but instructive example to show that the travel time
functional does not satisfy the Palais--Smale condition in the space of
curves satisfying an $H^2$-regularity condition.

\end{section}


\begin{section}{The Functional Spaces and the Variational Setup}
\label{sec:setup}
Throughout this paper we will denote by $(\M,g)$ a stationary Lorentzian manifold,
with $g$ a Lorentzian metric tensor on $\M$, and
$Y$ is a smooth timelike Killing vector field on $\M$, which is assumed
to be complete. 

The symbol $\iip\cdot\cdot$ will denote the bilinear
form induced by $g$ on the tangent spaces of $\M$; the usual nabla symbol $\nabla$ will
denote the covariant derivative relative to the Levi--Civita connection
of $g$. Given a smooth function $\phi$ on $\M$, for $q\in\M$ we denote 
by $\nabla\phi(q)$ the gradient of $\phi$ at $q$ with respect to $g$, which
is the vector in $T_q\M$ defined by $\iip{\nabla\phi(q)}\cdot={\rm d}\phi(q)[\,\cdot\,]$;
the Hessian $H^\phi(q)$ of $\phi$ at $q$ is the symmetric bilinear form
on $T_q\M$ given by $H^\phi(q)[v_1,v_2]=\iip{\nabla_{v_1}\nabla\phi}{v_2}$,
for $v_1,v_2\in T_q\M$.

We denote by $\psi:\M\times\R\longmapsto\M$ the flow of $Y$, i.e., for $q\in\M$
and $t\in\R$, $\psi(q,t)$ is the value
$\gamma_q(t)$, where $\gamma_q$ is the maximal integral line of $Y$ satisfying
$\gamma_q(0)=q$. The Killing
property of
$Y$, which is crucial in most of the results presented in this paper, will be used
systematically in our computations through the following three facts:
\begin{enumerate}
\item\label{itm:costante} the quantity $\iip YY$ is constant along the flow
lines of $Y$,
\item\label{itm:derivataisometrica} the differential
${\rm d}_x\psi(q,t_0):T_q\M\longmapsto T_{\psi(q,t_0)}\M$
of the map $\psi(\cdot,t_0)$ is an isometry for all $t_0$, or,
equivalently, for all $t_0$ the map $q\longmapsto\psi(q,t)$
is a local isometry of $\M$;
\item\label{itm:antisim} $\iip{\nabla_v Y}{w}=-\iip{\nabla_w Y}v$ for all
pair of vectors $v$ and $w$; in particular, for all $v\in T\M$, we have
$\iip{\nabla_v Y}v=0$.
\end{enumerate}
Observe that the second or the third
condition above is in fact {\em
equivalent\/} to the Killing property of $Y$ 
(see \cite[Proposition~9.25]{ON}).

We set: \[m={\rm dim}(\M);\]
the physical interesting case is $m=4$.

We denote by $R$ the {\em curvature tensor\/} of the Lorentzian metric
$g$, with the following sign convention:
\begin{equation}\label{eq:convention}
R(X,Y)=\nabla_X\nabla_Y-\nabla_Y\nabla_X-\nabla_{[X,Y]}.
\end{equation}
for $X,Y$ vector fields on $\M$. 

As customary, for $1\le p\le+\infty$, $L^p([0,1],\R)$ will denote
the space of Lebesgue $p$-integrable real functions; for $n\in\N$, $H^n([0,1],\R)$
will denote the Sobolev space of functions of class $C^{n-1}$ and having weak
$n$-th derivative in $L^2([0,1],\R)$.

\noindent
We introduce for convenience the auxiliary Riemannian metric $\gr$ on $\M$, given~by:

\begin{equation}\label{eq:defgr}
\gr(p)[v_1,v_2]=\rip{v_1}{v_2}=\iip{v_1}{v_2}-2\frac{\iip{v_1}{Y(q)}\cdot
\iip{v_2}{Y(q)}}{\iip{Y(q)}{Y(q)}},
\end{equation}
for $q\in\M$ and  $v_1,\,v_2\in T_q\M$. 
It is easy to see that $Y$ is Killing also in the metric $\gr$; moreover, the
restriction of $g$ and $\gr$ on the orthocomplement of $Y$ coincide.

We define the space $L^2([0,1],T\M)$
of square integrable $T\M$-valued functions:
\begin{equation}\label{eq:defl2tm}
L^2([0,1],T\M)=\Big\{\zeta:[0,1]\longmapsto T\M\
\text{measurable}:
\int_0^1\rip{\zeta(t)}{\zeta(t)}\;{\rm d}t<+\infty\Big\}.
\end{equation}
Let $\pi:T\M\longmapsto\M$ be the canonical projection.
Given any curve $\sigma:I\subseteq\R\longmapsto A$, a {\em vector field along $\sigma$\/} is
a map $\zeta:I\longmapsto T\M$ such that $\pi\circ\zeta=\sigma$. 
Let $A$ be any open set of $\M$;
the Sobolev space $H^1([0,1],A)$ is defined by:
\begin{equation}\label{eq:defH1}
H^1([0,1],A)=\Big\{\sigma:[0,1]\longmapsto A:\sigma\ \text{absolutely continuous,}\ 
\dot\sigma\in L^2([0,1],T\M)\Big\}.
\end{equation}

For $A\subseteq\M$, the symbol $C^1([0,1],A)$ will denote the set of $C^1$-curves
defined $[0,1]$ and with image in $A$; we also define the Sobolev
space $H^2([0,1],A)$ as:
\begin{eqnarray}\label{eq:defh2}
H^2([0,1],A)=\Big\{\sigma\in C^1([0,1],A)&:&\dot\sigma\ \text{is absolutely continuous, and}
\nonumber\\ &&\nabla_{\dot\sigma}\dot\sigma\in L^2([0,1],T\M)\Big\}.
\end{eqnarray}
It is not too difficult to prove that the definition of the spaces $H^i([0,1],A)$ does {\em
not\/} indeed depend on the choice of the Riemannian metric $\gr$, nor on the choice of the
linear connection $\nabla$ that appears in (\ref{eq:defh2}). As a matter of fact, the
spaces $H^i([0,1],A)$ can be defined intrinsically for any differentiable
manifold $A$ using local charts (see \cite{Pa}) or, equivalently, using
auxiliary structures on $A$, like for instance a Riemannian metric. 
In the sequel, we will use the spaces $H^i([0,1],A)$, $i=1,2$, where
$A$ will be an open subset of $\M$ or $T\M$. 

If $A$ is a smooth submanifold of $\M$, 
in particular if $A$ is an open subset, then
$H^i([0,1],A)$ has the structure of an infinite dimensional Hilbertian manifold, 
modeled on the Sobolev space $H^i([0,1],\R^m)$; for $\sigma\in H^i([0,1],A)$, the
tangent space $T_\sigma H^i([0,1],A)$ can be identified with the Hilbert space:
\begin{equation}\label{eq:tsh2}
T_\sigma H^i([0,1],A)=\Big\{\zeta\in H^i([0,1], T\M):\zeta\ \text{vector field along}\
\sigma\Big\}.
\end{equation}
The inner product in $T_\sigma H^1([0,1],A)$ is given by:
\begin{equation}\label{eq:cazzarola}
\iip\zeta\zeta_{*}=\int_0^1\Big(\rip\zeta\zeta+\rip{\nabla_{\dot\sigma}\zeta}%
{\nabla_{\dot\sigma}\zeta}\Big)
\;{\rm d}t,
\end{equation}
while the inner product in $T_\sigma H^2([0,1],A)$ is given by:
\begin{equation}\label{eq:1innpr}
\iip\zeta\zeta_{**}=\int_0^1\Big(\rip\zeta\zeta+\rip{\nabla_{\dot\sigma}\zeta}%
{\nabla_{\dot\sigma}\zeta}+\rip{\nabla^2_{\dot\sigma}\zeta}{\nabla^2_{\dot\sigma}\zeta}\Big)
\;{\rm d}t,
\end{equation}
where $\nabla^2_{\dot\sigma}\zeta=\nabla_{\dot\sigma}(\nabla_{\dot\sigma}\zeta)$. 

Let $k$ be a fixed positive constant, with
$-k^2<\sup\limits_{\M}\iip{Y(q)}{Y(q)}$, and $U_k$
be the open set:
\begin{equation}\label{eq:defUk}
U_k=\Big\{q\in\M:\iip{Y(q)}{Y(q)}+k^2>0\Big\}.
\end{equation} 
Since $Y$ is Killing, the quantity
$\iip YY$ is constant along the integral lines of $Y$, hence $U_k$ is
invariant by the flow of $Y$.

We will denote by $p$ a fixed event of $U_k$ and by $\gamma:\R\longmapsto U_k$ a given
integral line of $Y$ which does not pass through $p$. We introduce the spaces
\begin{equation}\label{eq:defopg} 
\opga{i}=\opga{i}(U_k)=\Big\{w\in
H^i([0,1], U_k):
w(0)=p,\ w(1)\in\gamma(\R)\Big\},\quad i=1,2.
\end{equation}
It is well known that $\opga{i}$ is a smooth submanifold of $H^i([0,1],U_k)$;
for $w\in\opga{i}$, the tangent space $T_w\opga{i}$
is given by:
\begin{equation}\label{eq:tansp}
T_w\opga{i}=\Big\{\zeta\in T_w H^i([0,1],U_k):
\zeta(0)=0,\ \zeta(1)\in\R\cdot Y(w(1))\Big\}.
\end{equation}
For $w\in\opga{i}$,
$T_w\opga{i}$ is a Hilbert space with respect to the inner products:
\begin{equation}\label{eq:innpr1}
\iip\zeta\zeta_{1}=\int_0^1 \rip{\nabla_{\dot w}\zeta}%
{\nabla_{\dot w}\zeta}
\;{\rm d}t 
\end{equation}
in the case of $T_w\opga{1}$ and
\begin{equation}\label{eq:innpr}
\iip\zeta\zeta_{2}=\int_0^1\Big(\rip{\nabla_{\dot w}\zeta}%
{\nabla_{\dot w}\zeta}+\rip{\nabla^2_{\dot w}\zeta}{\nabla^2_{\dot w}\zeta}\Big)
\;{\rm d}t
\end{equation}
for $T_w\opg$.
Observe that, since $\zeta(0)=0$ for all $\zeta\in T_w\opga{i}$,
then the inner products
$\iip\cdot\cdot_{*}$ and 
$\iip\cdot\cdot_{**}$ of formulas (\ref{eq:cazzarola}) and (\ref{eq:1innpr})
are equivalent in $T_w\opg$, respectively,
to the products $\iip\cdot\cdot_1$ and $\iip\cdot\cdot_2$ 
of formulas (\ref{eq:innpr1}) and (\ref{eq:innpr}). 

We consider the {\em action\/} functional $F$ on $\opga{i}$, given by:
\begin{equation}\label{eq:action}
F(\sigma)=\frac12\int_0^1\iip{\dot\sigma}{\dot\sigma}\;{\rm d}t.
\end{equation}
It is well known that $F$ is smooth; for $\sigma\in\opga{i}$ and $V\in T_\sigma\opga{i}$, the
Gateaux derivative ${\rm d}F(\sigma)[V]$ is given by:
\begin{equation}\label{eq:derivataF}
{\rm d}F(\sigma)[V]=\int_0^1\iip{\nabla_{\dot\sigma}V}{\dot\sigma}\;{\rm d}t.
\end{equation}
Finally, for all positive constant $k\in\R^+$, we introduce the spaces $\bpga{i}$,
$i=1,2$, by:
\begin{equation}\label{eq:defbpg}
\bpga{i}=\Big\{\sigma\in\opga{i}:\exists\,{\Tsigma}\in\R^+\ \text{such that}\
\iip{\dot\sigma}Y\equiv -k\,{\Tsigma}\ \text{and}\
\iip{\dot\sigma}{\dot\sigma}\equiv-{\Tsigma}^2\Big\}.
\end{equation}
We define the {\em travel time functional\/} $T$ on $\bpga{i}$ by:
\begin{equation}\label{eq:defT}
T(\sigma)={\Tsigma}.
\end{equation}

The main goal of this section is to establish an infinite
dimensional differentiable structure on the sets
$\bpg$ and $\bpga{1}$.  
The case of $\bpg$ is easier, and its regularity is
proven in the next Proposition. For the set $\bpga1$,
we are only able to establish its regularity around some
special points; this second case is treated at the end of this section.
\begin{regolaritabpg}\label{thm:regolaritabpg}
$\bpg$ is a smooth submanifold of $\opg$. For $\sigma\in\bpg$, the tangent space
$T_\sigma\bpg$ can be identified with the Hilbert space:
\begin{equation}\label{eq:sptgbpg}
\begin{split}
T_\sigma\bpg=\Big\{\zeta\in T_\sigma\opg:\;&\exists\;C_\zeta\in\R\ \text{such that}
\\ 
&\iip{\nabla_{\dot\sigma}\zeta}Y-\iip\zeta{\nabla_{\dot\sigma}Y}\equiv
C_\zeta\ \text{and}\ 
\iip{\nabla_{\dot\sigma}\zeta}{\dot\sigma}\equiv\frac{{\Tsigma} C_\zeta}k\Big\},
\end{split}
\end{equation}
endowed with the inner product $\iip\cdot\cdot_2$ of formula (\ref{eq:innpr}).
\end{regolaritabpg}
\begin{proof}
For $\sigma\in\opg$, the maps $\iip{\dot\sigma}Y$, $\iip{\dot\sigma}Y^2$ and
$\iip{\dot\sigma}{\dot\sigma}$ are in $H^1([0,1],\R)$. Let $k\in\R^+$ be a fixed
constant. We consider
the following map:
\begin{equation}\label{eq:mapf}
{\mathcal F}:\opg\longmapsto H^1([0,1],\R)\times H^1([0,1],\R)
\end{equation}
given by:
\begin{equation}\label{eq:mapf2}
{\mathcal F}(\sigma)=(\iip{\dot\sigma}Y, \iip{\dot\sigma}Y^2+k^2\iip{\dot\sigma}{\dot\sigma}).
\end{equation}
It is not difficult to prove that ${\mathcal F}$ is a smooth map  and that, for
$\sigma\in\opg$ and $V\in T_\sigma\opg$, the Gateaux derivative ${\rm d}{\mathcal F}(\sigma)%
[V]$ is given by:
\begin{eqnarray}\label{eq:derf}
&&{\rm d}{\mathcal F}(\sigma)[V]= \nonumber\\ &&\quad
(\iip{\nabla_{\dot\sigma}V}Y-\iip
V{\nabla_{\dot\sigma}Y}, 2\iip{\dot\sigma}Y(\iip{\nabla_{\dot\sigma}V}Y-\iip
V{\nabla_{\dot\sigma}Y})+2k^2
\iip{\nabla_{\dot\sigma}V}{\dot\sigma}).
\end{eqnarray}
Here we have used the fact that $Y$ is Killing, thus $\iip{\dot\sigma}{\nabla_VY}=-
\iip V{\nabla_{\dot\sigma}Y}$.
\smallskip

Let ${\mathcal C}$ denote the subspace of $H^1([0,1],\R)$ given by the constant 
functions, and let ${\mathcal C}^-$ denote the open submanifold of ${\mathcal C}$
consisting of negative functions:
\begin{equation}\label{eq:defcalc}
\begin{split}
&{\mathcal C}=\Big\{h\in H^1([0,1],\R):h\equiv h_0\ (\text{const.})\
\ \text{a.\ e.}\Big\},\\
&{\mathcal C}^-=\Big\{h\in{\mathcal C}:h<0\ \ \text{a.\ e.}\Big\}.
\end{split}
\end{equation}
 It is easy to see that
$\bpg={\mathcal F}^{-1}({\mathcal C}^-\times\{0\})$. 

Let $\tilde H^1([0,1],\R)$ denote the quotient space $H^1([0,1],\R)/{\mathcal C}$,
which is naturally identified with the set of functions with null average in
$[0,1]$.

Let $\Pi$ be the map:
\begin{equation}\label{eq:defPi}
\Pi:H^1([0,1],\R)\times H^1([0,1],\R)\longmapsto
\tilde H^1([0,1],\R)\times H^1([0,1],\R)
\end{equation}
given by the quotient map on the first
factor and the identity on the second factor.

To prove the Proposition we use the 
{\em Inverse Mapping Theorem\/} (see \cite{L}). According to this Theorem, 
$\bpg$ is a smooth submanifold of $\opg$ provided that the map ${\mathcal F}$ be
{\em transversal\/} over ${\mathcal C}^-\times\{0\}$, i.e., if for all $\sigma\in\bpg$
the composite map:
\begin{equation}\label{eq:transv}
\Pi\circ {\rm d}{\mathcal F}(\sigma):T_\sigma\opg\longmapsto
\tilde H^1([0,1],\R)\times H^1([0,1],\R)
\end{equation}
is surjective. This amounts to saying that, for all $h_1,h_2\in H^1([0,1],\R)$ there
exists a constant $c\in\R$ such that the
system of differential equations:
\begin{eqnarray}\label{eq:prima}
&& \iip{\nabla_{\dot\sigma}V}Y-\iip V{\nabla_{\dot\sigma}Y}=h_1+c\\
\label{eq:seconda}
&& 2\iip{\dot\sigma}Y(\iip{\nabla_{\dot\sigma}V}Y-\iip V{\nabla_{\dot\sigma}Y})+
2k^2\iip{\nabla_{\dot\sigma}V}{\dot\sigma}=h_2
\end{eqnarray}
has at least one solution $V\in T_\sigma\opg$. Using the fact that 
$\iip{\dot\sigma}Y\equiv -k{\Tsigma}$, we can rewrite (\ref{eq:seconda}) as:
\begin{equation}\label{eq:terza}
\iip{\nabla_{\dot\sigma}V}{\dot\sigma}=h_3,
\end{equation}
where
\[h_3=\frac{h_2+2k{\Tsigma}(h_1+c)}{2k^2}\]
is  in $H^1([0,1],\R)$.

Let $Z\in H^2([0,1],T\M)$ be a vector field along $\sigma$ satisfying 
\begin{equation}\label{eq:proprZ}
\iip YZ\equiv0,\quad\text{and}\quad\iip Z{\dot\sigma}\ne0.
\end{equation}
To prove the existence of such a vector field $Z$, consider first the
vector field along $\sigma$ given by  $\dot\sigma^\perp$, which is the orthogonal
projection of $\dot\sigma$ onto the  distribution $\Delta=Y^\perp$ orthogonal
to $Y$. Formally, we have:
\begin{equation}\label{eq:defsigmaerp}
\dot\sigma^\perp=\dot\sigma-\frac{\iip{\dot\sigma}Y}{\iip YY}\,Y=
\dot\sigma+\frac{k\,{\Tsigma}}{\iip YY}\,Y.
\end{equation}
Obviously, we have:
\begin{equation}\label{eq:positivo}
\iip{\dot\sigma^\perp}{\dot\sigma}=-{\Tsigma}^2\,\frac{k^2+\iip YY}{\iip YY}\ne0,
\end{equation}
because $k^2+\iip YY\ne0$ in $U_k$.

Observe that $\dot\sigma^\perp\in H^1$, and it does not have the required
$H^2$-regularity.
Now, let $Z$ be any section of class $H^2$ of $\Delta$ which is {\em uniformly\/}
close to $\dot\sigma^\perp$, in such a way that $\iip Z{\dot\sigma}\ne0$ as well.%
\footnote{For the approximation theorem, we can use an $H^2$ parallel
referential of $\Delta$ along $\sigma$, so that sections of $\Delta$ along
$\sigma$ will be identified with curves in the Euclidean space, and standard approximation
results apply.}

\noindent
Observe in particular that, since $\iip Z{\dot\sigma}$ is continuous, then
$\iip Z{\dot\sigma}^{-1}$ is in $L^\infty([0,1],\R)$. 

In order to solve equations (\ref{eq:prima}) and (\ref{eq:terza}), we set
\[V=aY+bZ,\]
where $a,b\in H^2([0,1],\R)$ are to be determined. Observe that such a $V$ belongs
to $T_\sigma\opg$ provided that $a$ and $b$ satisfy the boundary conditions:
\begin{equation}\label{eq:condition}
a(0)=b(0)=0,\quad\text{and}\quad b(1)=0.
\end{equation}
Since $\iip ZY=0$, equations (\ref{eq:prima}) and (\ref{eq:terza}) are translated into:
\begin{eqnarray}\label{eq:quarta}
&& a'\iip YY+2b\iip{\nabla_{\dot\sigma}Z}Y=h_1+c\\
\label{eq:quinta}
&& -a'k{\Tsigma} +b'\iip Z{\dot\sigma}+b\iip{\nabla_{\dot\sigma}Z}{\dot\sigma}=h_3.
\end{eqnarray}
We solve for $a'$ equation (\ref{eq:quarta}) obtaining:
\begin{equation}\label{eq:sesta}
a'=\iip YY^{-1}\left[h_1+c-2b\iip{\nabla_{\dot\sigma}Z}Y\right];
\end{equation}
substituting (\ref{eq:sesta}) in (\ref{eq:quinta}) gives:
\begin{equation}\label{eq:settima}
b'+\alpha b=\beta+c\gamma,
\end{equation}
where 
\[\alpha=\frac{\iip YY\iip{\nabla_{\dot\sigma}Z}{\dot\sigma}+2k{\Tsigma}
\iip{\nabla_{\dot\sigma}Z}Y}{\iip Z{\dot\sigma}\iip YY},\]
and
\[\beta=\frac{%
k{\Tsigma} h_1+h_3\iip YY}{\iip Z{\dot\sigma}\iip YY},\quad
\gamma=\frac{k{\Tsigma}}{\iip Z{\dot\sigma}\iip YY}.\]
Observe that $\alpha$, $\beta$ and $\gamma$ are in $H^1([0,1],\R)$. 
Thus, the unique solution $b$ of (\ref{eq:settima}) satisfying $b(0)=0$, given by:
\begin{equation}\label{eq:ottava}
b(t)=e^{-\int_0^t\alpha}\left[\int_0^t\beta e^{\int\!\alpha}+c\int_0^t\gamma
e^{\int\!\alpha}\right],
\end{equation}
is in $H^2([0,1],\R)$. Observe that $\gamma\ne0$ in $[0,1]$, and so 
$\int_0^1\gamma e^{\int\alpha} \ne0$. In particular, there exists $c\in\R$ such
that
$b(1)=0$. 

Finally, $a$ can be chosen as the unique solution of (\ref{eq:sesta}) satisfying
$a(0)=0$. Observe that the right hand side of (\ref{eq:sesta}) is in
$H^1([0,1],\R)$, so $a\in H^2([0,1],\R)$ and ${\mathcal F}$ is transversal over
${\mathcal C}^-$. Hence, $\bpg$ is a smooth submanifold of $\opg$.

By the Inverse Mapping Theorem, for $\sigma\in\bpg$, the tangent space $T_\sigma\bpg$
is identified with the kernel of the map $\Pi\circ{\rm d}{\mathcal F}(\sigma)$,
which consists of the vector fields $\zeta\in T_\sigma\opg$ such that 
${\rm d}{\mathcal F}(\sigma)[\zeta]\in {\mathcal C}\times\{0\}$. 

Recalling (\ref{eq:prima}) and (\ref{eq:seconda}), we have that $\zeta\in T_\sigma\opg$
belongs to $T_\sigma\bpg$ if and only if there exists $C_\zeta\in\R$
such that $\zeta$ satisfies the equations:
\begin{eqnarray}\label{eq:penultima}
&&\iip{\nabla_{\dot\sigma}\zeta}Y-\iip\zeta{\nabla_{\dot\sigma}Y}=C_\zeta,\\
\label{eq:ultima}
&&-2k{\Tsigma} C_\zeta+2k^2\iip{\nabla_{\dot\sigma}\zeta}{\dot\sigma}=0.
\end{eqnarray}
From (\ref{eq:penultima}) and (\ref{eq:ultima}) we easily obtain (\ref{eq:sptgbpg})
and we are done.
\end{proof}
Given a curve $\sigma\in\bpg$, a vector field $\zeta\in T_\sigma\bpg$
will be called a {\em variational vector field\/} along $\sigma$.\smallskip

In some parts of the paper (see Section~\ref{sec:morse}) we will need to consider
variations of curves in $\bpg$ by curves $\sigma$
satisfying the conditions (\ref{eq:nbb}) and (\ref{eq:nbb2}), but not necessarily
with endpoints in $p$ and $\gamma$. For this reason, 
for $i=1,2$ we introduce the sets:
\begin{equation}\label{eq:defB}
\Bpa i=\bigcup_{\gamma\subset U_k}\bpga{i},\quad\text{and}\quad
\Ba i=\bigcup_{p,\gamma\subset U_k}\bpga{i},
\end{equation}
where the unions in (\ref{eq:defB}) are taken over all  $\gamma$'s that are
integral lines of $Y$ having image in $U_k$.

Using the same argument of Proposition~\ref{thm:regolaritabpg}, it is an easy exercise
to prove that both $\Bp$ and $\B$ are smooth Hilbert submanifolds of $H^2([0,1],U_k)$ and
that, for
$\sigma\in\Bp$ or $\sigma\in\B$, the tangent spaces $T_\sigma\Bp$ and
$T_\sigma\B$ are
Hilbert subspaces of
$T_\sigma H^2([0,1],U_k)$ given by:
\begin{equation}\label{eq:tanspBp}
\begin{split}
T_\sigma\Bp=\Big\{\zeta\in T_\sigma H^2([0,1],U_k) 
 &:\zeta(0)=0,\;\exists\;C_\zeta\in\R\
\text{such that}
 \\ 
\iip{\nabla_{\dot\sigma}\zeta}Y &- \iip\zeta{\nabla_{\dot\sigma}Y}\equiv
C_\zeta\ \text{and}\ 
\iip{\nabla_{\dot\sigma}\zeta}{\dot\sigma}\equiv\frac{{\Tsigma} C_\zeta}k\Big\}.
\end{split}
\end{equation}
and

\begin{eqnarray}\label{eq:tanspB}
T_\sigma\B=\Big\{\zeta\in T_\sigma H^2([0,1],U_k)\!\!\!\!\!\!
\!\!\!&&:\exists\;C_\zeta\in\R\
\text{such that}
\nonumber\\ 
\iip{\nabla_{\dot\sigma}\zeta}Y\!\!\!\!&-&\!\!\!\!\iip\zeta{\nabla_{\dot\sigma}Y}\equiv
C_\zeta\ \text{and}\ 
\iip{\nabla_{\dot\sigma}\zeta}{\dot\sigma}\equiv\frac{{\Tsigma} C_\zeta}k\Big\}.
\end{eqnarray}
We restrict the action functional $F$ of (\ref{eq:action}) to $\bpg$, 
obtaining the following:
\begin{corTsigma}\label{thm:corTsigma}
The Gateaux derivative ${\rm d}T(\sigma)[\zeta]$ of the travel
time functional on $\bpg$ is given by:
\begin{equation}\label{eq:derivataT}
{\rm d}T(\sigma)[\zeta]=-\frac{C_\zeta}k.
\end{equation}
\end{corTsigma}
\begin{proof}
Since $\bpg$ is a smooth submanifold of $\opg$, then the restriction of the action
functional $F$ to $\bpg$ is smooth. For $\sigma\in\bpg$, we have:
\begin{equation}\label{eq:F}
F(\sigma)=-\frac12{\Tsigma}^2<0,
\end{equation}
hence $T(\sigma)=\sqrt{-2F(\sigma)}$ is also smooth. 

Equality (\ref{eq:derivataT}) follows
easily by differentiating the expression ${\Tsigma}=-k^{-1}{\iip{\dot\sigma}Y}$
and using the equality
$C_\zeta=\iip{\nabla_{\dot\sigma}\zeta}Y-\iip\zeta{\nabla_{\dot\sigma}Y}$.
\end{proof}

After setting up our variational framework, we are ready to give the
following definition:
\begin{defbrach}\label{thm:defbrach}
A {\em brachistochrone\/} of energy $k$ between $p$ and $\gamma$ is a stationary point
for the travel time functional $T$ on $\bpg$. A brachistochrone curve $\sigma$ is
said to be {\em minimal\/} if $\sigma$ is a minimum point for $T$ on $\bpg$.
\end{defbrach}
From Corollary~\ref{thm:corTsigma} and Definition~\ref{thm:defbrach} we obtain immediately:
\begin{carbrach}\label{thm:carbrach}
A curve $\sigma\in\bpg$ is a brachistochrone of energy $k$ between $p$ and $\gamma$
if and only if for every $\zeta\in T_\sigma\bpg$ it is 
$C_\zeta=\iip{\nabla_{\dot\sigma}\zeta}Y-\iip\zeta{\nabla_{\dot\sigma}Y}=0$.\qed
\end{carbrach}
\begin{comm1}\label{thm:comm1}
Observe that the definition of brachistochrone of energy $k$ 
given in Definition~\ref{thm:defbrach}
is different from the one given in (\cite{GPV}, Definition~1.1) and used in
Reference~\cite{GP}. Namely, in these articles, it was not established a differentiable
structure in the set of admissible curves for the variational problem, and
the brachistochrones of energy $k$ were defined as curves 
locally minimizing their travel time. The equivalence of the two 
definitions will be given in Section~\ref{sec:diffeq}, where we prove that the 
two approaches lead to exactly the same solutions.
\end{comm1}

\begin{comm2}\label{thm:comm2}
Since $T$ is strictly positive on $\bpg$, then its critical points coincide
with the critical points in $\bpg$ of the restriction of the action functional
$F=-\frac12T^2$. The minimal brachistochrones of energy $k$ are {\em maximum\/} points
of $F$ on $\bpg$.
\end{comm2}
\begin{remregolarita}\label{thm:remregolarita}
The proof of the regularity of the space $\bpga{2}$ presented
in Proposition~\ref{thm:regolaritabpg} does {\em not\/}
apply to the space $\bpga{1}$; more precisely, the failure of the
proof is in the existence of the vector field $Z$ satisfying (\ref{eq:proprZ}).
Observe indeed that, in the case of $\bpga{1}$, the derivative $\dot \sigma$
is only defined as an  $L^2$-function, and in general it is not a continuous
curve. 

This fact is the reason why we have to introduce here our global
variational setup using the space $\bpg$. 

However, the same proof of Proposition~\ref{thm:regolaritabpg} can be 
adapted to prove that, if $\sigma$ is $C^1$, then a suitable
neighborhood of $\sigma$ in $\bpga{1}$ has the structure of a smooth
Hilbert manifold. The proof is more delicate; we omit the details
that can be found in a forthcoming paper (see \cite{GPb2}).

As a matter of facts, we will see that the the solutions to our 
variational problem as given in Section~\ref{sec:diffeq} are  
indeed smooth curves (see Proposition~\ref{thm:improve}). 
This fact will allow
us to  to work in the space $\bpga{1}$ in the second part of the paper
(starting from Section~\ref{sec:diffeq}), 
when we will be studying the {\em local\/}  
properties of the brachistochrones, i.e., the properties of
objects that are defined only around the brachistochrones,
like for instance the second variation of $T$, the Jacobi fields,
conjugate points and the Morse Index Theorem for brachistochrones
(Theorem~\ref{thm:finaleMorse}).
\end{remregolarita}

We summarize the main properties of the set $\bpga{1}$
as follows:
\begin{regolaritabpg1}\label{thm:regolaritabpg1}
$\bpga{1}$ is a metric space with the metric induced
by $H^1([0,1],\M)$. 
The inclusion $\iota:\bpg\longmapsto\bpga{1}$ is continuous
and it has dense image.

If $\sigma\in\bpga{1}$ is a map of class $C^1$, then there exists 
a neighborhood ${\mathcal V}_\sigma$ of $\sigma$ in $\bpga1$ that has the structure
of an infinite dimensional Hilbertian manifold. In particular, 
$\bpga1$ has a dense open subset that is a smooth Hilbert manifold.

If $\sigma\in\bpga1$ is a curve of class $C^1$, then, for all
$\sigma_1\in{\mathcal V}_{\sigma}$, the tangent space
$T_{\sigma_1}{\mathcal V}_\sigma$ can be identified with
the Hilbert subspace of $T_{\sigma_1}\opga1$ given by:
\begin{equation}\label{eq:tanspbpg1}
\begin{split}
T_{\sigma_1}{\mathcal V}_\sigma=\Big\{\zeta\in T_\sigma\opga1:\;&\exists\;C_\zeta\in\R\
\text{such that}
\\ 
&\iip{\nabla_{\dot\sigma}\zeta}Y-\iip\zeta{\nabla_{\dot\sigma}Y}\equiv
C_\zeta\ \text{and}\ 
\iip{\nabla_{\dot\sigma}\zeta}{\dot\sigma}\equiv\frac{{\Tsigma} C_\zeta}k\Big\}.
\end{split}
\end{equation}
The restriction of the travel time functional $T$ to each
neighborhood of the form ${\mathcal V}_{\sigma}$, for
some $\sigma\in\bpga1$ of class $C^1$, is smooth, and the same
result of Corollary~\ref{thm:corTsigma} holds.
\end{regolaritabpg1}
\begin{proof}
Convergence in the space $\bpg$ clearly
implies the convergence in $\bpga1$, which implies that
the inclusion $\iota:\bpg\longmapsto\bpga1$ is continuous.

For the second part of the thesis, it suffices to adapt the proof
of Proposition~\ref{thm:regolaritabpg}, and the details will be omitted.
\end{proof}
We can give the following definition:
\begin{defpticrit1}\label{thm:defpticrit1}
A curve $\sigma\in\bpga1$ is said to be a {\em regular\/}
point of $\bpga1$ if $\bpga1$ has the structure of a smooth
Hilbert manifold in a neighborhood ${\mathcal V}_\sigma$
of $\sigma$. By Proposition~\ref{thm:regolaritabpg1}, every curve
$\sigma\in\bpga1$ of class $C^1$ is a regular point of $\bpga1$.

\noindent
A {\em critical point\/} of $T$ in $\bpga1$ is a regular
point $\sigma$ of $\bpga1$ such that ${\rm d}T(\sigma)=0$~%
in~$T_\sigma{\mathcal V}_{\sigma}$.
\end{defpticrit1}
To conclude this section, we remark that, in perfect
analogy with Proposition~\ref{thm:regolaritabpg1}, 
if $\sigma$ is a regular
point in $\Bpa1$ or in $\Ba1$, then these two sets
have the structure of smooth manifolds around $\sigma$.
Their tangent spaces are given by:
\begin{equation}\label{eq:tanspBp1}
\begin{split}
T_\sigma\Bpa1=\Big\{\zeta\in T_\sigma H^1([0,1],U_k) 
 &:\zeta(0)=0,\;\exists\;C_\zeta\in\R\
\text{such that}
 \\ 
\iip{\nabla_{\dot\sigma}\zeta}Y &- \iip\zeta{\nabla_{\dot\sigma}Y}\equiv
C_\zeta\ \text{and}\ 
\iip{\nabla_{\dot\sigma}\zeta}{\dot\sigma}\equiv\frac{{\Tsigma} C_\zeta}k\Big\}.
\end{split}
\end{equation}
and

\begin{eqnarray}\label{eq:tanspB1}
T_\sigma\Ba1=\Big\{\zeta\in T_\sigma H^1([0,1],U_k)\!\!\!\!\!\!
\!\!\!&&:\exists\;C_\zeta\in\R\
\text{such that}
\nonumber\\ 
\iip{\nabla_{\dot\sigma}\zeta}Y\!\!\!\!&-&\!\!\!\!\iip\zeta{\nabla_{\dot\sigma}Y}\equiv
C_\zeta\ \text{and}\ 
\iip{\nabla_{\dot\sigma}\zeta}{\dot\sigma}\equiv\frac{{\Tsigma} C_\zeta}k\Big\}.
\end{eqnarray}
\end{section}


\begin{section}{An Abstract Approach to the Lagrange Multiplier Method.\\
The First Variation of the Travel Time}
\label{sec:LM}
In this Section we use the Lagrange multiplier technique to
derive a system of differential equation satisfied by the
brachistochrones, and to extend the variational principle
proven in \cite{GPV}.

In order to use this technique,
we need a global Banach differentiable structure on our set of maps,
and for this reason we will work in the space $\bpg$
rather than $\bpga1$ (see Remark~\ref{thm:remregolarita}
and Proposition~\ref{thm:regolaritabpg1}).
This approach has the unpleasant drawback of  making
our notations and calculations much heavier  then one would
expect. This is due to the fact that the duality in the Sobolev spaces 
$H^1$ and $\tilde H^1$, which are the natural images for the map ${\mathcal F}$
defined by (\ref{eq:mapf}), 
involves also products of the first  derivatives of the maps and of
the Lagrange multipliers, resulting in very lengthy formulas that make  it a
complicated task to determine an explicit form of the
Euler--Lagrange equation satisfied by the critical points of
our functional.

To overcome this difficulty, the authors have decided to
use the formalism of {\em generalized functions\/} and {\em distributions\/}
on Sobolev spaces, which will make the computations formally similar
to the {\em naive\/} calculations made by the classical variationalists
of the last century. Unfortunately, our problem does not fit perfectly
in the theory of distributions on Sobolev spaces presented in standard
textbooks, and we are forced to develop our own theory from
scratch. Hopefully, the formalism developed here will be adaptable
to the study of other variational problems involving several constraints
and requiring a {\em high\/} degree of regularity for the trial
maps.

The first part of this section is devoted to this aim, and it is of rather
technical nature. A first time reader who wants to avoid technicalities
and who is willing to make an act of faith, after reading
formula (\ref{eq:functionals}) can  just skip everything that comes
before formula (\ref{eq:***1}) without seriously jeopardizing
his/her general understanding of the subject.

Keeping Remark~\ref{thm:comm2} in mind, in the notation of Section~\ref{sec:setup}
(recall in particular formulas (\ref{eq:mapf2}), (\ref{eq:defcalc}) and
(\ref{eq:defPi}),  we want to extremize the action functional
$F(\sigma)=\frac12\int_0^1\iip{\dot\sigma}{\dot\sigma}\,{\rm d}s$ in the space
of all curves $\sigma\in\opg$ subject to the constraint
${\mathcal F}(\sigma)\in {\mathcal C}^-\times\{0\}$.

Then, $\sigma\in\bpg$ is a solution to our variational problem if and only if there
exists an element $\Lambda$ in the dual space of $\tilde H^1([0,1],\R)\times
H^1([0,1],\R)$ such that 
\begin{equation}\label{eq:primaLM}
{\rm d}F(\sigma)-\Lambda\circ(\Pi\circ{\rm d}{\mathcal F}(\sigma)):T_\sigma\opg\longmapsto\R
\end{equation}
vanishes identically. In this case, $\Lambda$ is unique,
and it is the {\em Lagrange multiplier\/} of $\sigma$; from a physical point
of view, $\Lambda$ represents the constraint forces acting on the particle moving along
$\sigma$.

A Lagrangian multiplier for our variational problem is of the form $\Lambda=(\lambda,\mu)$,
where $\lambda\in \tilde H^1([0,1],\R)^*$ and $\mu\in H^1([0,1],\R)^*$; here
the $*$ means the dual space in the sense of Banach spaces. Observe that 
$\tilde H^1([0,1],\R)^*$ can be identified with the closed subspace of
$H^1([0,1],\R)^*$ consisting of functionals vanishing on constant functions. 

It is convenient to write the duality in the spaces $H^1([0,1],\R)$ and
$\tilde H^1([0,1],\R)$ in the form:
\begin{equation}\label{eq:functionals}
\lambda(f)=\int\lambda f
\quad\text{and}\quad\mu(f)=\int\mu f,
\end{equation} 
where $\lambda$ and $\mu$ are seen as {\em
generalized functions}. Indeed, we know that not all the continuous functionals 
on the spaces $H^1$ and $\tilde H^1$ are of the form (\ref{eq:functionals}) for
some function $\lambda$ and $\mu$.

This argument introduces a substantial simplification in the computations
involving dual spaces, but it needs a concrete formalization, which is rather technical
and it is done once and for all in the following.
\smallskip

We introduce the following formalism. Let $I$ denote the interval $[0,1]$.
For each $i\in\N$, let $D^i$ be the dual space $H^i(I,\R)^*$.
If $\pi_E:E\longmapsto\M$ is any fiber bundle over $\M$ with
projection $\pi_E$, given $\sigma\in H^i(I,\M)$,
let $H^i(I,\sigma,E)$ denote the set of maps $\omega\in H^i(I,E)$ such that
$\pi_E\circ\omega=\sigma$. We will consider in particular the tangent bundle
$T\M$ and the cotangent bundle $T\M^*$ with their canonical projections onto $\M$.

Finally, we denote by
$D^i_\sigma$  the dual space $H^i(I,\sigma,T\M^*)^*$.

We remark that there are canonical inclusions $D^i\subset D^{i+1}$ and $D^i_\sigma\subset
D^{i+1}_\sigma$ given by restriction of the linear functionals.
By convention, keeping in mind the Riesz representation theorem for Hilbert spaces,
we define $H^0=D^0=L^2(I,\R)$ and
$D_\sigma^0=L^2(I,\sigma, T\M)$.

\smallskip

We consider the following operations in the spaces $D^i$, $D^i_\sigma$ and $H^i$:
\begin{itemize}
\item[(a)] For $\lambda\in D^i$, $i\ge1$, 
and $f\in H^i(I,\R)$, $(\lambda f)\in D^i$ is defined by
$(\lambda f)(\phi)=\lambda(f \phi)$. Observe that this product is well defined
because, $H^i(I,\R)$ is an algebra (i.e., closed with respect to products) and the
product $(f,\phi)\longmapsto f\phi$ is continuous in $H^i(I,\R)$. The validity
of the operations defined in the other items is checked by similar arguments.
\smallskip

\item[(b)]  For $\lambda\in D^i$, $i\ge1$, 
and $V\in H^i(I,\sigma, T\M)$, $(\lambda V)\in D^i_\sigma$ is defined by 
$(\lambda V)(\alpha)=\lambda(\alpha(V))$, for
$\alpha\in H^i(I,\sigma,T\M^*)$.
\smallskip

\item[(c)] For $f\in H^i(I,\R)$, $i\ge1$, and $\nu\in D^i_\sigma$, $(f\nu)\in
D_\sigma^i$ is defined by $(f\nu)(\alpha)=\nu(f\alpha)$, for $\alpha\in
H^i(I,\sigma,T\M^*)$.
\smallskip

\item[(d)] For $V\in H^i(I,\sigma,T\M)$ and $\nu\in D^i_\sigma$,
$i\ge1$, the inner product
$\iip\nu V\equiv\iip V\nu\in D^i$ is defined by $\iip\nu V(\phi)=\nu(\iip{\phi
V}{\cdot})$ for $\phi\in H^i(I,\R)$.
\smallskip

\item[(e)] For $\lambda\in D^i$, $i\ge0$, we define
$\int_I\lambda=\lambda(1)\in\R$. 
\smallskip

\item[(f)] For $\lambda\in D^i$, $i\ge0$, we denote by $\tilde\lambda$ the element
in
$D^{i+1}$ defined by $\tilde\lambda(\phi)=-\lambda(\phi')$ for $\phi\in
H^{i+1}(I,\R)$. Observe that $\tilde\lambda$ is well defined because the map
$\phi\longmapsto\phi'$ from
$H^{i+1}(I,\R)$ to $H^i(I,\R)$ is linear and continuous. Note also that $\lambda$
is a sort of {\em distributional  derivative}, but keeping in mind that $\lambda$
is an element of a dual space of functions which {\em do not\/} vanish on the boundary.
In particular, even for differentiable functions $\lambda$, it is not true that
$\tilde\lambda=\lambda'$.
An explicit form of $\tilde\lambda$  is given in part~\ref{itm:derf} of
Proposition~\ref{thm:formalizzazione}.
\smallskip

\item[(g)] For $\nu\in D_\sigma^i$, $i\ge0$, the element $\tilde \nu\in
D_\sigma^{i+1}$ is defined by $\tilde\nu(\alpha)=-\nu(\nabla_{\dot\sigma}\alpha)$,
where $\alpha$ belongs to
$H^{i+1}(I,\sigma,T\M^*)$ and $\nabla_{\dot\sigma}\alpha$ is the covariant
derivative of the covector
$\alpha$ along $\sigma$.  This means that, if $\alpha$ is the covector given by
$\iip V\cdot$ for some
$V\in H^{i+1}(I,\sigma,T\M)$, then $\nabla_{\dot\sigma}\alpha
=\iip{\nabla_{\dot\sigma}V}\cdot$. The element $\tilde\nu$ is the
distributional derivative for covectors, analogue to formula (f) (see
part~\ref{itm:derV} of Proposition~\ref{thm:formalizzazione}).
\end{itemize}
For $t_0\in I$, we denote by $\delta_{t_0}\in D^1$ the {\em Dirac delta\/} at
$t_0$, which is the element defined by
$\delta_{t_0}(\phi) =\phi(t_0)$ for all $\phi\in H^i(I,\R)$; moreover, for $A\in
T_{\sigma(t_0)}\M$, $\delta_{t_0}^A\in D^i_\sigma$ will denote the element defined by
$\delta_{t_0}^A(\alpha)=\alpha(t_0)(A)$.

For $V\in H^i(I,\sigma,T\M)$, $t_o\in I$ and $A\in T_{\sigma(t_0)}\M$, we have:
\begin{equation}\label{eq:81}
\iip V{\delta_{t_0}^A}=\iip{V(t_0)}A\, \delta_{t_0}.
\end{equation}
Namely, using property (d) above, for $\phi\in H^i(I,\R)$ we have:
\begin{eqnarray*}
\iip V{\delta_{t_0}^A}(\phi)&=&\delta_{t_0}^A(\iip{\phi V}\cdot)=\iip{%
\phi(t_0)\,V(t_0)}A=\phi(t_0)\iip{V(t_0)}A=\\&=&\iip{V(t_0)}A\cdot\delta_{t_0}(\phi).
\end{eqnarray*}
\begin{formalizzazione}\label{thm:formalizzazione}
The following statements hold true:
\begin{enumerate}
\item\label{itm:integrale} for $\lambda\in D^i$ and $\phi\in H^i(I,\R)$,
$i\ge1$, it is
$\lambda(\phi)=\int_I\lambda\phi$;
\item\label{itm:dual} the dual space $\tilde H^1(I,\R)^*$ is identified
with the closed subspace of $D^1$ consisting of elements $\lambda$ satisfying
$\int_I(\lambda\cdot1)=0$;
\item\label{itm:zero} if $\tilde\lambda=0$, then $\lambda=0$;
\item\label{itm:dualint} for $\nu\in D_\sigma^i$ and $V\in H^{i+1}(I,\sigma,T\M)$, 
$i\ge0$, it
is $\int_I\iip\nu{\nabla_{\dot\sigma}V}=-\int_I\iip{\tilde\nu}V$;
\item\label{itm:incl1} there exists a continuous linear injection of $L^1(I,\R)$
into
$D^i$, $i\ge1$, given by the map $\lambda\in L^1(I,\R)\longmapsto\hat\lambda\in
D^i$, where
$\hat\lambda(\phi)=\int_I\lambda(s)\phi(s)\;{\rm d}s$ for all $\phi\in H^i(I,\R)$;
\item\label{itm:incl2} if $L^1(I,\sigma,T\M)$ denotes the set of vector fields
along $\sigma$ whose Riemannian length (\ref{eq:defgr}) is Lebesgue integrable,
then there is a continuous linear injection of $L^1(I,\sigma,T\M)$ into
$D^i_\sigma$, $i\ge1$, given by $\nu\in L^1(I,\sigma,T\M)\longmapsto\hat\nu\in
D^i_\sigma$, where $\hat\nu(\alpha)=\int_I\alpha(t)\nu(t)\;{\rm d}t$ for $\alpha\in
H^i(I,\sigma,T\M^*)$;
\item\label{itm:l2} if $\psi\in D^1$ is such that $\tilde\psi\in D^2$ is also in $D^1$
(recall the inclusion $D^1\subset D^2$),
then $\psi\in L^2(I,\R)$; similarly, if $\psi,\tilde\psi\in D^1_\sigma$, then
$\psi\in L^2(I,\sigma, T\M)$;
\item\label{itm:derprod1}for $\lambda\in D^i$ and $f\in H^{i+1}(I,\R)
\subset H^{i}(I,\R)$, $i\ge0$, it is
$\widetilde{(\lambda f)}=\tilde\lambda f+\lambda f'$;
\item\label{itm:derprod2} for $\lambda\in D^i$ and $V\in H^{i+1}(I,\sigma,T\M)
\subset H^i(I,\sigma,T\M)$, $i\ge0$, it
is $\widetilde{(\lambda V)}=\tilde\lambda V+\lambda\nabla_{\dot\sigma}V$;
\item\label{itm:derf} for $f\in H^1(I,\R)$, it is $\tilde f=f(0)\,\delta_0-f(1)\,\delta_1+f'$;
\item\label{itm:derV} for $V\in H^1(I,\sigma,T\M)$, it is $\tilde V=\delta_0^{V(0)}-
\delta_1^{V(1)}+\nabla_{\dot\sigma}V$.
\end{enumerate}
\end{formalizzazione}
\begin{proof}
For part~\ref{itm:integrale}, it is $\lambda(\phi)=\lambda(\phi\cdot1)$. By (a),
it is $\lambda(\phi\cdot1)=(\lambda\phi)(1)$ and by (e) $(\lambda\phi)(1)=\int_I\lambda\phi$.

Part~\ref{itm:dual} is simply the fact that elements in the dual space
of $\tilde H^1(I,\R)$ are characterized by the property of vanishing on constant
functions.

For part \ref{itm:zero}, it suffices to observe that the map $\phi\longmapsto\phi'$
is surjective from $H^{i+1}(I,\R)$ to $H^i(I,\R)$.

For part \ref{itm:dualint}, using (d) and (e), we have:
\[\int_I\iip\nu{\nabla_{\dot\sigma}V}=\iip\nu{\nabla_{\dot\sigma}V}(1)=\nu(\iip{1\cdot%
\nabla_{\dot\sigma}V}\cdot).\] 
On the other hand, by (e) and (g) we have:
\[\int_I\iip{\tilde\nu}V=\iip{\tilde\nu}V(1)=\tilde\nu(\iip{1\cdot V}\cdot)=-
\nu(\iip V\cdot')=-\nu(\iip{\nabla_{\dot\sigma}V}\cdot),\]
which proves the claim.

For part \ref{itm:incl1}, observe that $\hat\lambda$ is a well defined
element in the dual of $H^i(I,\R)$. The linearity of the map $\lambda\longmapsto\hat\lambda$
is trivial; the continuity depends on the fact that convergence in $H^1$ implies uniform
convergence. Finally, the injectivity is simply the Fundamental Theorem of Calculus
of Variations.

Part \ref{itm:incl2} is proven analogously. Namely, using an orthonormal frame
along $\sigma$, one reduces the problem to the case $\M=\R^m$. In this case
the proof of part~\ref{itm:incl1} can be repeated {\em verbatim\/} 
for each component
of $\nu$.

Using part~\ref{itm:incl1} and \ref{itm:incl2}, we will identify
each $\lambda\in L^1(I,\R)$ with $\hat\lambda\in D^i$ and
every $V\in L^1(I,\sigma,T\M)$ with $\hat V\in D^i_\sigma$.
Suppressing the symbol $\hat{\phantom{a}}$, for all
$\lambda\in L^1(I,\R)$, $V\in L^1(I,\sigma,T\M)$,
$f\in H^i(I,\R)$ and $\alpha\in H^i(I,\sigma,T\M^*)$ we will write
concisely:
\begin{equation}\label{eq:freccia}
\lambda(f)=\int_I\lambda(t)\,f(t),\quad\text{and}\quad
V(\alpha)=\int_I\alpha(t)\,V(t).
\end{equation}

To prove part~\ref{itm:l2}, observe that the map $f\longmapsto f'$ from $H^1(I,\R)$
to $L^2(I,\R)$ is continuous and surjective. Hence, if $\tilde\psi\in D^1$,
then the map $f'\longmapsto
\int_I\psi f'=-\int_I\tilde\psi f$, where $f$ is the unique primitive of $f'$
such that $f(0)=0$,
gives a continuous linear functional on $L^2(I,\R)$, and the conclusion
follows by Riesz Theorem. The second half is proven similarly.

The formulas in part \ref{itm:derprod1} and \ref{itm:derprod2} are the product rules for the
$\tilde{\phantom{a}}$-derivative. We prove \ref{itm:derprod2} as follows.
For $\alpha\in H^{i+1}(I,\sigma,T\M^*)$, we have:
\[\widetilde{(\lambda V)}(\alpha)=-(\lambda
V)(\nabla_{\dot\sigma}\alpha)=-\lambda(\nabla_{\dot\sigma}\alpha(V)).\] On the
other hand, we compute:
\begin{eqnarray*}
(\tilde\lambda V+\lambda\nabla_{\dot\sigma}V)(\alpha)&=&(\tilde\lambda V)(\alpha)+(\lambda%
\nabla_{\dot\sigma}V)(\alpha)=\tilde\lambda(\alpha(V))+\lambda(\alpha(\nabla_{\dot\sigma}V))=\\
&=& -\lambda(\alpha(V)')+\lambda(\alpha(\nabla_{\dot\sigma}V))=\\&=&
-\lambda(\nabla_{\dot\sigma}\alpha(V))-\lambda(\alpha(\nabla_{\dot\sigma}V))+
\lambda(\alpha(\nabla_{\dot\sigma}V))=
-\lambda(\nabla_{\dot\sigma}\alpha(V)).
\end{eqnarray*}
Part \ref{itm:derprod1} is proven similarly.

We omit the proof of part~\ref{itm:derf} and we prove part~\ref{itm:derV}.
For $\alpha\in H^{i+1}(I,\sigma,T\M^*)$, using (\ref{eq:freccia}), 
we have:
\begin{eqnarray*}
\tilde
V(\alpha)&=&-V(\nabla_{\dot\sigma}\alpha)=-\int_I(\nabla_{\dot\sigma}\alpha(t))V(t)=
-\int_I\Big[(\alpha(t)V(t))'-\alpha(t)\nabla_{\dot\sigma}V(t)\Big]=\\&=&
-\alpha(t)V(t)\Big\vert_0^1+\int_I\alpha(t)\nabla_{\dot\sigma}V(t)=
\alpha(0)(V(0))-\alpha(1)(V(1))+
\nabla_{\dot\sigma}V(\alpha)=\\&=&\delta_0^{V(0)}(\alpha)-
\delta_1^{V(1)}(\alpha)+\nabla_{\dot\sigma}V(\alpha).
\end{eqnarray*}
This concludes the proof of Proposition~\ref{thm:formalizzazione}.
\end{proof}
We now present three preliminary results that will be needed in
the computation of the first
variation for the travel time functional:
\begin{calcvar1}\label{thm:calcvar1}
Let $\nu\in D^i_\sigma$ and suppose that 
$\int_I\iip V\nu=0$ for all $V\in H^i(I,\sigma,T\M)$ such that
$V(0)=0$ and $V(1)$ is parallel to $Y(\sigma(1))$. Then, we have
$\nu=\delta_0^A+\delta_1^B$ for some $A\in T_{\sigma(0)}\M$ and $B\in T_{\sigma(1)}\M$ with
$\iip B{Y(\sigma(1))}=0$.
\end{calcvar1}
\begin{proof}
Under the hypotheses, it is $\nu(\alpha)=0$ for all $\alpha\in H^i(I,\sigma, T\M^*)$
such that $\alpha(0)=0$ and such that $\alpha(1)$ is a multiple of the covector
$\iip{Y(\sigma(1))}\cdot$. The subspace $H$ of such $\alpha$'s has codimension
equal to $(2m-1)$ in $H^i(I,\sigma, T\M^*)$. 
Then, the {\em annihilator\/} $H^o$ of $H$ in $D^i_\sigma$ has dimension $(2m-1)$.
The subspace $N$ of $D^i_\sigma$ consisting
of elements $\nu$ of the form $\delta_0^A+\delta_1^B$ for some $A\in T_{\sigma(0)}\M$ 
and $B\in T_{\sigma(1)}\M$ with
$\iip B{Y(\sigma(1))}=0$ clearly has dimension $(2m-1)$ and it is contained in the
annihilator of $H$. Thus, $N=H^o$ and we are done.
\end{proof}
\begin{calcvar2}\label{thm:calcvar2}
Let $\lambda\in D^1$ be fixed. If $\tilde\lambda=c_0\,\delta_0+c_1\,\delta_1$ for some
$c_0,c_1\in\R$, then necessarily $c_0=-c_1$ and $\lambda\equiv c_0$ is constant, i.e.,
$\lambda(\phi)=\int_Ic_0\phi(t)\,{\rm d}t$ for all $\phi\in H^1(I,\R)$.
\end{calcvar2}
\begin{proof}
First of all, observe that there exists no $\lambda\in D^1$ such that
$\tilde\lambda=\delta_0$. Namely, if $\tilde\lambda=\delta_0$ and $\phi\in H^1(I,\R)$,
then it would be $\tilde\lambda(\phi)=\phi(0)$, and so $\lambda(\phi')=-\phi(0)$.
On the other hand, for all constants $c\in\R$, it would be $\tilde\lambda(\phi+c)=
\phi(0)+c$, and $\tilde\lambda(\phi+c)=-\lambda(\phi')=-\phi(0)$, which is a
contradiction. 

It follows that there exists no $\lambda\in D^1$ such that $\tilde\lambda=c_0\,\delta_0
+c_1\,\delta_1$ with $c_0\ne -c_1$. Indeed, if such $\lambda$ existed, then the element
$\lambda_1=(c_0+c_1)^{-1}(\lambda+c_1) $ would satisfy $\tilde\lambda_1=\delta_0$.

Finally, suppose that $\tilde\lambda=c_0\,\delta_0-c_0\,\delta_1$. Then,
$\widetilde{(\lambda-c_0)}=0$, and by part~\ref{itm:zero} of 
Proposition~\ref{thm:formalizzazione}, $\lambda\equiv c_0$.
\end{proof}
The following simple result states the well known fact that the Dirac delta's
are not given by any $L^1$-function:
\begin{calcvar3}\label{thm:calcvar3}
If $\lambda\in L^1(I,\R)$ is such that $\hat \lambda\in D^i$ is of the form
$c_0\,\delta_0+c_1\,\delta_1$ for some $c_0,c_1\in\R$, then $\lambda\equiv0$ and
$c_0=c_1=0$. Similarly, if $\nu\in L^1(I,\sigma,T\M)$ is such that $\hat\nu\in
D_\sigma^i$ is of
the form $\delta_0^A+\delta_0^B$ for some vectors $A\in T_{\sigma(0)}\M$ and
$B\in T_{\sigma(1)}\M$, then $\nu\equiv0$ and $A=B=0$.
\end{calcvar3}
\begin{proof}
If $\hat\lambda=c_0\,\delta_0+c_1\,\delta_1$, then $\int_I\lambda(t)\phi(t)\,{\rm d}t=0$
for all smooth function $\phi$ with support contained in $]0,1[$. This implies
$\lambda\equiv0$. The proof of the second part of the Lemma is analogous.
\end{proof}
We are now ready to determine the Euler--Lagrange equation
for the critical points of the travel time functional
in $\bpg$. Recalling (\ref{eq:derf}),
(\ref{eq:primaLM}) and part~\ref{itm:dual} of
Proposition~\ref{thm:formalizzazione}, we now fix a curve
$\sigma\in\bpg$. 
Recall from the definition (\ref{eq:defbpg}) of $\bpg$ that there exists
${\Tsigma}>0$ such that:
\begin{equation}\label{eq:conssigma}
\iip{\dot\sigma}Y\equiv
-k\,{\Tsigma},\quad\text{and}\quad\iip{\dot\sigma}{\dot\sigma}=-{\Tsigma}^2.
\end{equation}
 We assume that there exist $\lambda,\mu\in D^1$ (see part~\ref{itm:dual}
of Proposition~\ref{thm:formalizzazione}), with
$\int_I\lambda=0$, such that the equation:
\begin{eqnarray}\label{eq:calcln1}
0&=&\int_I\iip{\nabla_{\dot\sigma}V}{\dot\sigma}\;{\rm d}s-\lambda\Big(
\iip{\nabla_{\dot\sigma}V}Y-\iip V{\nabla_{\dot\sigma}Y}\Big)+\\
&&-\mu\Big(
2\iip{\dot\sigma}Y(\iip{\nabla_{\dot\sigma}V}Y-\iip
V{\nabla_{\dot\sigma}Y})+2k^2
\iip{\nabla_{\dot\sigma}V}{\dot\sigma}\Big)\nonumber
\end{eqnarray}
is satisfied for all $V\in T_\sigma\opg$. Using the formalism introduced in
the first part of the Section, we rewrite equation (\ref{eq:calcln1}) as:
\begin{eqnarray}\label{eq:calcln2}
0&=&\int_I\iip{V}{\lambda\,\nabla_{\dot\sigma}Y+2\mu\iip{\dot\sigma}Y\,\nabla_{\dot\sigma}Y}+\\
&&+\int_I\iip{\nabla_{\dot\sigma}V}{\dot\sigma-\lambda\,Y-2\mu\,\iip{\dot\sigma}Y\,Y
-2\mu \,k^2\,\dot\sigma}.\label{eq:secint}
\end{eqnarray}
In the above formula, the "products" between the dual maps
$\lambda$ and $\mu$ with functions or vector fields along $\sigma$ have to
be interpreted in the sense of the operations (a)---(c) above;
moreover, the inner product $\iip\cdot\cdot$ in (\ref{eq:calcln2}) and
(\ref{eq:secint}) is meant in the sense of (d).

\noindent
Observe that the elements:
\begin{equation}\label{eq:phiepsi}
\phi=\lambda\,\nabla_{\dot\sigma}Y+2\mu\iip{\dot\sigma}Y\,\nabla_{\dot\sigma}Y,
\quad\text{and}\quad
\psi=\dot\sigma-\lambda\,Y-2\mu\iip{\dot\sigma}Y\,Y
-2\mu k^2\,\dot\sigma\end{equation} are in $D^1_\sigma$. 

We need the following {\em regularity\/} result for the Lagrangian multipliers
$\lambda$ and $\mu$:
\begin{regolaritaL2}\label{thm:regolaritaL2}
The Lagrangian multipliers $\lambda$ and $\mu$ are indeed $L^2$-functions, i.e.,
there exist $f_\lambda, f_\mu\in L^2(I,\R)\subset L^1(I,\R)$ such that $\lambda=\hat
f_\lambda$ and $\mu=\hat f_\mu$.
\end{regolaritaL2}
\begin{proof}
From (\ref{eq:calcln2}), (\ref{eq:secint}) and (\ref{eq:phiepsi}), we have 
$\int\iip V\phi+\int\iip{\nabla_{\dot\sigma}V}\psi=\int\iip V{\phi-\tilde\psi}=0$ for
all $V\in H^2(I,\sigma, T\M)$ such that $V(0)=0$ and $V(1)$ is parallel to $Y(\sigma(1))$.
From Lemma~\ref{thm:calcvar1} it follows that $\phi-\tilde\psi$ is a linear combination of
delta's, and in particular,  $\phi-\tilde\psi$ is in $D^1_\sigma$. Hence, $\tilde\psi$
is in $D^1_\sigma$, and, by part~\ref{itm:l2} of Proposition~\ref{thm:formalizzazione},
$\psi\in L^2(I,\sigma, T\M)$.  Since $\dot\sigma\in H^1(I,TM)$, then $\iip\psi{\dot\sigma}$
is in $L^2(I,\R)$; computing explicitly, we have:
\[\iip\psi{\dot\sigma}=-{\Tsigma}^2+\lambda\, k\, {\Tsigma}-2\,\mu\,
k^2{\Tsigma}^2+2\,\mu\, k^2{\Tsigma}^2=-{\Tsigma}^2+
\lambda\, k\, {\Tsigma}\in L^2(I,\R),\]
hence $\lambda\in L^2(I,\R)$. Then, from the definition (\ref{eq:phiepsi}) of $\psi$, 
we obtain that $\mu\,{\Tsigma} Y-\mu\,k\,\dot\sigma\in L^2(I,T\M)$; multiplying by
$Y$ we have:
\[\mu\,{\Tsigma}\iip YY+\mu\,k^2{\Tsigma}=\mu\,{\Tsigma}\left(\iip YY+k^2\right)\in
L^2(I,\R).\] Since $(\iip YY+k^2)^{-1}\in L^\infty(I,\R)$
(because $\sigma$ has image in $U_k$), it follows that $\mu\in L^2(I,R)$ and
the proof is concluded.
\end{proof}

We use the operation (f) to "integrate by parts" (\ref{eq:secint}),
and, keeping in mind parts~\ref{itm:derprod1} and \ref{itm:derprod2} of 
Proposition~\ref{thm:formalizzazione}, we obtain
\begin{eqnarray}\label{eq:***1}
\quad0&=&\int_I\iip{V}{\lambda\,\nabla_{\dot\sigma}Y+2\mu\iip{\dot\sigma}Y\,\nabla_{\dot\sigma}Y}+
\nonumber\\
&-&\int_I\iip{V}{\tilde{\dot\sigma}-\tilde\lambda Y-\lambda\,\nabla_{\dot\sigma}Y
-2\tilde\mu\iip{\dot\sigma}YY-2\mu\iip{\dot\sigma}Y \nabla_{\dot\sigma}Y}+\\ 
&+&\int_I\iip
V{2\,\tilde\mu\,k^2\dot\sigma+2\mu\,k^2\,\nabla_{\dot\sigma}\dot\sigma},
\nonumber
\end{eqnarray}
for all $V\in T_\sigma\opg$.
Observe that, when using parts~\ref{itm:derprod1} and \ref{itm:derprod2} of
Proposition~\ref{thm:formalizzazione}, if the functions involved are only in $H^1$
(like in this particular case the function $\dot\sigma$)
then they must be multiplied by distributions in $D^0=L^2$ for the rule to apply.
This is where we use Lemma~\ref{thm:regolaritaL2}.
\smallskip

We substitute
\[\tilde{\dot\sigma}=\nabla_{\dot\sigma}\dot\sigma+\delta_0^{\dot\sigma(0)}-
\delta_1^{\dot\sigma(1)}\]
in (\ref{eq:***1}), and, from Lemma~\ref{thm:calcvar1}, we have:
\begin{eqnarray}\label{eq:***2}
&&\lambda\,\nabla_{\dot\sigma}Y+4\mu\,\iip{\dot\sigma}Y\,\nabla_{\dot\sigma}Y-
\nabla_{\dot\sigma}\dot\sigma+\tilde\lambda\,Y+\lambda\,\nabla_{\dot\sigma}Y+
2\tilde\mu\iip{\dot\sigma}Y\,Y+\nonumber\\ &&\qquad\qquad+2\tilde\mu\,k^2\dot\sigma
+2\mu\,k^2\,\nabla_{\dot\sigma}\dot\sigma=\delta_0^A+\delta_0^{\dot\sigma(0)}+\delta_1^B-
\delta_1^{\dot\sigma(1)},
\end{eqnarray}
for some $A\in T_{\sigma(0)}\M$ and some $B\in T_{\sigma(1)}\M$ such that $\iip
B{Y(\sigma(1))}=0$.

Now, we multiply equation (\ref{eq:***2}) by $\dot\sigma$, and since
$\iip{\nabla_{\dot\sigma}Y}{\dot\sigma}=\iip{\nabla_{\dot\sigma}\dot\sigma}{\dot\sigma}=0$
and $\iip{\dot\sigma}Y=-k{\Tsigma}$, $\iip{\dot\sigma}{\dot\sigma}=-{\Tsigma}^2$,
using (\ref{eq:81}), we get:
\begin{equation}\label{eq:comblin}
\tilde\lambda\,k\,{\Tsigma}=\iip{\dot\sigma(0)}A\,\delta_0-{\Tsigma}^2\delta_0+
\iip{\dot\sigma(1)}B\,\delta_1+T_{\sigma}^2\delta_1.
\end{equation}
This means that $\tilde\lambda$ is a linear combination of $\delta_0$ and
$\delta_1$. By Lemma~\ref{thm:calcvar2}, $\lambda$ is constant and
$ 
\iip{\dot\sigma(0)}A=-\iip{\dot\sigma(1)}B 
$.
But $\lambda$ constant and $\int_I\lambda=0$ imply immediately:
\begin{equation}\label{eq:l0}
\lambda=0.
\end{equation}
In particular, we have:
\begin{equation}\label{eq:qrosso}
\iip{\dot\sigma(0)}A=-\iip{\dot\sigma(1)}B={\Tsigma}^2.
\end{equation}
We now substitute $\lambda=\tilde\lambda=0$ in (\ref{eq:***2}); multiplying
the resulting equation by $Y$, using (\ref{eq:qrosso})
and recalling that $\iip{\dot\sigma}Y$ is constant and
that $\iip{\nabla_{\dot\sigma}\dot\sigma}Y=0$, we obtain:
\begin{eqnarray*}
&&-4k{\Tsigma}\mu\iip{\nabla_{\dot\sigma}Y}Y-2k{\Tsigma}\tilde\mu\iip%
YY-2k^3{\Tsigma}\tilde\mu=\\&&\qquad(\iip{Y(p)}A-k{\Tsigma})\,\delta_0+
(\iip{Y(\sigma(1))}B+k{\Tsigma})\,\delta_1,
\end{eqnarray*}
which can be written as:
\begin{equation}\label{eq:comblin2}
-2k{\Tsigma}{(\widetilde{\mu\iip
YY}+k^2\tilde\mu)}=(\iip{Y(p)}A-k{\Tsigma})\,\delta_0+
(\iip{Y(\sigma(1))}B+k{\Tsigma})\,\delta_1.
\end{equation}
Again, by Lemma~\ref{thm:calcvar2}, we have that:
\begin{equation}\label{eq:YAB}
\iip{Y(p)}A=-\iip{Y(\sigma(1))}B=0,
\end{equation}
and 
\begin{equation}\label{eq:muconst}
\mu\,\Big(\iip YY+k^2\Big)\equiv c
\end{equation}
for some constant $c\in\R$. Finally, from (\ref{eq:comblin2}) and (\ref{eq:YAB}) we
compute easily $c=\frac12$, and
\begin{equation}\label{eq:esprmu}
\mu=\frac1{2(k^2+\iip YY)}.
\end{equation}
From (\ref{eq:esprmu}) we compute easily:
\begin{equation}\label{eq:esprtildemu}
\tilde\mu=-\frac{\iip{\nabla_{\dot\sigma}Y}Y}{(\iip%
YY+k^2)^2}+\mu(0)\,\delta_0-\mu(1)\,\delta_1;\end{equation}
substituting (\ref{eq:conssigma}), (\ref{eq:l0}), (\ref{eq:esprmu})
and (\ref{eq:esprtildemu}) into (\ref{eq:***2})
gives:
\begin{eqnarray}\label{eq:diffeqcond}
&&-\frac{\iip YY}{\iip YY+k^2}\,\nabla_{\dot\sigma}\dot\sigma-2k^2\frac{%
\iip{\nabla_{\dot\sigma}Y}Y}{(\iip YY+k^2)^2}\,\dot\sigma-
\frac{2k{\Tsigma}}{\iip YY+k^2}\,\nabla_{\dot\sigma}Y+\nonumber\\
&&\qquad+2k\,{\Tsigma}\,\frac{%
\iip{\nabla_{\dot\sigma}Y}Y}{(\iip YY+k^2)^2}\,Y=\\
&&=\delta_0^{A+\dot\sigma(0)}+
\delta_1^{B-\dot\sigma(1)}-2k\,(-{\Tsigma}
Y(p)\,\mu(0)+k\,\dot\sigma(0))\,\delta_0+
\nonumber\\ &&\qquad+2k\,(-{\Tsigma}
Y(\sigma(1))\,\mu(1)+k\,\dot\sigma(1))\,\delta_1.
\nonumber\end{eqnarray}
Observe that for $t_0\in I$ and $v_0\in T_{\sigma(t_0)}\M$, it is
$v_0\,\delta_{t_0}=\delta_{t_0}^{v_0}$, hence, the second member of the equality
(\ref{eq:diffeqcond}) can be written as:
\[\delta_0^{A_1}+\delta_1^{B_1},\]
where
\begin{eqnarray*}
&&A_1=A+\dot\sigma(0)-2k\,(-{\Tsigma} Y(p)\,\mu(0)+k\,\dot\sigma(0)),\\
&&B_1=B-\dot\sigma(1)+2k\,(-{\Tsigma} Y(\sigma(1))\,\mu(1)+k\,\dot\sigma(1)).
\end{eqnarray*}
Hence, by Lemma~\ref{thm:calcvar3}, the first member of the equality (\ref{eq:diffeqcond})
is null, and also $A_1=B_1=0$. Therefore, we obtain the following differential
equation for $\sigma$:
\begin{equation}\label{eq:diffeq}
\begin{split}
\nabla_{\dot\sigma}\dot\sigma+2k^2&\,\frac{\iip{\nabla_{\dot\sigma}Y}Y}{%
\iip YY\,(k^2+\iip YY)}\,\dot\sigma+\frac{2k\,{\Tsigma}}{\iip
YY}\,\nabla_{\dot\sigma}Y+
\\ & -2k\,{\Tsigma}\frac{\iip{\nabla_{\dot\sigma}Y}Y}{%
\iip YY\,(k^2+\iip YY)}\,Y=0.
\end{split}
\end{equation}
We have proven the following:
\begin{diffeqbrach}\label{thm:diffeqbrach}
Let $\sigma\in\bpg$. Then, $\sigma$ is a brachistochrone of energy $k$ between
$p$ and $\gamma$ if and only if $\sigma$ is a curve of class $C^2$ and there exists
${\Tsigma}>0$ such that $\sigma$
satisfies the differential equation
(\ref{eq:diffeq}).\qed
\end{diffeqbrach} 
Observe that any curve $\sigma$ in $H^2(I,\M)$ that satisfies (\ref{eq:diffeq})
almost everywhere is automatically smooth.
\smallskip

Besides determining the differential equation~(\ref{eq:diffeq}),
the importance of Proposition~\ref{thm:diffeqbrach} lies in the fact
that, due to the smoothness of the brachistochrones,we will be able to
work in the space $\bpga1$ when we are in the vicinity
of such a curve (recall Remark~\ref{thm:remregolarita} and
Proposition~\ref{thm:regolaritabpg1}).
This will be done systematically starting from the next Section.

Recalling Definition~\ref{thm:defpticrit1}, we have the following:
\begin{pticritici12}\label{thm:pticritici12}
A curve $\sigma$ is a brachistochrone of energy $k$ between $p$
and $\gamma$ if and only if it is a critical point for
$T$ in $\bpga1$.
\qed
\end{pticritici12}
\end{section}
\begin{section}{The Brachistochrone Differential Equation\\ and the First Order
Variational Principle Revisited}\label{sec:diffeq}
In this section we will take a closer look at the differential equation
(\ref{eq:diffeq}) and we will prove that it characterizes the brachistochrones
between $p$ and $\gamma$ among all the curves in $\opga1$ satisfying suitable
initial conditions. 

Proposition~\ref{thm:diffeqbrach} can be improved as follows:
\begin{improve}\label{thm:improve}
A curve $\sigma\in\opga1$ is a brachistochrone of energy $k$ between $p$ and $\gamma$
if and only if $\sigma$ is smooth and there exists a ${\Tsigma}>0$  such that
$\sigma$ satisfies (\ref{eq:diffeq}), with initial velocity $\dot\sigma(0)$
satisfying:
\begin{equation}\label{eq:IC}
\iip{\dot\sigma(0)}{\dot\sigma(0)}=-{\Tsigma}^2,\quad\text{and}\quad
\iip{\dot\sigma(0)}{Y(p)} =-k\,{\Tsigma}.
\end{equation}
\end{improve}
\begin{proof}
From Proposition~\ref{thm:diffeqbrach}, all we need to prove is that any smooth curve
$\sigma\in\opga1$ that satisfies the differential equation~(\ref{eq:diffeq})
and whose initial velocity $\dot\sigma(0)$ satisfies~(\ref{eq:IC}) is in
$\bpga1$. 

To this aim, it suffices to show that the functions $\eta(t)=\iip{\dot\sigma(t)}%
{\dot\sigma(t)}+{\Tsigma}^2$ and
$\theta(t)=\iip{\dot\sigma(t)}{Y(\sigma(t))}+k\,{\Tsigma}$ are constant.

If we multiply (\ref{eq:diffeq}) by $Y$, we obtain:
\[
\iip{\nabla_{\dot\sigma}\dot\sigma}Y+\frac{2k^2\iip{\nabla_{\dot\sigma}Y}Y}{%
\iip YY\,(k^2+\iip YY)}\,\left(k\,{\Tsigma}+\iip{\dot\sigma}Y\right)=0,
\]
that can be written as:
\begin{equation}\label{eq:derivata1}
\theta'+u\,\theta=0,
\end{equation}
with
\[u=\frac{2k^2\iip{\nabla_{\dot\sigma}Y}Y}{%
\iip YY\,(k^2+\iip YY)}.\]
Since $\theta(0)=0$, then, the uniqueness of the solution for equation~(\ref{eq:derivata1})
implies $\theta\equiv0$. Now, if we multiply (\ref{eq:diffeq}) by $\dot\sigma$,
knowing that $\iip{\dot\sigma}Y=-k\,{\Tsigma}$ is constant
and $\iip{\nabla_{\dot\sigma}Y}{\dot\sigma}=0$, we obtain:
\[\iip{\nabla_{\dot\sigma}\dot\sigma}{\dot\sigma}+\frac{2k^2\iip{\nabla_{\dot\sigma}Y}Y}{%
\iip YY\,(k^2+\iip YY)}\,\left(\iip{\dot\sigma}{\dot\sigma}+{\Tsigma}^2\right)=0,\]
that can be written as:
\begin{equation}\label{eq:derivata2}
\frac12\,\eta'+g\,\eta=0.
\end{equation}
Again, since $\eta(0)=0$, equation~(\ref{eq:derivata2}) implies $\eta\equiv0$ and
we are done.
\end{proof}
We give two more different descriptions of the brachistochrone curves.
We first characterize them as curves that minimize {\em locally\/}
their travel time.

If $q$ is any point in $U_k$, we denote by $\gamma_q$ the maximal
integral line of $Y$ through $q$. Moreover, if $I=[a,b]\subseteq[0,1]$ is
any interval, and if $q_1,q_2$ are any two points in $U_k$, we define
${\mathcal B}^{\scriptscriptstyle{(1)}}%
_{q_1,\gamma_{q_2}}(k,I)$ as the space of curves $\tau\in H^1(I,U_k)$
such that $\tau(a)=q_1$, $\tau(b)\in\gamma_{q_2}(\R)$, and satisfying
$\iip{\dot\tau}Y\equiv-k\,{\mathcal T}_\tau$,
$\iip{\dot\tau}{\dot\tau}\equiv-{\mathcal T}_\tau^2$ for some 
${\mathcal T}_\tau\in\R^+$.

Observe that if $\sigma\in\bpga1$, then, for every $I=[a,b]\subseteq[0,1]$, 
the restriction of $\sigma$ to $I$ is a curve in 
${\mathcal B}^{\scriptscriptstyle{(1)}}_{\sigma(a),\gamma_{\sigma(b)}}(k,I)$.

\begin{deflocmin}\label{thm:deflocmin}
A curve $\sigma\in\bpga1$ is said to be a {\em local minimizer\/} for
the travel time if, for all $0\le a<b\le 1$ such that $b-a$ is sufficiently small,
the restriction of $\sigma$ to the interval $I=[a,b]$ is a minimum point
for the travel time functional in the space 
${\mathcal B}^{\scriptscriptstyle{(1)}}_{\sigma(a),\gamma_{\sigma(b)}}(k,I)$
\end{deflocmin}
Note that Definition~\ref{thm:deflocmin} is essentially the definition
of brachistochrones of energy $k$ given in \cite{GPV}. For curves that
are local minimizers of the travel time, the differential equation (\ref{eq:diffeq})
was established in \cite{GPV} by means of a variational principle, that
we can now state in a more complete form.
\smallskip

We denote by $\Delta$ the smooth distribution on $\M$ given by the orthocomplement
of the vector field $Y$. Observe that, since $Y$ is timelike, the wrong way Schwartz's
inequality implies that $\Delta$ is {\em spacelike}, i.e., the restriction of
the Lorentzian metric $g$ on $\Delta$ is positive definite. 

Let $\psi:\M\times\R\longmapsto\M$ be the flow of $Y$. Recall that, since $Y$ is
Killing, then $\psi(\cdot,t)$ is a local isometry for all $t\in\R$; moreover, it
is easy to see that the distribution
$\Delta$ is $\psi$-invariant, which means that
$\psi_x(q,t_0)(\Delta_q)=\Delta_{\psi(q,t_0)}$, where $\psi_x(q,t_0)$ denotes the
differential of the map $\psi(\cdot,t_0)$ at the point $q$. A function
$\phi:\M\longmapsto\R$ is said to be $Y$-invariant if it is constant along the
flow lines of $Y$; if $\phi$ is $C^1$, this amounts to saying that $\iip
Y{\nabla\phi}\equiv0$.

We define $\opga1(\Delta)$ to be the subset of $\opga1$ consisting of curves 
with tangent vector at each point lying in $\Delta$:
\begin{equation}\label{eq:defopgdelta}
\opga1(\Delta)=\Big\{w\in\opga1:\dot w(t)\in\Delta_{w(t)},\;\forall\,t\in[0,1]\Big\}.
\end{equation} 
Using the language of sub-Riemannian geometry, we will call {\em horizontal\/}
the curves in $\opga1(\Delta)$.

By the same arguments of Proposition~\ref{thm:regolaritabpg}, one checks immediately
that, since $\iip YY$ is never vanishing, $\opga1(\Delta)$ is a smooth submanifold
of
$\opga1$, and that, for
$w\in\opga1(\Delta)$, the tangent space $T_w\opga1(\Delta)$ is given by:
\begin{equation}\label{eq:tanspdelta}
T_w\opga1(\Delta)=\Big\{V\in T_w\opga1:\iip{\nabla_{\dot w}V}Y-\iip V{\nabla_{\dot
w}Y}=0\Big\}.
\end{equation}
It will also be useful, as in the case of the spaces $\bpg$ and $\Bp$ (see formula~%
(\ref{eq:defB}), to introduce the spaces $\Opa1$ and $\Opa1(\Delta)$, by:
\begin{equation}\label{eq:defOp}
\Opa1=\bigcup_{\gamma\subset U_k}\opga1,\quad\text{and}\quad\Opa1(\Delta)=
\bigcup_{\gamma\subset U_k}\opga1(\Delta).
\end{equation}
We single out the following simple fact:
\begin{stessipunti}\label{thm:stessipunti}
Let $\phi$ be a smooth $Y$-invariant positive function. Then, the functional 
\begin{equation}\label{eq:functphi}
E_\phi(w)=
\frac12\int_0^1\phi(w)\,\rip{\dot w}{\dot w}\,{\rm d}t
\end{equation} 
on $\opga1$ and
its restriction to $\opga1(\Delta)$ have the same critical points.
These critical points are geodesics in $\M$ with respect to the Riemannian metric
$\phi\cdot\gr$ that join $p$ and $\gamma$ and that are orthogonal to $\gamma$. 
\end{stessipunti}
\begin{proof}
The critical points of $E_\phi$ in $\opga1$ are precisely the geodesics
in $\M$ with respect to $\phi\cdot\gr$ that  join $p$ and $\gamma$ and 
that are orthogonal to $\gamma$, i.e., $\rip{\dot w(1)}{Y(w(1))}=0$. 
Since $\phi$ is $Y$-invariant, then 
$Y$ is Killing in the metric $\phi\cdot\gr$, thus, for every such geodesic $w$,
the quantity $\rip{\dot w}Y=$ is constant. Hence $\rip{\dot w}Y\equiv0$ and
$w$ is horizontal. Therefore, the critical points of $E_\phi$ on $\opga1$ belong
to $\opga1(\Delta)$, and clearly they are critical points of the restriction
of $E_\phi$ to $\opga1(\Delta)$.

Conversely, if $w$ is a critical point of the restriction of $E_\phi$ to
$\opga1(\Delta)$, then the Gateaux derivative ${\rm d}E_\phi(w)[V]$ vanishes
for all $V\in T_w\opga1(\Delta)$. Let's define:
\begin{equation}\label{eq:defmT}
{\mathbf T}_w=\Big\{V\in T_w\opga1:V=\tau\cdot Y,\;\text{for some}\ \tau\in
H^1(I,\R)\ 
\text{with}\ \tau(0)=\tau(1)=0\Big\}.
\end{equation}
Since $Y$ is Killing in the metric $\gr$, an easy calculation shows that for all
$w\in\opga1$, the Gateaux derivative ${\rm d}E_\phi(w)[V]$ vanishes for all
$V\in{\mathbf T}_w$. 

Moreover, for all $w\in\opga1(\Delta)$ it is (see \cite{GPV}):
\[T_w\opga1={\mathbf T}_w+T_w\opga1(\Delta),\]
which implies ${\rm d}E_\phi(w)[V]=0$ for all $V\in T_w\opga1$. This concludes the proof.
\end{proof}
The functional $E_\phi$ of (\ref{eq:functphi}) is called the {\em energy\/} functional
relative to the metric $\phi\cdot\gr$.
The critical points of $E_\phi$ in $\opga1$ (or equivalently in $\opga1(\Delta)$,
see~\cite{GPV}) will
be called {\em horizontal geodesics\/} between $p$ and $\gamma$ with respect to the
Riemannian metric $\phi\cdot\gr$.\medskip

In order to state properly our variational principle, we introduce an
operator ${\mathcal D}$ that {\em deforms\/} curves in $\opg$ into horizontal curves
using the flow of $Y$. 

Let $\mathcal D$ be the map:
\begin{equation}\label{eq:defdeformation}
{\mathcal D}:\opga1\longmapsto\opga1(\Delta)
\end{equation}
defined by ${\mathcal D}(\sigma)=w$, where
\begin{equation}\label{eq:defD}
w(t)=\psi(\sigma(t),{\tsig}(t)),
\end{equation}
and ${\tsig}$ is the unique solution on $[0,1]$ of the Cauchy problem:
\begin{equation}\label{eq:deftausigma}
{\tsig}'=-\frac{\iip{\dot\sigma}Y}{\iip YY},
\qquad {\tsig}(0)=0.
\end{equation}
Using the Killing property of $Y$ it is easily checked that ${\mathcal D}$
is well defined, i.e., the maximal solution of (\ref{eq:deftausigma}) is defined 
on the entire interval $[0,1]$ and the corresponding curve $w$ given by (\ref{eq:defD})
is horizontal. Namely, using the fact that the differential
${\rm d}_x\psi$ is an isometry, we compute easily:
\begin{equation}\label{eq:maserve}
\begin{split}
\iip{\dot w}{Y(w)}&=\iip{{\rm
d}_x\psi(\sigma,{\tsig})[\dot\sigma]}{Y(\psi(\sigma,{\tsig}))}
+ {\tsig}'\iip{Y(\psi(\sigma,{\tsig}))}{Y(\psi(\sigma,{\tsig}))}
=\\ &=\iip{\dot\sigma}{Y(\sigma)}+{\tsig}'\iip{Y(\sigma)}{Y(\sigma)}=0.
\end{split}
\end{equation}

Observe that, if $\sigma\in\bpga1$, then (\ref{eq:deftausigma}) 
gives:
\begin{equation}\label{eq:nuovatsigma}
{\tsig}'=\frac{k\,{\Tsigma}}{\iip YY}.
\end{equation}

In Section~\ref{sec:morse} we will need to use the differential ${\rm d}{\mathcal D}$
of $\mathcal D$ on brachistochrones; the differentiability of $\mathcal D$ and a formula
for ${\rm d}{\mathcal D}$ is established in the next:
\begin{smoothnessD}\label{thm:smoothnessD}
The map $\mathcal D$ is smooth around the regular points
of $\bpga1$. If $\sigma$ is a curve
of class $C^1$ in $\bpga1$ and $\zeta\in T_\sigma\bpga1$, the
Gateaux derivative ${\rm d}{\mathcal D}(\sigma)[\zeta]$ is given by:
\begin{equation}\label{eq:derivataD}
{\rm d}{\mathcal D}(\sigma)[\zeta]={\rm d}_x\psi(\sigma,{\tsig})\left[\zeta+\tau_\zeta\cdot%
Y(\sigma)\right],
\end{equation}
where $\tau_\zeta:[0,1]\longmapsto\R$
is the function:
\begin{equation}\label{eq:tauzetacompl}
\tau_\zeta(t)=-\int_0^t\frac{C_\zeta\,\iip YY+2k\,{\Tsigma}\iip{\nabla_\zeta
Y}Y}{\iip YY^2}\;{\rm d}r,
\end{equation}
where $C_\zeta$ is the constant
$\iip{\nabla_{\dot\sigma}\zeta}Y-\iip\zeta{\nabla_{\dot\sigma}Y}$. 
In particular, if
$\sigma$ is a brachistochrone, then $\tau_\zeta$ takes the following form:
\begin{equation}\label{eq:tauzeta}
\tau_\zeta(t)=-2k\,{\Tsigma}\int_0^t\frac{\iip{\nabla_\zeta Y}Y}{\iip YY^2}\;{\rm
d}r.
\end{equation}
\end{smoothnessD}
\begin{proof}
The smooth dependence on
$\sigma$ of the solution
${\tsig}$ of (\ref{eq:deftausigma}) proves that $\mathcal D$ is a smooth map.
Formulas (\ref{eq:derivataD}), (\ref{eq:tauzetacompl}) and (\ref{eq:tauzeta})
are easily obtained by differentiating (\ref{eq:defD}) using (\ref{eq:derivataT}), 
and keeping in mind that
${\rm d}_x\psi(\sigma,{\tsig})[Y(\sigma)]=Y(\psi(\sigma,{\tsig}))$.
In particular, formula (\ref{eq:tauzeta}) follows immediately from (\ref{eq:tauzetacompl})
and Corollary~\ref{thm:carbrach}.
\end{proof}
Observe that formula (\ref{eq:defD}) allows to extend the definition of
the map $\mathcal D$ to the space $\Bpa1$ and with values in $\Opa1$;
these spaces were defined in (\ref{eq:defB}) and (\ref{eq:defOp}).
Obviously, Proposition~\ref{thm:smoothnessD} remains true for the extension.
\smallskip

Now everything is ready to state and prove the following:
\begin{first}[First Variational Principle for Brachistochrones]\label{thm:first}
\hfill\break
Let $\sigma\in\bpga1$ be fixed. The following are equivalent:
\begin{enumerate}
\item\label{itm:uno} $\sigma$ is a brachistochrone of energy $k$ between $p$ and $\gamma$;
\item\label{itm:due} $\sigma$ is a local minimizer for the travel time;
\item\label{itm:tre} $w={\mathcal D}(\sigma)\in\opga1(\Delta)$ is a horizontal geodesic between
$p$ and
$\gamma$ with respect to the Riemannian metric $\phi_k\cdot\gr$, where:
\begin{equation}\label{eq:defphik}
\phi_k=-\frac{\iip YY}{k^2+\iip YY}.
\end{equation}
\end{enumerate}
Moreover, if one of the conditions above is satisfied, then
$E_{\phi_k}(w)=\frac12{\Tsigma}^2$, where $E_{\phi_k}$ is the energy functional
relative to the metric $\phi_k\cdot\gr$, given by:
\begin{equation}\label{eq:defEphik}
\phantom{\quad\forall\,w\in\opga1.}
E_{\phi_k}(w)=\frac12\int_0^1\phi_k(w)\rip{\dot w}{\dot w}\;{\rm d}t,
\quad\forall\,w\in\opga1.
\end{equation}
\end{first}
\begin{proof}
The equivalence of conditions~\ref{itm:uno} and \ref{itm:due} follows from the
fact that the brachistochrones of energy $k$ between $p$ and $\gamma$ and the local
minimizers for the travel time are characterized by the same differential
equation (see Proposition~\ref{thm:diffeqbrach} and Ref.~\cite[Definition~1.1, %
Corollary~3.2]{GPV}).

The equivalence of condition~\ref{itm:due} and \ref{itm:tre} is based on the
fact that, for $\sigma\in\bpga1$ and $w={\mathcal D}(\sigma)$, using
(\ref{eq:conssigma}),  (\ref{eq:defD}) and (\ref{eq:deftausigma}), one computes easily:
\begin{equation}\label{eq:equiv23}
\begin{split}
&\phi_k(w)\iip{\dot w}{\dot w}= \\
&\quad=-\frac{\iip{Y(\sigma)}{Y(\sigma)}}{k^2%
+\iip{Y(\sigma)}{Y(\sigma)}}\,\left(\iip{\dot\sigma}{\dot\sigma}+
2\iip{\dot\sigma}{Y(\sigma)}\,{\tsig}'+({\tsig}')^2
\iip{Y(\sigma)}{Y(\sigma)}\right)={\Tsigma}^2.
\end{split}
\end{equation}
Here we have used the facts that $\iip YY$ is constant along the flow lines of $Y$,
that $\psi(\cdot,t_0)$ is an isometry for all $t_0\in\R$ and the conservation law
of the energy of the Riemannian geodesics. 
Observe that, since $Y$ is Killing in the metric $\phi_k\cdot\gr$, then
a critical point of $E_{\phi_k}$ in $\opga1(\Delta)$ is indeed
a geodesic with respect to $\phi_k\cdot\gr$ (see~\cite{GPV}).
It follows that the quantity $\phi_k(w)\iip{\dot w}{\dot w}$ is constant
along each horizontal geodesic $w$.
 
Recalling (\ref{eq:F}), integrating formula (\ref{eq:equiv23}) yields:
\begin{equation}\label{eq:relFEk}
F=-E_{\phi_k}\circ{\mathcal D}.
\end{equation}

From (\ref{eq:equiv23}) it follows that $\sigma$ is a local minimizer for the
travel time if and only if $w$ is a local minimizer for the energy functional
$E_{\phi_k}$ in $\opga1(\Delta)$, i.e., if and only if $w$ is a horizontal geodesic
between $p$ and $\gamma$ with respect to $\phi_k\cdot\gr$.

The last statement of the thesis follows easily by integrating (\ref{eq:equiv23}) over
$[0,1]$.
\end{proof}
The result of Proposition~\ref{thm:first} remains true for brachistochrones and
horizontal geodesics with free endpoints in $U_k$. The correct statement of
this fact is obtained by replacing the spaces $\bpga1$ and $\opga1(\Delta)$ respectively
with $\Bpa1$ and $\Opa1(\Delta)$, which were defined in formulas (\ref{eq:defB}) and
(\ref{eq:defOp}).
\end{section}

\begin{section}{The Second Variation of the Travel Time}
\label{sec:second}
In this section we want to investigate the problem of whether a given stationary
point $\sigma$ in $\bpga1$ for the travel time functional is a local minimum,
maximum or a saddle point. To this aim, we need a second order variational
formula for our variational problem.

In the first part of the Section we will discuss the abstract problem 
of relating the Hessians of smooth functions on Banach manifolds
that are intertwined by a Banach manifold morphism; 
then we use the first part to determine the relation between the
Hessian of the travel time $T$ 
and the Hessian of the Riemannian action $E_{\phi_k}$.
\smallskip

Let $M$ be a Banach manifold
and $f:M\longmapsto\R$ be a smooth map. If $x_0\in M$ is a critical point 
for $f$, i.e., ${\rm d}f(x_0)=0$, then it makes sense to define the
{\em Hessian\/}  of $f$ at $x_0$, denoted by $H^f(x_0)$, which is a continuous
symmetric  bilinear form on $T_{x_0}M$, in the following way.

Choose a coordinate system around $x_0$, $\phi:U\subset M\longmapsto U_0\subset E$,
where $E$ is some Banach space. Define:
\begin{equation}\label{eq:defabstrhess}
H^f(x_0)[v,w]={\rm d}^2(f\circ\phi^{-1})(\phi(x_0))[{\rm d}\phi(x_0)[v], {\rm d}\phi(x_0)[w]],
\end{equation}
for $v,w\in T_{x_0}M$. Using the fact that $x_0$ is critical for $f$, it is easy to see
that this definition will not depend on the chart $(U,\phi)$. 
Indeed, is is easily seen that for every smooth curve $s\longmapsto y_s\in M$
such that $y_0=x_0$ and $y'_0=v\in T_{x_0}M$, we have:
\begin{equation}\label{eq:forhess}
\frac{{\rm d}^2(f(y_s))}{{\rm d}s^2}\,\big\vert_{s=0}=H^f(x_0)[v,v].
\end{equation}

Formula (\ref{eq:forhess}) provides a simple way of computing $H^f(x_0)[v,v]$;
the general formula for $H^f(x_0)[v,w]$ is easily obtained by polarization.

We now prove the following:
\begin{abstract1}\label{thm:abstract1}
Let $M$ and $N$ be Banach manifolds and ${\mathcal D}:M\longmapsto N$ be a smooth map;
let $f:N\longmapsto\R$ be a smooth function. If $x_0\in M$ is such that
${\mathcal D}(x_0)$ a critical point for $f$, then $x_0$ is a critical point for 
$f\circ{\mathcal D}$, and the Hessians $H^f({{\mathcal D}(x_0)})$
and $H^{f\circ{\mathcal D}}(x_0)$ are related by:
\begin{equation}\label{eq:dan**}
H^f({{\mathcal D}(x_0)})\big[{\rm d}{\mathcal D}(x_0)[v],{\rm d}{\mathcal D}(x_0)[w]\big]=
H^{f\circ{\mathcal D}}(x_0)[v,w],
\end{equation}
for all $v,w\in T_{x_0}M$.  
\end{abstract1}
\begin{proof}
Since both sides of (\ref{eq:dan**}) are symmetric, it suffices to prove
the equality in the case $v=w$. Let $y(s)$, $s\in\,]-\varepsilon,\varepsilon\,[$ be a 
smooth curve in $M$ such that $y(0)=x_0$ and $y'(0)=v$. Then, clearly,
$\tilde y={\mathcal D}\circ y$ is a smooth curve in $N$ such that $\tilde y(0)=
{\mathcal D}(x_0)$ and $\tilde y'(0)={\rm d}{\mathcal D}(x_0)[v]$. Using (\ref{eq:forhess}),
we have:
\[
H^f({\mathcal D}(x_0))\big[{\rm d}{\mathcal D}(x_0)[v],{\rm d}{\mathcal D}(x_0)[v]\big]
=\frac{{\rm d}^2(f\circ{\mathcal D}\circ y)}{{\rm d}s^2}\,\big\vert_{s=0}=
H^{f\circ{\mathcal D}}(x_0)[v,v],
\]
which concludes the proof.
\end{proof}

From Lemma~\ref{thm:abstract1} and formula~(\ref{eq:relFEk}),
setting $f=E_{\phi_k}$, it follows immediately:
\begin{princsec}[Second order variational principle for brachistochrones]\label{thm:princsec}
\hfill\break 
Let $\sigma\in\bpga1$ be a brachistochrone and $w={\mathcal D}(\sigma)$. Then, for
all $\zeta_1,\zeta_2\in T_\sigma\bpga1$, we have:
\begin{equation}\label{eq:ughessiani}
H^F(\sigma)[\zeta_1,\zeta_2]=-H^{E_{\phi_k}}(w)\big[{\rm d}{\mathcal D}(w)[\zeta_1],
{\rm d}{\mathcal D}(w)[\zeta_2]\big].\qed
\end{equation}
\end{princsec}
From (\ref{eq:F}) and
(\ref{eq:forhess}) we obtain easily:
\begin{equation}\label{eq:relhessiani}
H^F(\sigma)=-{\Tsigma}\cdot H^T(\sigma)
\end{equation}
for all brachistochrone $\sigma\in\bpga1$.

From (\ref{eq:ughessiani}) and (\ref{eq:relhessiani})  we obtain also:
\begin{equation}\label{eq:ughessmigl}
H^T(\sigma)[\zeta_1,\zeta_2]={\Tsigma}^{-1}\cdot H^{E_{\phi_k}}(w)
\big[{\rm d}{\mathcal D}(w)[\zeta_1],
{\rm d}{\mathcal D}(w)[\zeta_2]\big],
\end{equation}
for all brachistochrone $\sigma\in\bpga1$ and all $\zeta_1,\zeta_2\in T_\sigma\bpga1$.
\end{section}
\begin{section}{The Riemannian Morse Index Theorem}
\label{sec:index}
For Riemannian geodesics, the classical Morse Index Theorem
(see References~\cite{A, Bol, Kal, M} for the different versions of this
Theorem) relates the index of
the action functional  with some geometrical properties of the geodesic.
The main ingredients for the theory are given by the curvature tensor
of the metric and the concepts of {\em Jacobi fields\/} and {\em conjugate\/} or 
{\em focal\/} points along a geodesic. 

In view to applications to the brachistochrone problem, in this section
we quickly review some known results about the Morse Index Theorem
for Riemannian geo\-desics joining a curve with a point, as presented, for instance,
in \cite{Kal}.  Then, 
we prove a different version of this theorem in the case of an orthogonal
geodesic between the integral line of a Killing vector field
and a point. 

In order to simplify the formulas, in this
section we interchange the role of $p$ and
$\gamma$, that is, we consider curves starting at 
the curve $\gamma$ and ending at the point $p$.
Clearly, the final results (Theorems~\ref{thm:MorseRiem}
and \ref{thm:secmorseindexth}) will not be affected by this change.
Moreover, all  the results and the formulas of the previous sections
remain true after changing the variable $t$ with $1-t$
in the interval $[0,1]$, and, in particular, the role
of  the endpoints $t=0$ and $t=1$ will be interchanged.
To avoid confusion, in this section we will use the symbols
$\ogpa1$ and $\ogpa1(\Delta)$ to indicate the spaces
of curves in $U_k$ of class $H^1$ from $\gamma$ to $p$.
If we denote by ${\mathcal O}$ the {\em direction reversing map\/}
for curves $w:[0,1]\longmapsto\M$, i.e.,
\begin{equation}\label{eq:dirrev}
{\mathcal O}(w)(t)=w(1-t),
\end{equation}
then clearly $\ogpa1={\mathcal O}(\opga1)$ and $\ogpa1(\Delta)={\mathcal O}
(\opga1(\Delta))$. Observe that, for all $i\in\N$, the restriction
of ${\mathcal O}$ to the Sobolev manifold $H^i([0,1],\M)$
is smooth, and its differential is formally given by:
\[\phantom{\quad V\in\,H^i([0,1],T\M).}{\rm d}{\mathcal O}[V](t)=V(1-t),\quad
V\in\,H^i([0,1],T\M).\]  
Observe also that the energy functional $E_{\phi_k}$ can be defined
in $\ogpa1$ by the same formula \eqref{eq:defEphik}; obviously,
a curve $w$ is a critical point for $E_{\phi_k}$ in $\opga1$
if and only if ${\mathcal O}(w)$ is a critical point for $E_{\phi_k}$
in $\ogpa1$. In this case, we have:
\begin{equation}\label{eq:servedopo}
\phantom{\forall\,V,W\in T_w}
H^{E_{\phi_k}}(w)[V,W]=H^{E_{\phi_k}}({\mathcal O}(w))[{\rm d}
{\mathcal O}[V],{\rm d}{\mathcal O}[W]],\quad
\forall\,V,W\in T_w\opga1.
\end{equation}

By Lemma~\ref{thm:stessipunti}, we know that the critical points
of the Riemannian energy functional $E_{\phi_k}$ corresponding to
the metric $\phi_k\cdot\gr$ on the spaces $\ogpa1$ and $\ogpa1(\Delta)$ are
the same. However, given a horizontal geodesic $w$ between $p$ and $\gamma$,
the Morse index of $E_{\phi_k}$ at $w$ (see Definition~\ref{thm:defmorseindex})
in the Hilbert manifold $\ogpa1(\Delta)$ may be strictly less then the Morse
index of $E_{\phi_k}$ at $w$ 
in the manifold $\ogpa1$. The purpose of this section is to prove
that the two indices are indeed equal; we accomplish this result by
proving an index theorem for the Morse index $m(w,{E_{\phi_k}})$
restricted the space $T_w\ogpa1(\Delta)^\perp$, defined by:
\begin{equation}\label{eq:tanspdeltaperp}
T_w\ogpa1(\Delta)^\perp=\Big\{V\in T_w\ogpa1(\Delta)\;\big\vert\;
\phi_k(w)\cdot\rip{V}{\dot w}\equiv C_V\ \text{(const.)}\Big\}.
\end{equation}
Observe that $T_w\ogpa1(\Delta)^\perp$ is a (closed) Hilbert subspace
of $T_w\ogpa1(\Delta)$; moreover, if $w$ is a horizontal
geodesic with respect to the metric $\phi_k\cdot\gr$ in $\ogpa1$,
then, for a vector field $V\in T_w\ogpa1(\Delta)$ we have:
\begin{equation}\label{eq:conderivata}
V\in T_w\ogpa1(\Delta)^\perp\quad\iff\quad \rip{\nablak_{\dot w}V}{\dot w}=0,
\end{equation}
where $\nablak$ is the covariant derivative of the
Levi--Civita connection of the Riemannian metric $\phi_k\cdot\gr$.

Indeed, if $V\in T_w\ogpa1(\Delta)^\perp$, then, since $\nablak_{\dot w}\dot w=0$,
it is \[0=\ddt\big[\phi_k(w)\cdot \rip{V}{\dot w}\big]=\phi_k(w)\cdot
\rip{\nablak_{\dot w}V}{\dot w}.\]
On the other hand, if $0=\phi_k(w)\cdot
\rip{\nablak_{\dot w}V}{\dot w}=\ddt\big[\phi_k(w)\cdot \rip{V}{\dot w}\big]$,
the quantity $\phi_k(w)\cdot \rip{V}{\dot w}$ is constant and
(\ref{eq:conderivata}) is proven.

In particular, since $V(1)=0$ (recall that we are considering
curves ending at the fixed point $p$), if $w$ is a
horizontal geodesic and $V\in T_w\ogpa1(\Delta)^\perp$, then $C_V=0$.
Hence, a vector field $V\in T_w\ogpa1(\Delta)$ belongs to $T_w\ogpa1(\Delta)^\perp$
if and only if it is everywhere perpendicular to $w$, which is 
the reason for the notation.

\begin{commentoinu}\label{thm:commentoinu}
From (\ref{eq:conderivata}) it is easy to see that, if we think of
 the elements in $T_w\ogpa1$ as variational vector fields relative
to variations $w_s$ of the horizontal geodesic
$w$, then the condition $V\in
T_w\ogpa1(\Delta)^\perp$ means that, up to infinitesimals of order larger than
$1$, the  curves $w_s$ are horizontal, and they are
parameterized by a constant multiple of 
arclength:
\[\ddso\,\Big[\phi_k(w_s)\rip{\dot w_s}{Y}\Big]=\phi_w(w)\left(\rip{\nablak_{\dot
w}V}{Y}-\rip{V}{\nablak_{\dot w}Y}\right)=0,\]
\[\ddso\,\Big[\phi_k(w_s)\rip{\dot w_s}{\dot w_s}\Big]=
2\phi_k(w)\cdot
\rip{\nablak_{\dot w}V}{\dot w}=0.\]
\end{commentoinu}
We can easily write (\ref{eq:conderivata}) in terms of the Lorentzian
structure, by differentiating the above expression using
the Lorentzian covariant derivative. Given
a horizontal geodesic $w$ and $V\in T_w\ogpa1(\Delta)$, we have
that $V\in T_w\ogpa1(\Delta)^\perp$ if and only if the
following equation holds:
\begin{equation}\label{eq:conderivatabis}
\iip{\nabla\phi_k(w)}V\cdot
\iip{\dot w}{\dot w}+2\,\phi_k(w)\cdot\iip{\nabla_{\dot w}V}{\dot w}=0.
\end{equation}

\smallskip

We recall the basic facts concerning the Morse Index Theorem for Riemannian
geodesics between a point and a curve, as it is presented, for instance,
in Ref.~\cite{Kal}.

Given a horizontal geodesic $w$ in between $p$ and $\gamma$ with respect to the
Riemannian metric $\phi_k\cdot\gr$, let $\nabla^{\{k\}}$ and $R^{\{k\}}$
denote respectively the Levi--Civita connection and the curvature
tensor (chosen with the same sign convention as in (\ref{eq:convention}))
of the metric $\phi_k\cdot\gr$, and  let
${\mathcal J}_w^{\{k\}}$ be the finite dimensional vector space of all the Jacobi fields $J$
along $w$ with respect to $\phi_k\cdot\gr$, i.e., all smooth vector fields satisfying the
second order differential equation:
\begin{equation}\label{eq:eqJacobi}
\nabla^{\{k\}}_{\dot w}\nabla^{\{k\}}_{\dot w}J-R^{\{k\}}(\dot w,J)\,\dot
w=0.\end{equation}
We recall that, in analogy with the Riemannian case, given a submanifold
$\Sigma$ of $\M$ whose tangent bundle $T\Sigma$ is non degenerate, i.e., the
restriction of $g$ to the tangent space $T_q\Sigma$ is non degenerate
for all $q\in\Sigma$, one can define
the {\em second fundamental form\/} $S^\Sigma$ (also known as the {\em shape tensor\/} of
$\Sigma$) as follows.  For each $q\in\Sigma$ and each vector $n\in T_q\Sigma^\perp$, 
the second fundamental form of $\Sigma$ in the direction of $n$ is the bilinear
form
$S^\Sigma_n:T_q\Sigma\times T_q\Sigma\longmapsto\R$ defined by:
\begin{equation}\label{eq:2ff}
S^\Sigma_n(v_1,v_2)=\iip n{\nabla_{v_1}V_2},
\end{equation}
where $V_2$ is any smooth vector field on $\Sigma$ that takes value $v_2$ at $q$.
One can show that $S^\Sigma_n$ is well defined (i.e., formula (\ref{eq:2ff}) does
not indeed depend on the choice of the extension $V_2$ of $v_2$), and it is {\em
symmetric\/} (see for instance \cite{BEE} and \cite{ON}).

In the following, we will denote by $S^\gamma$ the second fundamental
form of the timelike submanifold $\gamma(\R)$ of $\M$.

Let ${\mathcal J}_w^{\{k\}}(\gamma)$ denote the subspace of 
${\mathcal J}_w^{\{k\}}$ consisting of all {\em $\gamma$-Jacobi fields}  
i.e., all the Jacobi fields $J$ along $w$ satisfying:
\begin{enumerate}
\item\label{itm:b1} $J(0)\parallel Y(w(0))$;
\item\label{itm:orto} $\iip{\nabla_{\dot w(0)}J}{Y}+S_{\dot w(0)}^\gamma(J(0),Y)=
\iip{\nabla_{\dot w(0)}J}{Y}+\iip{\dot w(0)}{\nabla_{J(0)}Y}=0$.
\end{enumerate}
Finally, for $t_0\in\,]\,0,1]$, we denote by ${\mathcal J}_w^{\{k\}}(\gamma,t_0)$
the set of $\gamma$-Jacobi fields $J$ along $w$ that vanish at $t_0$:
\begin{enumerate} 
\item[3.] $J(t_0)=0$.
\end{enumerate}

A point $w(t_0)$ along
$w$ is said to be a {\em $\gamma$-focal point\/}
if ${\rm dim}({\mathcal J}_w^{\{k\}}(t_0))>0$; the {\em multiplicity\/} of the
a $\gamma$-focal point $w(t_0)$ is the dimension of ${\mathcal J}_w^{\{k\}}(t_0)$
(which is clearly finite, because the Jacobi fields are solutions of
a second order linear system of differential equations).
\begin{remfocali}\label{thm:remfocali}
It is well known that the set of $\gamma$-focal points
along every Riemannian 
geodesic is discrete, hence there is only a finite number of
$\gamma$-focal points along each compact portion of a geodesic. 
For the reader's convenience, we sketch a simple proof of this fact based
on~\cite[Ex.\ 8, p.\ 299]{ON}). The set of $\gamma$-Jacobi field along a given
geodesic
$w$ has dimension equal to $m={\rm dim}(M)$. If $J_1,J_2,\ldots, J_m$
is a family of linearly independent $\gamma$-Jacobi fields and
$E_1,E_2,\ldots,E_m$ is a parallely transported orthonormal basis
along $w$, then one considers the smooth
function $g(t)={\rm
det}(\iip{J_i(t)}{E_j(t)})$. Using elementary arguments, one proves that $t_0$ is
a zero of order
$d$ for $g$, i.e., $g(t_0)=g'(t_0)=\ldots g^{(d-1)}(t_0)=0$ and
$g^{(d)}(t_0)\ne0$, if and only if $w(t_0)$ is a $\gamma$-focal
point of multiplicity $d$. In particular, the set of $\gamma$-focal
points is discrete, as is the set of simple zeroes of a smooth function.
\end{remfocali}

Equation~(\ref{eq:eqJacobi}) is obtained by linearizing the geodesic equation
in the metric $\phi_k\cdot\gr$; hence, it is satisfied by vector fields
along $w$ that correspond to variations $w_s$, $s\in\,]-\varepsilon,\varepsilon\,[$
for some $\varepsilon>0$,
of $w$ consisting of geodesics. Loosely speaking, the arrow-head of $J$ traces out
infinitesimally close  neighboring geodesics to $w$. 

The condition~\ref{itm:b1} means
that, in a first order approximation, these geodesics start on $\gamma$;
condition~3 means that they pass through $w(t_0)$.
Condition~\ref{itm:orto} means that these geodesics start orthogonally at
$\gamma$; observe that orthogonality to the vector field $Y$ is equivalent in the
three metrics $g$, $\gr$ and
$\phi_k\cdot\gr$, and for this reason it is possible to write this condition
using the Lorentzian Levi--Civita connection $\nabla$ and 
the Lorentzian second fundamental form $S^\gamma$ of $\gamma$.
Using the Riemannian metric $\phi_k\cdot\gr$,
condition~\ref{itm:orto} can also be written as:
\begin{enumerate}
\item[2b.]\indent $\rip{\nablak_{\dot w(0)}J}Y+\rip{\dot w(0)}{\nablak_{J(0)}Y}=0$.
\end{enumerate}

\begin{remarkKillJac}\label{thm:remarkKillJac}
Observe that, since $Y$ is Killing, we obtain
easily that, if $J$ satisfies the differential equation~\ref{eq:eqJacobi},
then the condition $\iip{\nabla_{\dot w}J}{Y}+\iip{\dot w}{\nabla_{J}Y}=0$
is satisfied identically on $[0,1]$ provided that it is satisfied at one
single point $t_0\in[0,1]$. Indeed, using the fact that Killing vector fields
satisfy the Jacobi equation (see~\cite[Lemma~26, p.\ 252]{ON}), it is easy to see
that the quantity
$\iip{\nabla_{\dot w}J}Y+\iip{\dot w}{\nabla_JY}=
\iip{\nabla_{\dot w}J}Y-\iip{J}{\nabla_{\dot w}Y}$ is constant:
\begin{equation}
\begin{split}
\ddt\left(\iip{\nabla_{\dot w}J}Y-\iip{J}{\nabla_{\dot w}Y}\right)&=
\iip{\nabla_{\dot w}^2J}{Y}-\iip J{\nabla_{\dot w}^2 Y}=\\
&=\iip{R(\dot w,J)\,\dot w}Y-\iip J{R(\dot w,Y)\,\dot w}=0,
\end{split}
\end{equation}
where the last equality follows easily from well known symmetry properties
of the curvature tensor $R$.
\end{remarkKillJac}
From Remark~\ref{thm:remarkKillJac} and formula
(\ref{eq:tanspdelta}), we obtain immediately the following characterization
of the $\gamma$-Jacobi fields along a horizontal geodesic $w$:
\begin{cargammaJacobi}\label{thm:cargammaJacobi}
Let $w$ be a horizontal geodesic in $\ogpa1(\Delta)$ and
$W$ a Jacobi field along $w$. Then, $W$ is a $\gamma$-Jacobi field
if and only if $W\in T_w\ogpa1(\Delta)$.\qed
\end{cargammaJacobi}
Given a horizontal geodesic $w$, we denote by $\Ik$ the {\em index form\/}
on $T_w\ogpa1$, or more in general on $T_wH^1([0,1],\M)$,
given by the symmetric bilinear form:
\begin{equation}\label{eq:defindexform}
\Ik(V_1,V_2)=\int_0^1\phi_k(w)\left(%
\rip{\nablak_{\dot w}V_1}{\nablak_{\dot w}V_2}+\rip{\Rk(\dot w,V_1)\,\dot w}{%
V_2}\right)\;{\rm
d}t.
\end{equation}
The symmetry of $\Ik$ follows easily from the symmetry properties
of the curvature tensor $\Rk$; moreover, from the fundamental Lemma
of Calculus of Variations, a simple integration by parts in (\ref{eq:defindexform})
shows that a vector field $W$ along $w$ is a Jacobi field
if and only if
\begin{equation}\label{eq:condJacobi}
\Ik(W,V)=0
\end{equation}
for all smooth vector field $V$ along $w$ such that $V(0)=V(1)=0$.\smallskip

We recall the definition of the Morse index at a critical point of a 
$C^2$-functional on a Hilbert manifold:
\begin{defmorseindex}\label{thm:defmorseindex}
Let $M$ be a Hilbert manifold, $f:M\longmapsto\R$ be a map of class $C^2$
$x_0$ a critical point for $f$ in $M$ and $X$ a Hilbert
subspace of $T_{x_0}M$. The Morse index $m(x_0,f, X)$ of $f$ at $x_0$ in $X$
is the dimension of a maximal subspace of $X$ on which the Hessian
$H^f(x_0)$ is {\em negative\/} definite. Whenever there is no danger of confusion,
we will denote by $m(x_0,f)=m(x_0,f,T_{x_0}M)$ the Morse index of $f$ at $x_0$ in
the entire tangent space $T_{x_0}M$.

The {\em kernel\/} of $H^f(x_0)$, denoted by ${\rm Ker}\left(H^f(x_0)\right)$
is the Hilbert subspace of $T_{x_0}M$ consisting of vectors $X$ such that
$H^f(x_0)[X,Y]=0$ for all $Y\in T_{x_0}M$.
\end{defmorseindex}
Roughly speaking, the Morse index $m(x_0,f)$ gives the number of {\em essentially
different\/} directions in which the value of the functional $f$ increases from
the value $f(x_0)$. Clearly, if $m(x_0,f)=0$, then $x_0$ is a local
maximum for $f$.
\begin{commentoindici}\label{thm:commentoindici}
Observe that, for all subspace $X\subset T_{x_0}M$, we have
\begin{equation}\label{eq:dis*}
m(x_0,f,X)\le m(x_0,f).
\end{equation} 
On the other hand, suppose that $X$ is a closed subspace of $T_{x_0}M$
and that the restriction of $H^f(x_0)$ to $X$ is nondegenerate.
Let $X_1$ be the orthogonal space to $X$ relatively to the bilinear
form $H^f(x_0)$, which is the closed subspace of $T_{x_0}M$ defined by:
\[X_1=\Big\{V_1\in T_{x_0}M:H^f(x_0)[V,V_1]=0\ \forall\;V\in X\Big\}.\]
If the restriction of $H^f(x_0)$ to $X_1$ is positive semidefinite, then
$m(x_0,f,X)=m(x_0,f)$.
\end{commentoindici}

If $w$ is a horizontal geodesic between $p$ and $\gamma$ with respect to
the Riemannian metric $\phi_k\cdot\gr$, or equivalently, $w$ is a critical
point for $E_{\phi_k}$ in $\ogpa1$, then the Hessian $H^{E_{\phi_k}}(w)$ is 
computed easily in terms of the metric $\phi_k\cdot\gr$ as:
\begin{equation}\label{eq:hessEphikgr}
H^{E_{\phi_k}}(w)[V,V]=\Ik(V,V)-\phi_k(w(0))\rip{\nablak_{V(0)}V}{\dot w(0)}.
\end{equation}
Since $V(0)$ is tangent to the curve
$\gamma$ and $\dot w(0)$ is orthogonal to $\gamma$, then the term
\[\phi_k(w(0))\rip{\nablak_{V(0)}V}{\dot w(0)}\] is tensorial in $V$,
i.e., it only depends on the value $V(0)$. This is precisely the
second fundamental form of the curve $\gamma$ in the direction of
the normal vector $\dot w(0)$ with respect to the metric $\phi_k\cdot\gr$.

We can give a different expression of the Hessian $H^{E_{\phi_k}}(w)$  in terms
of the Lorentzian metric $g$.  This is done by direct computation in the following:
\begin{hessianoE}\label{thm:hessianoE}
Let $w\in\opga1(\Delta)$ be a horizontal geodesic between $p$ and $\gamma$ with respect
to the Riemannian metric $\phi_k\cdot\gr$. Then, the Hessian $H^{E_{\phi_k}}(w)$ 
is given by the following symmetric bilinear map on $T_w\opga1(\Delta)$:
\begin{eqnarray}\label{eq:hessianoE}
H^{E_{\phi_k}}(w)[V,V]&=&\int_0^1\phi_k(w)\,\Big[
\iip{\nabla_{\dot w}V}{\nabla_{\dot w}V}+\iip{R(V,\dot w)\,V}{\dot w}\Big]\;{\rm d}t+
\nonumber\\+&&\!\!\!\!\!\!\!\!\!\int_0^1
\Big[2\iip{\nabla\phi_k(w)}V\,\iip{\nabla_{\dot w}V}{\dot w}+\frac12\iip{H^{\phi_k}(w)V}V
\iip{\dot w}{\dot w}
\Big]\;{\rm d}t+\\
&+& \phi(w(1))\cdot S_{\dot w(1)}^\gamma \big(V(1),V(1)\big).
\nonumber
\end{eqnarray}
\end{hessianoE}
\begin{proof}
The geodesic equation for the metric $\phi_k\cdot\gr$ is easily computed
as the Euler--Lagrange equation for the functional $E_{\phi_k}$, and it
is given by:
\begin{equation}\label{eq:her1a}
\nabla_{\dot w}\Big[\phi_k(w)\,\dot w\Big]=\frac12\,\nabla\phi_k(w)\,\iip{\dot w}{\dot w}.
\end{equation}
In analogy with the proof of Proposition~\ref{thm:hessianoF}, let $V$ be a fixed
vector field in $T_w\opga1(\Delta)$ and let $w_s$ denote a
variation of $w$ in $\opga1(\Delta)$ such that $V=\dds\,\big\vert_{s=0}w_s$.

Then, we compute as follows:
\begin{eqnarray}\label{eq:her1b}
H^{E_{\phi_k}}(w)[V,V]&=&\frac{{\rm d}^2}{{\rm d}s^2}\,\Big\vert_{s=0}E_{\phi_k}(w_s)=
\nonumber\\
=&&\!\!\!\!\!\!\!\!\!\!\!\!\int_0^1\left(\frac12\,\frac{{\rm d}^2}{{\rm
d}s^2}\,\Big\vert_{s=0}
\big[\phi_k(w_s)\big]\,\iip{\dot w}{\dot w}+2\iip{\nabla\phi_k(w)}V\,
\iip{\nabla_{\dot w}V}{\dot w}\right)\;{\rm d}t+\\
&+&\int_0^1\left(\phi_k(w)\,\iip{\Dds\,\Ddt\,\dds\,w_s}{\dot w}+
\phi_k(w)\,\iip{\nabla_{\dot w}V}{\nabla_{\dot w}V}\right)\;{\rm d}t.
\nonumber
\end{eqnarray}
Using (\ref{eq:her1a}) and the commutation relations (\ref{eq:commutation}), we have:
\begin{eqnarray}\label{eq:her2a}
\int_0^1 \phi_k(w)\,\iip{\Dds\,\Ddt\,\dds\,w_s}{\dot w}\;{\rm d}t&=&
\int_0^1\phi_k(w)\,\iip{R(V,\dot w)\,V}{\dot w}\;{\rm d}t+\nonumber\\
-\frac12\int_0^1\iip{\Dds\,\dds&&\!\!\!\!\!\!\!\!\!\!\!\!\!\!\!w_s}{\nabla\phi_k(w)}\,\iip{\dot
w}{\dot w}\;{\rm d}t +\phi_k(w)\,\iip{\Dds\,\dds\,w_s}{\dot w}\,\Big\vert_0^1.
\end{eqnarray}
Keeping in mind that $w_s(0)\equiv p$ and
arguing as in the proof of Proposition~\ref{thm:hessianoF}  (see formula~\ref{eq:b5}),
the boundary term in (\ref{eq:her2a}) can be computed as:
\begin{eqnarray}\label{eq:bordoE}
\phi_k(w)\,\iip{\Dds\,\dds\,w_s}{\dot w}\,\Big\vert_0^1&=&
\phi_k(w(1))\frac{\iip{V(1)}{Y(w(1))}}{\iip{Y(w(1))}{Y(w(1))}}\,\iip{\nabla_{V(1)}Y}{%
\dot w(1)}=
\nonumber\\
&=&\phi_k(w(1))\,S^\gamma_{\dot w(1)}\Big(V(1),V(1)\Big).
\end{eqnarray}
Finally, we have:
\begin{equation}\label{eq:her3a}
\int_0^1\frac{{\rm d}^2}{{\rm d}s^2}\Big\vert_{s=0}\Big[\phi_k(w_s)\Big]\,
\iip{\dot w}{\dot w}\;{\rm d}t=
\int_0^1\left[\iip{H^{\phi_k}(w)\,V}{V}+\iip{\nabla\phi_k(w)}{%
\Dds\,\dds\,w_s}\right]\;{\rm d}t.
\end{equation}
Formula (\ref{eq:hessianoE}) follows from (\ref{eq:her1b}), (\ref{eq:her2a}),
(\ref{eq:bordoE}) and (\ref{eq:her3a}).
\end{proof}

Let's now go back to the study of the second variation
of $E_{\phi_k}$ in terms of the Riemannian metric $\phi_k\cdot\gr$.
Using integration by parts in the Index formula (\ref{eq:defindexform}),
it is easy to see that the set of $\gamma$-Jacobi fields
${\mathcal J}_w^{\{k\}}(t_0)$ can be also described as the kernel of the Hessian
$H^{E_{\phi_k}}(w)$ restricted to the interval $[0,t_0]$; in particular:
\begin{equation}\label{eq:nucleoHE}
{\mathcal J}_w^{\{k\}}={\rm Ker}\left(H^{E_{\phi_k}}(w)\right).
\end{equation}

The {\em geometric index\/} $\mu^{\{k\}}(w)$ of the horizontal geodesic
$w$ is defined as the natural number:
\begin{equation}\label{eq:defmoltgeomriem}
\mu^{\{k\}}(w)=\sum_{t_0\in ]0,1]}{\rm dim}\left({\mathcal J}_w^{\{k\}}(t_0)\right).
\end{equation}
Recall from Remark~\ref{thm:remfocali} that the number of $\gamma$-focal
points along $w$ is finite, hence the sum in (\ref{eq:defmoltgeomriem})
is finite.
\smallskip

The Morse Index Theorem says that, if $p$ is not a $\gamma$-focal point
along $w$, the Morse index
$m(w,{E_{\phi_k}})$ of $E_{\phi_k}$ in the space $T_w\ogpa1$  is given by the
number of
$\gamma$-focal points along $w$, counted with multiplicity: 
\begin{MorseRiem}
\label{thm:MorseRiem} 
Let $w$ be a critical point of $E_{\phi_k}$  in $\ogpa1$,
i.e., a geodesic from $\gamma$ to $p$ in the metric
$\phi_k\cdot\gr$ that starts orthogonally to $\gamma$. 
Then, the Morse index
$m(w,{E_{\phi_k}})$ is finite; moreover, if $p$ is not a $\gamma$-focal point
along $w$, we have:
\begin{equation}\label{eq:ugual}
m(w,{E_{\phi_k}})=\mu^{\{k\}}(w).\qed
\end{equation} 
\end{MorseRiem}
Theorem~\ref{thm:MorseRiem} is obtained as a special case of 
\cite[The Index Theorem, p.\
342]{Kal}.
Observe that Theorem~\ref{thm:MorseRiem} holds without any assumption
that $\gamma$ be the integral line of a Killing vector field.

In the rest of this section we will prove that, given
a horizontal geodesic $w$ in $\ogpa1$, then $m(w,{E_{\phi_k}})$
is equal to the Morse index $\bar m(w,{E_{\phi_k}})$ 
of the restriction of the Hessian $H^{E_{\phi_k}}$ on the space
$T_w\ogpa1(\Delta)^\perp$. Observe that, by (\ref{eq:dis*}), we have
\[\bar m(w,{E_{\phi_k}})=
m(w,{E_{\phi_k}},T_w\ogpa1(\Delta)^\perp)\le m(w,{E_{\phi_k}}).\]

The desired result will follow immediately from our next theorem,
that we state in a general form for future reference:
\begin{secmorseindexth}[Second Morse Index Theorem for Horizontal
Geodesics]\label{thm:secmorseindexth}\hfill\break
Let $(\M,\tilde g)$ be a complete Riemannian manifold,
$Y$ a never vanishing complete Killing vector field on $\M$,
$\gamma:\R\longmapsto\M$ an integral curve of $Y$,
and $  p\in \M$ be a point in $ \M\setminus \gamma(\R)$.

Let $\tilde\Delta=Y^\perp$ be the orthogonal distribution to $Y$;
moreover let $\Ogpa1$, $\Ogpa1(\tilde\Delta)$ denote the spaces:
\[\begin{split}
&\Ogpa1=\Big\{w\in H^1([0,1], \M)\;\big\vert\;w(0)\in \gamma(\R),\
w(1)=  p\Big\},\\
&\Ogpa1(\tilde\Delta)=\Big\{w\in\Ogpa1\;\big\vert\;\tilde g(\dot w,Y)\equiv0\Big\};
\end{split}
\] 
and, for $w\in\Ogpa1(\Delta)$, let $T_w\Ogpa1$, $T_w\Ogpa1(\tilde\Delta)$ and
$T_w\Ogpa1(%
\tilde\Delta)^\perp$
be defined in the obvious way (see formulas~(\ref{eq:tansp}), (\ref{eq:tanspdelta})
and (\ref{eq:tanspdeltaperp}) ). 

Let $\tilde E$ denote the energy functional of the metric $\tilde g$
in the space $\Ogpa1$; let $w$ be a critical point of $\tilde E$ in $\Ogpa1$
(or, equivalently, in $\Ogpa1(\tilde\Delta)$), and let $H^{\tilde E}(w)$ be the
Hessian of $\tilde E$ at $w$. 

Then, if $p$ is not a $\gamma$-focal point along $w$,
the three indices are equal:
\begin{equation}\label{eq:treuguali}
m(w,{H^{\tilde E}})=m(w,{H^{\tilde E}}, T_w\Ogpa1(\tilde\Delta))=
m(w,{H^{\tilde E}},T_w\Ogpa1(\tilde\Delta)^\perp).
\end{equation}
\end{secmorseindexth}
\begin{proof}
The condition that the Killing vector field $Y$ is never vanishing
is needed to prove that the space $\Omega_{p,\gamma}^{(1)}(\tilde\Delta)$ is a
smooth submanifold of $\Omega_{p,\gamma}^{(1)}$ (see
for instance Ref.~\cite{GPV}).

We start proving the second equality in (\ref{eq:treuguali});
we denote by $\tilde\nabla$ and $\tilde R$ respectively
the covariant derivative and the curvature tensor of the Levi--Civita
connection of $\tilde g$; moreover, let $\tilde I$ denote the
index form in $T_w\Ogpa1$ with respect to the metric
$\tilde g$, defined as in (\ref{eq:defindexform}).
Moreover, let $\tilde \mu(w)$ be the geometric index of the geodesic
$w$ in the metric $\tilde g$, defined as in (\ref{eq:defmoltgeomriem}).
\smallskip

Let $T_w\Ogpa1(\tilde\Delta)^\parallel$ be defined by:
\[T_w\Ogpa1(\tilde\Delta)^\parallel=
\Big\{V\in T_w\Ogpa1(\tilde\Delta):V=\lambda\cdot \dot w\ 
\text{for some}\ \lambda\in H^1([0,1],\R)\Big\}.\]
Clearly, $T_w\Ogpa1(\tilde\Delta)=T_w\Ogpa1(\tilde\Delta)^\perp\oplus
T_w\Ogpa1(\tilde\Delta)^\parallel$. Observe that, since $\dot w(0)$ is
orthogonal to $\gamma$, then $V^\parallel(0)=0$ for all
$V^\parallel\in T_w\Ogpa1(\tilde\Delta)^\parallel$.\smallskip

Let $V^\perp\in T_w\Ogpa1(\tilde\Delta)^\perp$ and $V^\parallel\in
T_w\Ogpa1(\tilde\Delta)^\parallel$ be fixed; using the fact
that $V^\parallel(1)=0$, $\tilde g(\tilde R(\dot w,\cdot)\,\dot w,\dot w)=0$  
and that $\tilde g(\tilde\nabla_{\dot w} V^\perp,\dot w)=
\ddt\, \tilde g(V^\perp,\dot w)=0$, it is easy to see that
$H^{\tilde E}(w)[V^\perp, V^\parallel]=0$. 

This implies
that $T_w\Ogpa1(\tilde\Delta)^\perp$ and $T_w\Ogpa1(\tilde\Delta)^\parallel$ are
orthogonal with respect to the bilinear form $H^{\tilde E}(w)$;
in particular, it is:
\[m(w,{H^{\tilde E}},T_w\Ogpa1(\tilde\Delta))=
m(w,{H^{\tilde E}},T_w\Ogpa1(\tilde\Delta)^\perp)+m(w,{H^{\tilde
E}},T_w\Ogpa1(\tilde\Delta)^\parallel).\] It is easy to see that $m(w,{H^{\tilde
E}}, T_w\Ogpa1(\tilde\Delta)^\parallel)=0$; indeed, for
$V^\parallel=\lambda\cdot\dot w$, since $V^\parallel(0)=0$, from
(\ref{eq:hessEphikgr}) we get:
\[
H^{\tilde E}(w)[V,V]=\tilde I(V,V)=\int_0^1\lambda'(t)^2\cdot\tilde g
(\dot w(t),\dot w(t))\;{\rm d}t\ge0,
\]
and since $\tilde g(\dot w,\dot w)>0$ and $\lambda(0)=\lambda(1)=0$,
the above inequality implies that $H^{\tilde E}(w)$ is  positive
definite in $T_w\Ogpa1(\tilde\Delta)^\parallel$, and so $m(w,{H^{\tilde E}},
T_w\Ogpa1(\tilde\Delta)^\parallel)=0$.

This proves the second equality in (\ref{eq:treuguali}).

To prove the first equality, we prove that
\begin{equation}\label{eq:ugindicitilde}
m(w,{H^{\tilde E}}, T_w\Ogpa1(\tilde\Delta))=\tilde\mu(w),
\end{equation}
and the conclusion will follow directly from Theorem~\ref{thm:MorseRiem}.

To this goal, we will use also some abstract arguments in functional
analysis on Hilbert spaces, and we introduce the following notation.

For all $t\in\,]0,1]$, let $(\Tt t,\hip t\cdot\cdot)$ be  a real Hilbert space
with relative inner product, defined by:
\begin{eqnarray}
\label{eq:defTt}
\Tt t=\Big\{\zeta\in H^1([0,t],T\M)\!\!\!\!\!\!\!\!\!\!&&\ \text{vector field
along}\ w\big\vert_{[0,t]}:\nonumber\\ \zeta(0)&&\!\!\!\!\!\!\!\!\!\!\parallel
Y(w(0)),\ 
\zeta(t)=0,\ 
\tilde g(\tilde\nabla_{\dot w}\zeta,Y)-\tilde
g(\zeta,\tilde\nabla_{\dot w}Y)\equiv0\Big\};
\\
\label{eq:defhip}
&&\hip t{\zeta_1}{\zeta_2}=\int_0^t\tilde g(\tilde\nabla_{\dot w}\zeta_1,
\tilde\nabla_{\dot w}\zeta_2)\;{\rm d}r.
\end{eqnarray}
Observe that $\hip t\cdot\cdot$ is non degenerate on $\Tt t$, because
of the condition $\zeta(t)=0$. Let $\Vert\cdot\Vert_t=\hip t\cdot\cdot^{\frac12}$
be the relative norm.

We also define a continuous symmetric bilinear form ${\mathbf H}_t$ on $\Tt t$,
by:
\begin{equation}\label{eq:defHt}
{\mathbf H}_t(\zeta_1,\zeta_2)=\int_0^t\big[\tilde g(\tilde\nabla_{\dot w}\zeta_1,
\tilde\nabla_{\dot w}\zeta_2)+\tilde g(\tilde R(\dot
w,\zeta_1)\,\dot w,\zeta_2))\big]\;{\rm d}r-\tilde g(\dot w(0),\tilde\nabla_%
{\zeta_1(0)}\zeta_2);
\end{equation}
observe that for $t=1$, the Hilbert space $(\Tt t,\hip t\cdot\cdot)$ coincide
with $T_w\Ogpa1(\Delta)$ and the bilinear form ${\mathbf H}_t$ is
precisely the Hessian $H^{\tilde E}(w)$. The symmetry of
${\mathbf H}_t$ is easily obtained using the symmetry of
the curvature tensor and of the second fundamental form of $\gamma$.
Observe also that, since $\zeta_1(0)$ and $\zeta_2(0)$ 
are multiples of the Killing field $Y$, then we have:
\begin{equation}\label{eq:serveperlacompattezza}
\tilde g(\dot w(0),\tilde\nabla_{\zeta_1(0)}\zeta_2)=
-\frac{\tilde g(\zeta_1(0),Y(w(0)))\cdot\tilde g(\zeta_2(0),Y(w(0)))}{\tilde
g(Y(w(0)),Y(w(0)))^2}\,\tilde g(Y(w(0)),\tilde\nabla_{\dot w(0)}Y).
\end{equation}

Using the Riesz representation theorem, we can write ${\mathbf H}_t$ as:
\begin{equation}\label{eq:Riesz}
{\mathbf H}_t(\zeta_1,\zeta_2)=\hip t{{\mathbf L}_t[\zeta_1]}{\zeta_2},
\end{equation}
where ${\mathbf L}_t$ is a self-adjoint linear operator on $\Tt t$.

Comparing (\ref{eq:defhip}) and (\ref{eq:defHt}), we see that we can write:
\begin{equation}\label{eq:I-K}
{\mathbf H}_t={\mathbf I}_t-{\mathbf K}_t,
\end{equation}
where ${\mathbf I}_t$ is the identity on $\Tt t$ and ${\mathbf K}_t$
is the self-adjoint operator on $\Tt t$ defined by:
\begin{equation}\label{eq:defKt}
\hip t{{\mathbf K}_t[\zeta_1]}{\zeta_2}=-\int_0^t\tilde g(\tilde R(\dot
w,\zeta_1)\,\dot w,\zeta_2))\;{\rm d}r+\tilde g(\dot w(0),\tilde\nabla_%
{\zeta_1(0)}\zeta_2).
\end{equation}
Since the inclusions of $H^1([0,t],\R^m)$ into $L^2([0,t],\R^m)$
and into $C^0([0,t],\R^m)$ are compact (see~\cite{Brezis})
and keeping in mind (\ref{eq:serveperlacompattezza}), formula
(\ref{eq:defKt}) tells us that ${\mathbf K}_t$ is a compact operator
for every $t\in ]0,1]$. For all $t$,  
let $\{\lambda_l(t)\}_{k\in\N}$ be the sequence of all the eigenvalues
of ${\mathbf K}_t$; they can be characterized by the following
{\em minimax\/} property:
\[\lambda_l(t)=\max_{{\rm dim}(V)=l}\,\min_{\begin{array}{c}
\xi\in V\\ \Vert\xi\Vert_t=1 \end{array}}\,\hip t{{\mathbf K}_t [\xi]}\xi,\]
where the first maximum is taken over all possible subspaces $V$
of $\Tt t$ having dimension equal to $l$.

By standard arguments (see for instance \cite{Ma}) using the above
characterization of the $\lambda_l$'s one proves that  the map
\[t\longmapsto\lambda_l(t)\]
is continuous. 

We now prove the following claims:
\begin{enumerate}
\item\label{itm:c1} for $t$ small enough, ${\mathbf H}_t$ is positive definite
in $\Tt t$;
\item\label{itm:c2} for all $t$, the kernel of ${\mathbf H}_t$ consists precisely
of all $\gamma$-Jacobi fields along $w\big\vert_{[0,t]}$ that vanish at
$t$;
\item\label{itm:c3} for all $k\in\N$, the map $t\longmapsto\lambda_l(t)$ is
 increasing on $]0,1]$; moreover, if for some $t_0\in\,]0,1[$
it is $\lambda_l(t_0)=1$, then $\lambda_l(t)>1$ for all $t\in\,]t_0,1]$.
\end{enumerate}
Observe that the proof will be concluded once the above claims
are proven. Indeed, by definition, 
a point $w(t_0)$ is a $\gamma$-focal point along $w$ with multiplicity
$d$ if and only if there exists $k>0$ such that $\lambda_l(t_0)=
\lambda_{l+1}(t_0)=\ldots=\lambda_{l+d-1}(t_0)=1$.
From (\ref{eq:I-K}), the Morse index
of $H^{\tilde E}(w)$ on $T_w\Ogpa1(\tilde\Delta)$ is given by the
sum of the dimensions of the eigenspaces of ${\mathbf K}_1$
corresponding to eigenvalues $\lambda_l(1)$ which are strictly
larger than one. By the claims~\ref{itm:c1}, \ref{itm:c2} and
\ref{itm:c3} above, such number is given by the sum of the dimensions
of the kernels of ${\mathbf H}_t$, the sum being taken  over all $t\in\,]0,1]$.
By definition, this number is equal to the geometric index 
$\tilde \mu(w)$ of $w$.\smallskip

Let's prove the claim~\ref{itm:c1}; observe that 
another way of stating this claim is that, for all $k\in\N$
and for $t>0$ small
enough, we have:
\[\lambda_l(t)<1.\]

For $\zeta\in\Tt t$, since
$\zeta(t)=0$, we have:
\[
\tilde g(\zeta(r),\zeta(r))=-2\int_r^t\tilde g(\zeta,\tilde\nabla\zeta)\;
{\rm d}r,
\]
hence, using Schwartz's inequality we have:
\begin{equation}\label{eq:pag2}
\Vert\zeta(r)\,\Vert^2\le 2\int_r^t\Vert\zeta\,\Vert\cdot\Vert\tilde\nabla_{\dot w}
\zeta\,\Vert\;{\rm d}r\le2\left(%
\int_0^t\Vert\zeta\,\Vert^2\;{\rm d}r\right)^{\frac12}\left(
\int_0^t\Vert\tilde\nabla_{\dot w}\zeta\,\Vert^2\;{\rm d}r\right)^{\frac12}.
\end{equation}
Integrating (\ref{eq:pag2}) on $[0,t]$ we obtain:
\[
\int_0^t\Vert\zeta\,\Vert^2\;{\rm d}r\le 2t\left(%
\int_0^t\Vert\zeta\,\Vert^2\;{\rm d}r\right)^{\frac12}\left(
\int_0^t\Vert\tilde\nabla_{\dot w}\zeta\,\Vert^2\;{\rm d}r\right)^{\frac12}
\]
from which we get:
\begin{equation}\label{eq:tipoPoin}
\int_0^tg(\tilde\nabla_{\dot w}\zeta,\tilde\nabla_{\dot w}\zeta)\;{\rm d}r\ge
\frac1{4t^2}\int_0^t\tilde g(\zeta,\zeta)\;{\rm d}r.
\end{equation}
Moreover, another application of Schwartz's inequality gives us:
\[\Vert\zeta(0)\,\Vert\le \int_0^t\Vert\tilde\nabla_{\dot w}\zeta\,
\Vert\;{\rm d}r\le\sqrt t\cdot\left(%
\int_0^t\Vert\tilde\nabla_{\dot w}\zeta\,\Vert^2\;{\rm d}r\right)^{\frac12},\]
from which we obtain the inequality:
\begin{equation}\label{eq:pag3}
\int_0^t\Vert\tilde\nabla_{\dot w}\zeta\,\Vert^2\;{\rm
d}r\ge\frac1t\,\Vert\zeta(0)\,
\Vert^2.
\end{equation}
The proof of claim~\ref{itm:c1} follows immediately from (\ref{eq:defHt}),
(\ref{eq:serveperlacompattezza}), (\ref{eq:tipoPoin}) and (\ref{eq:pag3}).

For the claim~\ref{itm:c2}, we need to show that $\zeta\in\Tt t$
and the equality
${\mathbf H}_t(\zeta,\zeta_1)=0$   holds for all $\zeta_1\in\Tt t$
if and only if $\zeta$ is a $\gamma$-Jacobi field along $w$, i.e.,
if and only if $\zeta$ satisfies the four conditions:
\begin{equation}\label{eq:treequazioni}
\begin{split}
&\tilde\nabla_{\dot w}^2\zeta-\tilde R(\dot w,\zeta)\,\dot w=0,\quad
\zeta(0)\parallel Y(w(0)),\\
&\zeta(t)=0,\quad\text{and}\quad
\tilde g(\tilde\nabla_{\dot w(0)}\zeta,Y(w(0)))+\tilde g(\dot
w(0),\tilde\nabla_{\zeta(0)}Y)=0.
\end{split}
\end{equation}
For the first part of the claim, it suffices to show that
if $\zeta$ is a vector field along $w\big\vert_{[0,t]}$ such that
(\ref{eq:treequazioni}) holds, then $\zeta\in \Tt t$. Indeed, for any vector 
field $\zeta$ that satisfies (\ref{eq:treequazioni}), the equality
${\mathbf H}_t(\zeta,\zeta_1)=0$ is easily verified using integration
by parts. Since $Y$ is Killing and $w$ is a geodesic
in the metric $\tilde g$, then
the quantity $\tilde g(\tilde\nabla_{\dot w}\zeta,Y)+\tilde g(%
\dot w,\tilde\nabla_\zeta Y)$ is constant along $w$, hence
(\ref{eq:treequazioni}) implies that $\zeta\in\Tt t$.

Conversely, let's assume that ${\mathbf H}_t(\zeta,\zeta_1)=0$ for all
$\zeta_1\in\Tt t$.  Let $V$ be an arbitrary smooth vector field
along $w\big\vert_{[0,t]}$ such that $V(0)=V(t)=0$.

Let us set:
\begin{equation}\label{eq:defLV}
L_V=\tilde g(\tilde\nabla_{\dot w}V,Y)+\tilde g(\dot w,\tilde\nabla_VY),
\end{equation}
and
\begin{equation}\label{eq:defzeta1}
\zeta_1=V-\mu\cdot Y,
\end{equation}
where
\begin{equation}\label{eq:defmuzeta1}
\mu(r)=-\int_r^t\frac{L_V}{\tilde g(Y,Y)}\;{\rm d}u.
\end{equation}
From the definition (\ref{eq:defmuzeta1}) of $\mu$ it is easily
checked that $\zeta_1\in\Tt t$; we compute as follows:
\begin{equation}\label{eq:promapag5}
\begin{split}
{\mathbf H}_t(\zeta,\zeta_1)=&\int_0^t\tilde g(
\tilde\nabla_{\dot w}\zeta,\tilde\nabla_{\dot w}V-\mu'\cdot
Y-\mu\cdot\tilde\nabla_{\dot w}Y)\;{\rm d}r\\
&+\int_0^t\tilde g(\tilde R(\dot w,\zeta)\,\dot w,V-\mu\cdot Y)\;{\rm d}r
+\mu(0)\cdot\tilde g(\dot w(0),\tilde\nabla_{\zeta(0)}\left( Y\right))=\\
&={\mathbf H}_t(\zeta,V)-{\mathbf H}_t(\zeta,\mu\cdot Y).
\end{split}
\end{equation}
We now show that ${\mathbf H}_t(\zeta,\mu\cdot Y)=0$.
Since $Y$ is Killing, then its restriction to $w$ is a Jacobi
field (see~\cite[Lemma~26, p.\ 252]{ON}), and so it satisfies:
\begin{equation}\label{eq:YKilJac}
\tilde\nabla_{\dot w}^2Y=\tilde R(\dot w,Y)\,\dot w.
\end{equation}
Integration by parts and (\ref{eq:YKilJac}) yield:
\begin{equation}\label{eq:contaccio}
\begin{split}
\int_0^t\mu\cdot\tilde g(&\tilde\nabla_{\dot w}\zeta,\tilde\nabla_{\dot w}Y)\;
{\rm d}r=\mu\cdot\tilde g(\zeta,\tilde\nabla_{\dot w}Y)\big\vert_0^t-
\int_0^t\tilde g(\zeta,\mu'\cdot\tilde\nabla_{\dot w}Y+\mu\cdot
\tilde\nabla_{\dot w}^2Y)\;{\rm d}r=\\
&=\mu(0)\cdot\tilde g(\dot w(0),\tilde\nabla_{\zeta(0)}Y)-\int_0^t\Big[
\mu'\cdot\tilde g(\zeta,\tilde\nabla_{\dot w}Y)+\mu\cdot
\tilde g(\zeta,\tilde R(\dot w,Y)\,\dot w)\Big]\;{\rm d}r,
\end{split}
\end{equation}
where in the last equality we have used the anti-symmetry 
of the map $(a,b)\to\tilde g(a,\tilde\nabla_bY)$.

By the symmetry of the curvature tensor, we have:
\[\tilde g(\tilde R(\dot w,\zeta)\,\dot w, Y)=\tilde g(\zeta,\tilde R(\dot w,Y))\]
hence, we have:
\begin{equation}
{\mathbf H}_t(\zeta,\mu\cdot Y)=\int_0^t\Big[\mu'\cdot
\tilde g(\tilde\nabla_{\dot w}\zeta,Y)-\mu'\cdot\tilde g(\zeta,\tilde\nabla_{\dot
w}Y)\Big]\;{\rm d}r=0,
\end{equation}
because $\zeta\in\Tt t$ (see formula~(\ref{eq:defTt})).

If we use the equality ${\mathbf H}_t(\zeta,%
\zeta_1)=0$ we get:
\begin{equation}\label{eq:ultimapag5}
\begin{split}
0={\mathbf H}_t(\zeta,\zeta_1)&={\mathbf H}_t(\zeta,V) =\int_0^t\Big[
\tilde g(\tilde\nabla_{\dot w}\zeta,\tilde\nabla_{\dot w}V)
+\tilde g(\tilde R(\dot w,\zeta)\,\dot w,V)\Big]\;{\rm
d}r=\\
&=-\int_0^t\tilde g(\tilde\nabla_{\dot w}^2\zeta-
\tilde R(\dot w,\zeta)\,\dot w,V)\;{\rm d}r.
\end{split}
\end{equation}
Since (\ref{eq:ultimapag5}) holds for all smooth vector field
$V$ along $w$ vanishing at the endpoints, the fundamental
lemma of Calculus of Variations tells us that:
\[\tilde\nabla_{\dot w}^2\zeta-
\tilde R(\dot w,\zeta)\,\dot w=0,\]
which is the first condition in (\ref{eq:treequazioni}).
The other three conditions of (\ref{eq:treequazioni}) are satisfied
by any vector field in $\Tt t$, hence claim~\ref{itm:c2} is proven.
\smallskip

Let's go now to the proof of claim~\ref{itm:c3}. Let's fix $0<t_1<t_2$ in $[0,1]$;
we prove first that, for all $l$, we have: 
\begin{equation}\label{eq:aereo2}
\lambda_l(t_1)\le\lambda_l(t_2).
\end{equation}
To this goal, let $l$ be fixed and let
$V_1$ be a $l$-dimensional subspace of $\Tt{t_1}$ such that:
\[\lambda_l(t_1)=\min_{\begin{array}{c}\xi\in
V_1\\\Vert\xi\Vert_{t_1}=1\end{array}} \hip{t_1}{{\mathbf K}_{t_1}[\xi]}\xi.\]
We define a linear and continuous
map $I_{t_1,t_2}:\Tt {t_1}\longmapsto\Tt {t_2}$
given by:
\[I_{t_1,t_2}(\xi)(r)=\left\{\begin{array}{ll}
\xi(r),&\text{if}\ r\le t_1;\\
0,&\text{if}\ r\in\,]t_1,t_2].\end{array}\right.\]
We observe that, with the above definition, $I_{t_1,t_2}(\xi)$
does indeed belong to $\Tt {t_1}$ (see formula~\ref{eq:defTt}));
observe also that $I_{t_1,t_2}$ is an isometry, and in particular
injective. Moreover, the following equality holds trivially:
\begin{equation}\label{eq:manco}
\hip{t_1}{{\mathbf
K}_{t_1}[\xi]}\xi=\hip{t_2}{{\mathbf
K}_{t_2}[I_{t_1,t_2}(\xi)]}{I_{t_1,t_2}(\xi)},\quad\forall\;\xi\in
\Tt{t_1}\ .
\end{equation}
Let $V_2$ be the $l$-dimensional subspace
of $\Tt {t_2}$ defined by:
\[V_2=I_{t_1,t_2}(V_1).\]
Then, by (\ref{eq:manco}), we have:
\begin{equation}\label{eq:1pag8}
\begin{split}
\lambda_l(t_1)&=\min_{\begin{array}{c}\xi\in
V_1\\\Vert\xi\Vert_{t_1}=1\end{array}} \hip{t_1}{{\mathbf K}_{t_1}[\xi]}\xi=
\min_{\begin{array}{c}\eta\in
V_2\\ \Vert\eta\Vert_{t_2}=1\end{array}}
\hip{t_2}{{\mathbf K}_{t_2}[\eta]}\eta\le\\
&\le\max_{{\rm dim}(W)=k}\,\min_{\begin{array}{c}\eta\in
W\\ \Vert\eta\Vert_{t_2}=1\end{array}}
\hip{t_2}{{\mathbf K}_{t_2}[\eta]}\eta=\lambda_l(t_2),
\end{split}
\end{equation}
which proves (\ref{eq:aereo2}).

To prove the second part of claim~\ref{itm:c3}, it suffices to observe
that if $\lambda_l(t_0)=1$ then $w(t_0)$ is a $\gamma$-focal
point along $w$. Since the set of $\gamma$-focal points along $w$ is discrete
(see Remark~\ref{thm:remfocali}),
it follows that, if $\lambda_l(t_0)=1$, then $\lambda_l(t)\ne1$ in a 
neighborhood of $t_0$. Finally, by the monotonicity of $\lambda_l$,
we conclude that $\lambda_l(t)>1$ in $]t_0,1]$, and we are done.
\end{proof}
Theorem~\ref{thm:secmorseindexth} can be applied to the Riemannian
metric $\tilde g=\phi_k\cdot\gr$ defined in $U_k$.
Recalling that $\bar m(w,{E_{\phi_k}})$ denotes the Morse Index
of the restriction of the Hessian $H^{E_{\phi_k}}$ on the space
$T_w\ogpa1(\Delta)^\perp$, we have thus proven the
equality:
\begin{equation}\label{eq:ugualiindici}
\bar m(w,{E_{\phi_k}})= m(w,{E_{\phi_k}}).
\end{equation}
\end{section}
\begin{section}{The Index Theorem for Brachistochrones}\label{sec:morse}
We want to study now the Morse index of the travel
time functional at a given brachistochrone
$\sigma$, which is defined as the index of the symmetric bilinear form
$H^T(\sigma)$ (see Definition~\ref{thm:defmorseindex}). 

\noindent
In this section we extend the classical the Morse Theory for Riemannian geodesics, in order
to obtain a weak version of the Morse Index Theorem for brachistochrones
(Theorem~\ref{thm:finaleMorse}),
by introducing the concepts of b-Jacobi fields and b-focal points along a
brachistochrone $\sigma$ (see Definitions~\ref{thm:defbJacobi} and \ref{thm:defbfoc}
below).

\smallskip

We now begin with the study of the Hessian of the travel time functional.

Let $\sigma\in\bpga1$ be a brachistochrone, since ${\Tsigma}>0$, formula
(\ref{eq:relhessiani}) tells us that:
\begin{equation}\label{eq:indker}
m(\sigma,T)=m(\sigma,{-F}),\quad\text{and}\quad{\rm Ker}\left(H^T(\sigma)\right)=
{\rm Ker}\left(H^F(\sigma)\right).
\end{equation}

We emphasize that from
now on we will consider  brachistochrone curves whose endpoints may vary in the
open set $U_k$, whereas the value of their energy constant $k$ is a {\em fixed}
positive number. For the sake of shortness, when speaking of brachistochrones
we will omit to specify the value of their energy constant without danger of
confusion.

\smallskip

In this section and in the rest of the paper
we will be speaking of {\em variations\/} of a
given curve in some fixed space, 
which will be a family of curves {\em of the same type}, in a
sense that will be clarified in the different situations, parameterized
by a suitable variable, denoted by $s$. Whenever not specified, we
will tacitly assume that $s$ varies in an interval of the form
$]-\varepsilon,\varepsilon\,[$ for some $\varepsilon>0$.
A formal definition of smooth variation of a given curve $z\in\opga1$
is given in Appendix~\ref{sec:explicit} (Definition~\ref{thm:defvariazione}).

We also warn the reader that, in the course of the section, we will 
switch back and forth among the three Hessians $H^T$, $H^F$ and $H^{E_{\phi_k}}$,
keeping in mind the basic relations among them given by formulas~(\ref{eq:ughessiani})
and (\ref{eq:ughessmigl}).\smallskip

We mimic the classical Morse theory 
and we proceed as follows.

Let $\sigma\in\bpga1$ be a fixed brachistochrone, and, recalling the definition
of the space $\Ba1$ given in (\ref{eq:defB}), we consider a variation $\sigma_s\in\Ba1$
of $\sigma$, depending smoothly on the parameter $s\in\,]-\varepsilon,\varepsilon\,[$
and such that $\sigma_0=\sigma$. Suppose
that each  curve $\sigma_s$ is a brachistochrone of energy $k$ between $\sigma_s(0)$
and $\gamma_{\sigma_s(1)}$, where $\gamma_{\sigma_s(1)}$ is the integral
line of $Y$ passing through $\sigma_s(1)$.

This means that each $\sigma_s$ satisfies the differential
equation (\ref{eq:diffeq}) and with initial tangent vector $\dot\sigma_s(0)$
satisfying the two conditions:
\begin{equation}\label{eq:inconds}
\iip{\dot\sigma_s(0)}{Y(\sigma_s(0))}^2+k^2\iip{\dot\sigma_s(0)}{\dot\sigma_s(0)}=0,
\quad\text{and}\quad \iip{\dot\sigma_s(0)}{Y(\sigma_s(0))}<0.
\end{equation}
\begin{defbJacobi}\label{thm:defbJacobi}
A vector field $V\in T_\sigma\Ba1$ along the brachistochrone $\sigma$ in $\bpga1$
is called a {\em b-Jacobi field\/} if there exists a variation  $\sigma_s\in\Ba1$
of $\sigma$ as above  such that $V=\frac{{\rm d}}{{\rm d}s}\,\Big\vert_{s=0}\sigma_s$.
\end{defbJacobi}
In other words, a b-Jacobi field along $\sigma$ is a variational vector field
corresponding to variations made of brachistochrones with the same energy constant
and, possibly, with different endpoints. By definition, the b-Jacobi fields
are characterized by the property of satisfying the linearized brachistochrone
equation; this second order differential equation has a rather ugly aspect
and it is presented only for the sake of completeness in the following Proposition.
 
\begin{diffeqbKacobi}\label{thm:diffeqbKacobi}
Let $\sigma\in\bpga1$ be a brachistochrone of travel time ${\Tsigma}$
and let $V\in T_\sigma\Ba1$ be
a variational vector field along $\sigma$, with constant
$C_V=\iip{\nabla_{\dot\sigma}V}{Y}-\iip{V}{\nabla_{\dot\sigma}Y}$. If\/
$V$ is a b-Jacobi field then  $V$ satisfies the second order linear 
differential equation:
\begin{eqnarray}\label{eq:diffeqbJacobi}
&&\!\!\!\!\!\!\!\!\nabla_{\dot\sigma}^2V-R(\dot\sigma,V)\,\dot\sigma+\frac{2k\,
{\Tsigma}}{\iip YY^2}
\Big(\nabla_{\dot\sigma}\nabla_V Y-\iip YY R(\dot\sigma,V)\,Y-2
\iip{\nabla_VY}Y \nabla_{\dot\sigma}Y\Big)+\nonumber\\
&& \qquad -2\frac{C_V}{\iip YY}\,\nabla_{\dot\sigma}Y+
\frac{2k^2\dot\sigma-2k\,{\Tsigma} Y}{\iip YY (k^2+\iip YY)}
\Big(\iip{\nabla_{\dot\sigma}\nabla_VY}{Y}+\iip{\nabla_VY}{\nabla_{\dot\sigma}Y}\Big)+\\
&&+\frac{2k^2\dot\sigma-2k\,{\Tsigma} Y}{\iip YY^2 (k^2+\iip
YY)^2}\times\nonumber\\
&&\qquad\qquad\qquad\times\Big(-4\iip{\nabla_{\dot\sigma}Y}{Y}\iip
YY\iip{\nabla_VY}Y-2k^2\iip{%
\nabla_{\dot\sigma}Y}Y\iip{\nabla_VY}Y\Big)+
\nonumber\\
&& \qquad +\frac{2\iip{\nabla_{\dot\sigma}Y}Y}{\iip YY(k^2+\iip YY)}\Big(
C_V Y-k\,{\Tsigma}\nabla_VY+k^2\nabla_{\dot\sigma}V\Big)=0,
\nonumber
\end{eqnarray}
and the initial condition:
\begin{equation}\label{eq:icbJacobi}
-{\Tsigma} C_V+k\iip{\nabla_{\dot\sigma}V(0)}{\dot\sigma(0)}=0.
\end{equation}
\end{diffeqbKacobi}
\begin{proof}
The equation (\ref{eq:diffeqbJacobi}) is obtained by patiently
linearizing the brachistochrone differential equation (\ref{eq:diffeq}), using
the following {\em dictionary}:\smallskip

\begin{itemize}
\item $\displaystyle-k\,\ddso\left({\mathcal T}_{\sigma_s}\right)=C_V$;\smallskip

\item $\displaystyle \Ddso\left(\dot\sigma_s\right)=\nabla_{\dot\sigma}V$;\smallskip

\item $\displaystyle\Ddso
\left(\nabla_{\dot\sigma_s}\dot\sigma_s\right)=\nabla_{\dot\sigma}^2V-R(%
\dot\sigma,V)\,\dot\sigma$;\smallskip

\item $\displaystyle\ddso\left(\iip{Y(\sigma_s)}{Y(\sigma_s)}\right)=2\iip{\nabla_VY}Y$;
\smallskip

\item $\displaystyle\Ddso\left(\nabla_{\dot\sigma_s}Y\right)=
\nabla_{\dot\sigma}\nabla_VY+R(V,\dot\sigma)\,Y=
\nabla_{\dot\sigma}\nabla_VY-R(\dot\sigma,V)\,Y$;\smallskip

\item $\displaystyle\ddso\left(\iip{\nabla_{\dot\sigma_s}Y}Y\right)=
\iip{\nabla_{\dot\sigma}\nabla_VY}{Y}+\iip{\nabla_VY}{\nabla_{\dot\sigma}Y} $;
\smallskip

\item $\displaystyle\ddso\left[\iip{Y(\sigma_s)}{Y(\sigma_s)}
(k^2+\iip{Y(\sigma_s)}{Y(\sigma_s)})\right]=
(4\iip YY+2k^2)\iip{\nabla_VY}Y$.
\end{itemize}
The formulas above are obtained by considering the basic
properties of the Levi--Civita connection and the curvature
tensor of $g$. In particular, in the sixth formula we have
used the fact that $\iip{R(\dot\sigma,Y)\,Y}Y=0$, by the
anti-symmetry  in the last two variables.

The initial condition (\ref{eq:icbJacobi}) is obtained by linearizing the first equation
of formula (\ref{eq:inconds}).
\end{proof}
A partial converse to Proposition~\ref{thm:diffeqbKacobi} is provided by the following
Proposition:
\begin{converse}\label{thm:converse}
Let $\sigma\in\Ba1$ be a brachistochrone and suppose that $V$ is
a smooth vector field along $\sigma$ satisfying the differential equation
(\ref{eq:diffeqbJacobi}), the initial condition
(\ref{eq:icbJacobi}) and  with $V(0)=0$. Then, $V$ is a b-Jacobi field along $\sigma$,
i.e., 
there exists a variation
$\sigma_s$ of $\sigma$ consisting of brachistochrones between $p$ and $\gamma_s$,
$s\in\,]-\varepsilon,\varepsilon\,[$, such that $V=\ddso\sigma_s$.
\end{converse}
\begin{proof}
We use a sort of {\em brachistochrone exponential map}, as follows. 

Given a vector $v_0\in T_p\M$ such that 
\begin{equation}\label{eq:datiiniz}
\iip{v_0}{Y(p)}^2+k^2\iip{v_0}{v_0}=0,\quad\text{%
and}\quad\iip{v_0}{v_0}<0,\end{equation} 
then there exists a unique brachistochrone
$\sigma_{v_0}\in\Bpa1$ 
and such that $\dot\sigma_{v_0}(0)=v_0$. This is obtained
by solving the differential equation~(\ref{eq:diffeq}) with
initial conditions $\sigma(0)=p$ and $\dot\sigma(0)=v_0$.

Moreover, the map $v_0\longmapsto \sigma_{v_0}\in\Bpa1$ is $C^1$, due to the
regular dependence on the data of the solution of the differential 
equation~(\ref{eq:diffeq}).

Let $S\subset T_p\M$ be the set of vectors $v_0$ satisfying the conditions
(\ref{eq:datiiniz}); $S$ is a submanifold of $T_p\M$. Indeed, the condition
$\iip{v_0}{v_0}<0$ is open; moreover, the gradient of the smooth
map $G:T_p\M\ni v_0\longmapsto
\iip{v_0}{Y(p)}^2+k^2\iip{v_0}{v_0}\in\R$ is easily computed as:
\begin{equation}\label{eq:DerG}
G'(v_0)=2\iip{v_0}{Y(p)}\cdot Y(p)+2k^2v_0.
\end{equation}
Multiplying by $Y(p)$ we obtain:
\[\iip{G'(v_0)}{Y(p)}=2\iip{v_0}{Y(p)}\left(\iip{Y(p)}{Y(p)}+k^2\right)\ne0,\]
where the last inequality depends on the fact that both $v_0$ and $Y(p)$ are timelike,
hence $\iip{v_0}{Y(p)}\ne0$, and $\iip{Y(p)}{Y(p)}+k^2>0$ in $U_k$.
This implies that $G'\ne0$, hence $G^{-1}(0)$ is a  smooth
submanifold of $T_p\M$. Clearly, $\dot\sigma(0)\in S$.

Let $v_0(s):]-\varepsilon,\varepsilon\,[\longmapsto S$ be a smooth map
such that $v_0(0)=\dot\sigma(0)\in S$ and $v_0'(0)=\nabla_{\dot\sigma(0)}V$.
Observe that $\nabla_{\dot\sigma(0)}V$ belongs to $T_{\dot\sigma(0)}S$, because,
from (\ref{eq:DerG}), we have:
\[
\iip{G'(\dot\sigma(0))}{\nabla_{\dot\sigma(0)}V}=2
\iip{\dot\sigma(0)}{Y(p)}\iip{Y(p)}{\nabla_{\dot\sigma(0)}V}+2k^2
\iip{\dot\sigma(0)}{\nabla_{\dot\sigma(0)}V}.
\]
Since $V(0)=0$, then $C_V=\iip{Y(p)}{\nabla_{\dot\sigma(0)}V}$, so we have:
\[
\iip{G'(\dot\sigma(0))}{\nabla_{\dot\sigma(0)}V}=2k\left(-{\Tsigma}
C_V+k\iip{\dot\sigma(0)}{\nabla_{\dot\sigma(0)}V}\right)=0,
\] where the 
last equality follows immediately from (\ref{eq:icbJacobi}).
Hence, $\nabla_{\dot\sigma(0)}V\in T_{\dot\sigma(0)}S$
and the curve $v_0(s)$ is well defined. 

Now, for all $s\in\,]-\varepsilon,\varepsilon\,[$,
let $\sigma_s$ be the unique brachistochrone in
$\Bpa1$ satisfying $\dot\sigma_s(0)=v_0(s)$; clearly, $\sigma_0=\sigma$, and
$\sigma_s$ is a smooth variation of $\sigma$. Observe that, since
$\sigma_0$ is defined on the closed interval $[0,1]$, then
we can assume that also $\sigma_s$ is defined on $[0,1]$ for all $s$.

In order to conclude the proof, we need to show that the variational field
$\tilde V=\ddso\sigma_s$ coincides with $V$. 

By Proposition~\ref{thm:diffeqbKacobi}, $\tilde V$ satisfies the
second order differential equation~(\ref{eq:diffeqbJacobi}),
while $V$ satisfies (\ref{eq:diffeqbJacobi}) by assumption, and $\tilde
V(0)=V(0)=0$, because we are fixing the initial point $p$. 
By uniqueness, in order to prove that $\tilde V=V$ along
$\sigma$ it suffices to show that $\nabla_{\dot\sigma(0)}\tilde
V=\nabla_{\dot\sigma(0)}V$.  This is easily established by the following
calculation, that concludes the proof:
\[\nabla_{\dot\sigma(0)}\tilde V=\Ddto\ddso\sigma_s=
\Ddso\ddto\sigma_s=\Ddso\dot\sigma_s(0)=v_0'(0)=\nabla_{\dot\sigma(0)}V.\]
\end{proof}
\begin{corA}\label{thm:corA}
If $\sigma$ is a brachistochrone and $V$ is a b-Jacobi field along $\sigma$
such that $V(0)=0$, then $V\in T_\sigma\Bpa1$.
\end{corA}
\begin{proof}
Following the proof of  Proposition~\ref{thm:converse}, $V$ is the variational vector
field corresponding to a variation $\sigma_s\in\Bpa1$ of $\sigma$.
\end{proof}
In general, it may not be true that a b-Jacobi field $V$
along a brachistochrone $\sigma$ satisfying $V(0)=0$ and $V(1)\in\R\cdot Y(\sigma(1))$
is the variational vector field corresponding to a family of brachistochrones
in $\bpga1$.  However, such vector fields belong to the tangent space
$T_\sigma\bpga1$, and they are in the kernel of the 
Hessian $H^F(\sigma)$:
\begin{corB}\label{thm:corB}
If $\sigma\in \bpga1$ is a brachistochrone and $V$ is a b-Jacobi field along $\sigma$
such that $V(0)=0$ and $V(1)$ is parallel to $Y(\sigma(1))$, then $V\in T_\sigma\bpga1$,
and $V\in {\rm Ker}\left(H^F(\sigma)\right)$. 
\end{corB}
\begin{proof}
By Corollary~\ref{thm:corA}, $V\in T_\sigma\Bpa1$; the first part of the statement
follows immediately by observing that a vector field $V\in T_\sigma\Bpa1$ belongs
to $T_\sigma\bpga1$ if and only if $V(1)$ is parallel to $Y(\sigma(1))$
(see formulas (\ref{eq:tansp}), (\ref{eq:tanspbpg1}) and (\ref{eq:tanspBp1})). 

To prove the second part of the thesis, we need to show that $H^F(\sigma)[V,W]=0$
for all $W\in T_\sigma\bpga1$. By Corollary~\ref{thm:princsec}, we have:
\begin{equation}\label{eq:estrestr}
H^F(\sigma)[V,W]=-H^{E_{\phi_k}}({\mathcal
D}(\sigma))[{\rm d}{\mathcal D}(\sigma)[V],{\rm d}{\mathcal D}(\sigma)[W]],
\end{equation}
hence, to conclude the proof it suffices to show that ${\rm d}{\mathcal D}(\sigma)[V]$
is in the kernel of the Hessian $H^{E_{\phi_k}}({\mathcal D}(\sigma))$.
By (\ref{eq:nucleoHE}), this amounts to proving that $X={\rm d}{\mathcal D}(\sigma)[V]$
is the variational vector field corresponding to a smooth variation $w_s$ of $w={\mathcal
D}(\sigma)$ consisting of horizontal geodesics in the metric $\phi_k\cdot\gr$ between $p$ and
some integral curve $\gamma_s$ of $Y$ lying in $U_k$ 
(recall that a vector field along a geodesic is Jacobi if and only if
it is the variational vector field corresponding to a variation by 
geodesics).

To see this, let $\sigma_s$ be a smooth variation of $\sigma$ consisting of
brachistochrones in $\Bpa1$ between $p$ and some curve $\gamma_s$ in $U_k$, and
with variational vector field $V$. Such a variation exists by 
Proposition~\ref{thm:converse}. 

Then, if we consider the curves $w_s={\mathcal D}(\sigma_s)$, by part~\ref{itm:tre}
of Proposition~\ref{thm:first}, each $w_s$ is a horizontal geodesic between
$p$ and $\gamma_s$; by Proposition~\ref{thm:smoothnessD}, $w_s$ is a smooth
variation of $w$. Finally, we have:
\[\ddso w_s=\ddso{\mathcal D}(\sigma_s)={\rm d}{\mathcal D}(\sigma)[\ddso\sigma_s]=
{\rm d}{\mathcal D}(\sigma)[V]=X,\]
which concludes the proof.
\end{proof}
We will see later (Proposition~\ref{thm:carat}) that the kernel of
the Hessian $H^F(\sigma)$ consists precisely of the b-Jacobi fields
along $\sigma$; this fact can also be checked directly using the
explicit formula for the Hessian $H^F(\sigma)$ given in Appendix~\ref{sec:explicit}
and the Lagrange multipliers technique.
\smallskip

We are now ready to define the notion of a b-focal point along
a brachistochrone. 
\begin{defbfoc}\label{thm:defbfoc} 
Let $\sigma\in\bpga1$ be a brachistochrone. A point $\sigma(t_0)$ of $\sigma$ 
is said to be a {\em b-focal point\/}  
if there exists a non zero b-Jacobi field $V$ along 
$\sigma\big\vert_{[t_0,1]}$ that vanish at $t_0$, that is, a non zero vector field $V$ along
$\sigma$ for which the quantity
$C_V=\iip{\nabla_{\dot\sigma}V}{Y}-\iip{V}{\nabla_{\dot\sigma}Y}$ is constant along $\sigma$,
such that
$V(t_0)=0$, satisfying the differential equation (\ref{eq:diffeqbJacobi}) and the
condition:
\begin{equation}\label{eq:icbJacobit0}
-{\Tsigma} C_V+k\iip{\nabla_{\dot\sigma}V(t_0)}{\dot\sigma(t_0)}=0.
\end{equation}
In the above situation, we will also say that $\sigma(t_0)$ is
{\em b-conjugate\/} to $\sigma(1)=p$ along~$\sigma$.

For every $t_0\in[0,1]$, the set ${\mathcal J}_\sigma(t_0)$
of vector fields $V$ satisfying the above
conditions in the interval $[t_0,1]$ is a vector field; if
$\sigma(t_0)$ is a b-focal point along $\sigma$, then {\em multiplicity\/} 
$\mu_\sigma(t_0)$
of $\sigma(t_0)$ is the dimension of ${\mathcal J}_\sigma(t_0)$. The
{\em geometric index\/} $\mu(\sigma)$ of the brachistochrone $\sigma$ is defined
to be the (possibly infinite) number:
\begin{equation}\label{eq:defindicegeom}
\mu(\sigma)=\sum_{t_0\in[0,1[}\mu_\sigma(t_0)\in\N\cup\{+\infty\}.
\end{equation}
\end{defbfoc}
Observe that every vector field along $\sigma\big\vert_{[t_0,1]}$ which is solution of the
linear differential  equation (\ref{eq:diffeqbJacobi}) in the interval
$[t_0,1]$, can be extended to a vector field along $\sigma$
satisfying the equation on the entire interval
$[0,1]$. Also, it follows easily from Propositions~\ref{thm:improve} and
\ref{thm:converse} that if   the quantity
$\iip{\nabla_{\dot\sigma}V}{Y}-\iip{V}{\nabla_{\dot\sigma}Y}$ is constant
on $[t_0,1]$ and if $V$ satisfies
(\ref{eq:diffeqbJacobi}) on $[0,1]$, then 
$\iip{\nabla_{\dot\sigma}V}{Y}-\iip{V}{\nabla_{\dot\sigma}Y}$ is constant on $[0,1]$.
In particular, from Proposition~\ref{thm:converse} we have that $\sigma(t_0)$ is
a b-focal point if and only if there exists a non trivial variation $\sigma_s$,
$s\in\,]-\varepsilon,\varepsilon\,[$ of brachistochrones of energy $k$ between
$\sigma(t_0)$ and $\gamma$, depending smoothly on $s$, and such that $\sigma_0=
\sigma\big\vert_{[t_0,1]}$.
\smallskip

We now want to relate the b-focal points along a brachistochrone $\sigma$ with
the $\gamma$-focal points along the corresponding Riemannian geodesic
$w={\mathcal D}(\sigma)$. This is done in Theorem~\ref{thm:finaleMorse} below,
which is preceded by some preliminary results, aimed to determine the relation
of the notions of Jacobi fields along $\sigma$ and $w$.

More precisely, we will show that the linear map $
{\rm d}{\mathcal O}\circ{\rm d}{\mathcal D}(\sigma)$
gives an isomorphism of the spaces ${\mathcal J}_\sigma(t_0)$ and
${\mathcal J}^{\{k\}}_w(\gamma,t_0)$ (recall that the map
${\mathcal O}$ is the direction reversing map defined in \eqref{eq:dirrev}).
\smallskip

Given a horizontal geodesic $w$, a Jacobi field along $w$ is a (smooth) vector field
$J$ along $w$ satisfying the differential equation~(\ref{eq:eqJacobi}). 
From (\ref{eq:tansp}) and (\ref{eq:tanspdelta}), such a vector field $J$ belongs to the
tangent space $T_w\ogpa1(\Delta)$ if and only if $J(1)=0$, $J(0)\in\R\cdot
Y(w(0))$ (recall that we are considering curves $w$ starting on $\gamma$ and
arriving at $p$), and
$\iip{\nabla_{\dot w}J}Y+\iip{\dot w}{\nabla_JY}\equiv0$. 
Recalling Remark~\ref{thm:remarkKillJac}, this last equality
is satisfied identically on $[0,1]$ provided that it is satisfied at some 
point $t_0\in[0,1]$.

Hence, recalling the definitions \ref{itm:b1},  \ref{itm:orto} and~3
of page~\pageref{itm:b1} and Remark~\ref{thm:remarkKillJac}, 
we have that the set of Jacobi fields in $T_w\ogpa1(\Delta)$ coincides with
the finite dimensional vector space ${\mathcal J}_w^{\{k\}}(\gamma,0)$:
\begin{equation}\label{eq:intersJ}
{\mathcal J}_w^{\{k\}}\cap T_w\ogpa1(\Delta)={\mathcal J}_w^{\{k\}}(\gamma,0).
\end{equation}
  
We introduce the following map:
\begin{equation}\label{eq:G}
{\mathcal G}:\opga1\longmapsto\opga1,
\end{equation}
given by:
\begin{equation}\label{eq:defG}
{\mathcal G}(w)(t)=\psi(w(t),h_w(t)),
\end{equation}
where 
\[h_w(t)=-k\int_0^t\frac{\sqrt{\phi_k(w(0))\rip{\dot w(0)}{\dot w(0)}}}{\iip
YY}\;{\rm d}r.\] As in the case of the map ${\mathcal D}$, it is easy to see that
${\mathcal G}$ is smooth; moreover, using (\ref{eq:equiv23}) one checks
that it is a left-inverse for ${\mathcal D}$ in $\bpga1$, i.e., for
all $\sigma\in\bpga1$, we have:
\begin{equation}\label{eq:inversosinistra}
{\mathcal G}({\mathcal D}(\sigma))=\sigma.
\end{equation}
\begin{propC}\label{thm:propC}
Let $\sigma$ be a brachistochrone and $w={\mathcal O}({\mathcal D}(\sigma))$. If
$J\in {\mathcal J}_w^{\{k\}}(\gamma,1)$,
then there exists $V\in {\mathcal J}_\sigma(0)$ a b-Jacobi field along $\sigma$ such that
${\rm d}{\mathcal O}\circ{\rm d}{\mathcal D}(\sigma)[V]=J$.
\end{propC}
\begin{proof}
Let $s\in\,]-\varepsilon,\varepsilon\,[$ and $w_s$ be a smooth variation of $w$ 
consisting of horizontal geodesics and such that $J=\ddso w_s$. Let
$\sigma_s={\mathcal G}({\mathcal O}(w_s))\in\Bpa1$; since ${\mathcal G}$ is smooth,
then
$\sigma_s$ is a smooth variation of $\sigma$. Moreover, ${\mathcal O}({\mathcal
D}(\sigma_s))=w_s$, and since $w_s$ is a horizontal geodesic, by
Proposition~\ref{thm:first}, $\sigma_s$ is a brachistochrone in $\Bpa1$ for all
$s$.   By Definition~\ref{thm:defbJacobi}, $V=\ddso\sigma_s$ is a b-Jacobi field
in ${\mathcal J}_\sigma(0)$. 
Note that $V(0)=0$ because $\sigma_s(0)=p$ for all $s$.

It is easily computed:
\[{\rm d}{\mathcal O}\circ{\rm d}{\mathcal D}(\sigma)[V]=\ddso
{\mathcal O}({\mathcal
D}(\sigma_s))=\ddso w_s=J,\] which concludes the proof.
\end{proof}
Proposition~\ref{thm:propC} gives the surjectivity of the map $
{\rm d}{\mathcal O}\circ{\rm d}{\mathcal D}(\sigma)$
restricted to the spaces of Jacobi fields
${\mathcal J}_\sigma(0)$ and ${\mathcal
J}_w^{\{k\}}(\gamma,0)$. The injectivity of ${\rm d}{\mathcal D}(\sigma)$,
and hence that of ${\rm d}{\mathcal O}\circ{\rm d}{\mathcal D}(\sigma)$, 
can be proven on the entire tangent space $T_\sigma\bpga1$:
\begin{propD}\label{thm:propD}
For all $\sigma\in\bpga1$, ${\rm d}{\mathcal D}(\sigma):T_\sigma\bpga1
\longmapsto T_{{\mathcal D}(\sigma)}\opga1$ is an injective map.
\end{propD}
\begin{proof}
It suffices to prove that ${\rm d}{\mathcal D}(\sigma)$ has a left inverse, i.e.,
that there exists a linear bounded operator $L:T_{{\mathcal
D}(\sigma)}\opga1\longmapsto T_\sigma\bpga1$ such that $L\circ{\rm d}{\mathcal
D}(\sigma)$ is the identity on $T_\sigma\bpga1$. Such a map $L$ is given by the
differential of the map
${\mathcal G}$ defined by (\ref{eq:defG}). Indeed, by (\ref{eq:inversosinistra}),
${\mathcal G}\circ {\mathcal D}$ is the identity on $\bpga1$,and  by differentiating we have
that ${\rm d}{\mathcal G}\circ{\rm d}{\mathcal D}(\sigma)$ is the identity
on $T_\sigma\bpga1$ for all $\sigma\in\bpga1$.
\end{proof}
We can indeed identify the image of ${\rm d}{\mathcal D}(\sigma)$ in
$T_{{\mathcal D}(\sigma)}\opga1$:
\begin{immagine}\label{thm:immagine}
Let $\sigma\in\bpga1$ be a brachistochrone and $w={\mathcal D}(\sigma)$. 
Then, the image of the differential ${\rm d}{\mathcal D}(\sigma)$ in 
$T_w\opga1$  is given by  $T_w\opga1(\Delta)^\perp$ (see formula~(\ref{eq:tanspdeltaperp})).
\end{immagine}
\begin{proof}
We first show that ${\rm d}{\mathcal D}(\sigma)\subset T_w\opga1(\Delta)^\perp$.
To this end, let $\zeta\in T_\sigma\bpga1$ be fixed; by (\ref{eq:tanspbpg1})
and Corollary~\ref{thm:carbrach}, it satisfies:
\begin{equation}
\iip{\nabla_{\dot\sigma}\zeta}{\dot\sigma}\equiv0. 
\end{equation}
Since ${\mathcal D}(\bpga1)
\subset \opga1(\Delta)$, then clearly ${\rm d}{\mathcal D}(T_\sigma \bpga1)\subset
T_w\opga1(\Delta)$. Moreover, let $V={\rm d}{\mathcal D}(\sigma)[\zeta]$.
For the inclusion ${\rm d}{\mathcal D}(T_\sigma \bpga1)\subset
T_w\opga1(\Delta)^\perp$ we need to show that (\ref{eq:conderivatabis})
is satisfied. Using formulas
(\ref{eq:conssigma}), (\ref{eq:defD}), (\ref{eq:deftausigma}),
 (\ref{eq:derivataD}), (\ref{eq:tauzeta}) and (\ref{eq:equiv23}), we compute easily:
\begin{equation}\label{eq:granca}
\begin{split}
\iip{\nabla&\phi_k(w)}V  
\iip{\dot w}{\dot w}+2\,\phi_k(w)\cdot\iip{\nabla_{\dot w}V}{\dot w}=\\
&=-\frac{2k^2{\Tsigma}^2}{\iip YY\,(k^2+\iip YY)}\,\iip{\nabla_YY}\zeta+
\frac{2k^2{\Tsigma}^2}{\iip YY\,(k^2+\iip YY)}\,\iip{\nabla_YY}\zeta=0.
\end{split}
\end{equation}
For the opposite inclusion, we argue as follows.
Let $V$ be fixed in $T_w\opga1(\Delta)^\perp$ and let $w_s\in\Opa1$
be a variation of $w$ with variational vector field $V$
such that $\iip{\dot w_s}{Y(w_s)}\equiv0$ and $\iip{\dot w_s}{\dot w_s}\equiv
c_s$ (constant). 
Such a variation exists,
\footnote{%
the point here is that the variational fields in $T_w\Opga1$ are
given by variations $w_s$ of $w$ that {\em not necessarily\/}
have endpoints on $\gamma(\R)$. The only thing that can be said
about such variations $w_s$ is that $w_s(1)$ is infinitesimally close
to $\gamma$ as $s\to0$ with an order of infinitesimal bigger than
$1$.} 
provided that we do not require the
condition $w_s(1)\in\gamma(\R)$.

For all $s$, define $\sigma_s={\mathcal G}(w_s)$ where ${\mathcal G}$ is
the map defined in (\ref{eq:defG}). Then, $\sigma_s$ is a variation
of $\sigma$ in $\Bpa1$; if $\zeta=\ddso \sigma_s\in T_\sigma\Bpa1$
is the corresponding variational vector field, then clearly ${\rm d}{\mathcal D}%
(\sigma)[\zeta]=V$. To conclude the proof, we need to show that $\zeta\in T_\sigma\bpga1$,
i.e., that $\zeta(1)$ is parallel to $Y(\sigma(1))$. 
Recalling (\ref{eq:derivataD}), his follows easily
from the fact that $V(1)$ is a multiple of $Y(w(1))$ 
and from formula~(\ref{eq:derivataD}).
This concludes the proof.
\end{proof}

In analogy with formula (\ref{eq:derivataD}),  for
all $a\in [0,1[$ we can define a linear map 
${\mathcal L}_{a}$ on the space of vector fields along $\sigma\big\vert_{[a,1]}$
satisfying the two conditions appearing
in (\ref{eq:tanspbpg1}) on the interval $[a,1]$,
and taking values in the space of vector fields along $w\big\vert_{[a,1]}$.

The map ${\mathcal L}_a$ is given by:
\begin{equation}
{\mathcal L}_{a}[\zeta](r)={\rm d}_x\psi(\sigma(r),t^a_\sigma(r))[\zeta(r)+\tau_\zeta^a
\cdot Y(\sigma(r))],
\end{equation}
where 
\[{\tsig}^a(r)=-\int_a^r\frac{\iip{\dot\sigma}Y}{\iip YY}\;{\rm d}u,\quad\text{and}\quad
\tau_\zeta^a(r)=-\int_a^r\frac{C_\zeta\iip YY+2k\,{\Tsigma}
\iip{{\nabla_\zeta}Y}Y}{\iip YY^2}\;{\rm d}u.\] 
In particular, ${\mathcal L}_0={\rm d}{\mathcal D}(\sigma)$; observe
also that, of $\zeta(a)=0$, then ${\mathcal L}_a[\zeta](a)=0$.

\noindent
The result of Propositions~\ref{thm:propC} and \ref{thm:propD} can be extended
immediately to the maps ${\rm d}{\mathcal O}\circ{\mathcal L}_{t_0}:{\mathcal
J}_\sigma(t_0)
\longmapsto{\mathcal J}_w^{\{k\}}(\gamma,t_0)$ for all $t_0\in[0,1[$:
\begin{corE}\label{thm:corE}
Let $\sigma\in\bpga1$ be a brachistochrone and $w={\mathcal D}(\sigma)$
the corresponding geodesic in $\opga1(\Delta)$. Then, for all $t_0\in[0,1[$, the
linear map ${\rm d}{\mathcal O}\circ{\mathcal L}_{t_0}$ gives an
isomorphism of the vector spaces of Jacobi fields
${\mathcal J}_\sigma(t_0)$ and ${\mathcal J}_w^{\{k\}}(\gamma,t_0)$.
\end{corE}
\begin{proof}
The proofs of Propositions~\ref{thm:propC} and \ref{thm:propD}
can be repeated {\em verbatim}, by replacing the initial point $p$ with
the point $\sigma(t_0)$. The only technical subtlety to worry about is
that, when replacing the initial point, it will not hold, in general, that
$\sigma(t_0)=w(t_0)$. Nevertheless, this fact is not essential, because
one can always reduce to this case by considering a suitable 
isometry of $U_k$ given by $x\longmapsto\psi(x,\overline t)$. 
\end{proof}
We now prove that the kernel of the Hessian $H^F(\sigma)$ 
in $T_\sigma\bpga1$ consists precisely of  b-Jacobi fields.
This gives an analytical characterization of the b-Jacobi fields
along a brachistochrone.
\begin{carat}\label{thm:carat}
Let $\sigma$ be a brachistochrone. A vector field
$V\in T_\sigma\bpga1$ is a b-Jacobi field along $\sigma$
if and only if $V\in {\rm Ker}\left(H^F(\sigma)\right)$ in
$T_\sigma\bpga1$.
\end{carat}
\begin{proof}
Corollary~\ref{thm:corB} proves that any b-Jacobi field along
$\sigma$ is in the kernel of $H^F(\sigma)$.

Conversely, let $\sigma$ be a fixed brachistochrone and
$\zeta\in {\rm Ker}\left(H^F(\sigma)\right)$.
From Corollary~\ref{thm:corE}, it suffices to prove that
the vector field $J={\rm d}{\mathcal O}\circ{\rm d}{\mathcal D}(\sigma)[\zeta]$
is a $\gamma$-Jacobi field with respect to the Riemannian 
metric $\phi_k\cdot\gr$
along the geodesic $w={\mathcal O}({\mathcal D}(\sigma))$.
Moreover, since $J\in T_w\ogpa1(\Delta)$, from Lemma~\ref{thm:cargammaJacobi}
it suffices to show that $J$ is a Jacobi field along $w$, i.e., that
it satisfies equation~(\ref{eq:eqJacobi}).
Observe that, by Proposition~\ref{thm:immagine}, $J$ 
is in $T_w\ogpa1(\Delta)^\perp$, hence it satisfies the two equations:
\begin{equation}\label{eq:Jtangente}
\begin{split}
&\rip{\nablak_{\dot w}J}{Y}-\rip{J}{\nablak_{\dot w}Y}=0,\\
&\rip{J}{\dot w}=\rip{\nablak_{\dot w}J}{\dot w}=0.
\end{split}
\end{equation}

To prove that $J$ is Jacobi, let
$V\in C^\infty_o([0,1],T\M)$
be any smooth vector field along $w$ vanishing at the endpoints. 
We set:
\begin{equation}\label{eq:defWmulambda}
W=V-\mu\cdot Y-\lambda\cdot \dot w,
\end{equation}
where $\lambda$ and $\mu$ are functions in $H^1([0,1],\R)$ to be determined
in such a way that the resulting vector field $W$ belongs to
$T_w\ogpa1(\Delta)$. Straightforward computations show this
condition is satisfied by setting:
\begin{equation}\label{eq:lambdaemu}
\begin{split}
&\mu(t)=-\int_t^1\phi_k(w)\cdot\frac{%
\rip{\nablak_{\dot w}V}Y+\rip{\dot w}{\nablak_VY}}{\rip YY}\;{\rm d}r,
\\
&\lambda(t)=-\int_t^1\frac{\rip{\nablak_{\dot w}V}{\dot w}}{\rip{\dot w}{\dot
w}}\;{\rm d}r.
\end{split}
\end{equation}
Observe that, with the definitions above, since $w$ is a geodesic
with respect to $\phi_k\cdot\gr$ one has:
\begin{equation}\label{eq:lmzeroextr}
\lambda(0)=\lambda(1)=\mu(1)=0.
\end{equation}
Arguing as in the proof of Theorem~\ref{thm:secmorseindexth}
since $Y$ is Killing in the metric
$\phi_k\cdot\gr$, then its restriction to
$w$ is a Jacobi field (see also (\ref{eq:YKilJac})):
\begin{equation}\label{eq:YJacobi}
\nablak_{\dot w}\nablak_{\dot w}Y=\Rk(\dot w,Y)\,\dot w.
\end{equation}
Recalling (\ref{eq:hessEphikgr}), keeping in mind (\ref{eq:lmzeroextr})
and the fact that $V(0)=V(1)=0$, we have:
\begin{equation}\label{eq:sommaI}
\begin{split}
H^{E_{\phi_k}}(w)[J,W]=&\;\Ik(J,V)-\Ik(J,\lambda\cdot\dot w)-\Ik(J,\mu\cdot Y)\\
&-
\mu(1)\,\phi_k(w(1))\,\rip{\nablak_{J(1)}Y}{\dot w(1)}.
\end{split}
\end{equation}
From (\ref{eq:defindexform}), the second equation of (\ref{eq:Jtangente})
and the anti-symmetry of the curvature tensor $\Rk$, the term
$\Ik(J,\lambda\cdot\dot w)$ is easily seen to vanish:
\begin{equation}\label{eq:primotermzero}
\Ik(J,\lambda\cdot\dot w)=\int_0^1\phi_k(w)\left(
\lambda'\rip{\nablak_{\dot w}J}{\dot w}+\lambda\,
\rip{\Rk(\dot w,J)\,\dot w}{\dot w}\right)\;{\rm
d}t=0.
\end{equation}

From (\ref{eq:defindexform}), integrating by parts and using 
formulas (\ref{eq:Jtangente}), (\ref{eq:YJacobi}) and the symmetry of the
curvature tensor $\Rk$,
we have:
\begin{equation}\label{eq:in1}
\begin{split}
\Ik(J,\mu\cdot Y)=&\;\int_0^1\phi_k(w)\left(\mu'\cdot\rip{\nablak_{\dot w}J}{Y}
+\mu\cdot\rip{\nablak_{\dot w}J}{\nablak_{\dot w}Y}\right)\;{\rm d}t \\
&+\int_0^1\phi_k(w)\,\mu\cdot\rip{\Rk(\dot w,J)\,\dot w}Y\;{\rm d}t=\\
=&\int_0^1\phi_k(w)\left(\mu'\cdot\rip{\nablak_{\dot w}J}{Y}
+\mu\cdot\rip{\Rk(\dot w,J)\,\dot w}Y\right) \\&-
\int_0^1\phi_k(w)\,\mu'\cdot\rip{J}{\nablak_{\dot w}Y}\;{\rm d}t \\
&-\int_0^1\phi_k(w)\,\mu\cdot\rip{J}{\nablak_{\dot w}\nablak_{\dot w}Y}\;{\rm d}t 
\\
&+\mu(1)\cdot\phi_k(w(1))\cdot\rip{J(1)}{\nablak_{\dot w(1)}Y}=\\=&
\int_0^1\phi_k(w)\,\mu'\left(\rip{\nablak_{\dot w}J}{Y}-
\rip{J}{\nablak_{\dot w}Y}\right)\;{\rm d}t\\
&+\int_0^1\phi_k(w)\,\mu\left(\rip{\Rk(\dot w,J)\,\dot w}Y-
\rip{\Rk(\dot w,Y)\,\dot w}J\right) \;{\rm d}t\\
&+\mu(1)\cdot\phi_k(w(1))\cdot\rip{J(1)}{\nablak_{\dot w(1)}Y}=\\
=&-\mu(1)\cdot\phi_k(w(1))\cdot\rip{\dot w(1)}{\nablak_{J(1)}Y}.
\end{split}
\end{equation}
Finally, from (\ref{eq:sommaI}), (\ref{eq:primotermzero}) and (\ref{eq:in1}), we have proven
the equality:
\[\Ik(J,V)=H^{E_{\phi_k}}(w)[J,W].\]
Since $W\in T_w\ogpa1(\Delta)$, then $W$ is in the image
of ${\rm d}{\mathcal O}\circ{\rm d}{\mathcal D}$, say $W=
{\rm d}{\mathcal O}\circ{\rm d}{\mathcal
D}(\sigma)[\zeta_1]$ for some $\zeta_1\in T_\sigma\bpga1$. Since $\zeta\in {\rm
Ker}\left(H^F(\sigma)\right)$ and $J={\rm d}{\mathcal O}\circ{\rm d}{\mathcal
D}(\sigma)[\zeta]$, then, by Corollary~\ref{thm:princsec}
and formula \eqref{eq:servedopo}, it is
$H^{E_{\phi_k}}(w)[J,W]=-H^F(\sigma)[\zeta,\zeta_1]=0$,
and, in particular, $\Ik(J,V)=0$.
Hence, we have that $\Ik(J,V)=0$ for all smooth vector field
along $w$ vanishing at the endpoints, and by (\ref{eq:condJacobi})
this implies that $J$ is a Jacobi field, concluding the proof.
\end{proof}
 
We are finally ready to state and prove the Morse Index Theorem for
the travel time brachistochrones:

\begin{finaleMorse}[Morse Index Theorem for Relativistic
Brachistochrones]\label{thm:finaleMorse}\hfill\break 
Let $\sigma\in\bpga1$ be a brachistochrone and
$w={\mathcal O}({\mathcal D}(\sigma))\in\ogpa1$ the  corresponding horizontal
geodesic. Then, a point
$\sigma(t_0)$ is a b-focal point along $\sigma$ if and only if $w(t_0)$ is a $\gamma$-focal
point along $w$, in which case the two focal points have the same multiplicity. In
particular, we have
\begin{equation}\label{eq:finMorse}
\mu(\sigma)=\mu^{\{k\}}(w).
\end{equation}
Moreover, if $p$ is not a b-focal point along $\sigma$, then the Morse
index $m(\sigma,T)$ is equal to the geometric index
$\mu(\sigma)$ of $\sigma$:
\begin{equation}\label{eq:diseqMorse}
m(\sigma,T)=\mu(\sigma).
\end{equation}
\end{finaleMorse}
\begin{proof}
By Corollary~\ref{thm:corE}, since isomorphisms preserve dimensions, 
for all $t_0\in[0,1\,[$ we have:
\[{\rm dim}\left({\mathcal J}_w^{\{k\}}(t_0)\right)=\mu_\sigma(t_0).\]
This implies that $\sigma(t_0)$ is a b-focal point along $\sigma$ if and only
if $w(t_0)$ is a $\gamma$-focal point; moreover, summing over all $t_0\in[0,1\,[$,
we obtain (\ref{eq:finMorse}). 

From Corollary~\ref{thm:princsec} and formulas 
(\ref{eq:ughessmigl}) and \eqref{eq:servedopo},  we have:
\begin{equation}\label{eq:strcaz1}
m(\sigma,T)=m(\sigma,{-F});
\end{equation}
from (\ref{eq:estrestr}) and Propositions~\ref{thm:propD} and \ref{thm:immagine}
we obtain:
\begin{equation}\label{eq:strcaz2}
m(\sigma,{-F})=\bar m(w,{E_{\psi_k}});
\end{equation}
finally, from
(\ref{eq:ugualiindici}) we have the equality:
\begin{equation}\label{eq:stessoindice}
\bar m(w,{E_{\phi_k}})=m(w,{E_{\phi_k}}).
\end{equation}
If $p$ is not a $\gamma$-focal point along $w$, or equivalently if
$p$ is not a b-focal point along $\sigma$, then, Theorem~\ref{thm:MorseRiem} 
implies:
\begin{equation}\label{eq:unadelle}
m(w,{E_{\phi_k}})=\mu^{\{k\}}(w);
\end{equation}
the equality (\ref{eq:diseqMorse}) follows at once from
(\ref{eq:finMorse}) and (\ref{eq:strcaz1})---(\ref{eq:unadelle}). This concludes
the proof.
\end{proof}
From finiteness of the index $m(\sigma,T)$ we get the following:
\begin{maimax}\label{thm:maimax}
Let $\sigma\in\bpga1$ be a brachistochrone. Then, $\sigma$ is never a local
maximum for $T$.\qed
\end{maimax}

From the equality (\ref{eq:diseqMorse}) we get that, if $\mu(\sigma)=0$, then
the Morse index of the travel time vanishes at $\sigma$, hence
$\sigma$ is a local minimum for $T$. Therefore, we have:
\begin{minimilocali}\label{thm:minimilocali}
Let $\sigma\in\bpga1$ be a brachistochrone and $w={\mathcal O}({\mathcal
D}(\sigma))$. Suppose that there are no $\gamma$-focal points along $w$. Then, 
$\sigma$ is a local minimum for the arrival time functional $T$. \qed
\end{minimilocali}
\end{section}

\begin{section}{The Global Morse Relations}
\label{sec:global}
In this section we will use the infinite dimensional
Morse theory to prove some equalities relating the
differential structure of the travel time brachistochrone
problem and the topological structure carried by the set of
continuous paths joining $p$ and $\gamma$ in $U_k$.

Most of the technical results needed are obtained using the same
ideas and techniques employed in Reference~\cite{GM}, where the authors
prove the Morse relations for geodesics in a convex subset of
a stationary Lorentzian manifold.  In order to keep our exposition
short, we will omit some of the proofs that can be deduced easily from
analogous proofs presented in details in \cite{GM}.

Throughout the section, we will make the following assumptions:
\begin{enumerate}
\item\label{itm:hp1m} the vector field $Y$ is complete in $U_k$, i.e.,
its integral lines are defined over the entire real line;
\item\label{itm:hp2m} $\gamma:\R\longmapsto U_k$ is an integral
line of $Y$ without self-intersection;
\item\label{itm:hp3m} $p$ is an event in $U_k$;
\item\label{itm:hp4m} $k^2$ is a regular value for the function $-\iip YY$;
\item\label{itm:hp5m} $\overline U_k=U_k\bigcup\partial U_k$ is complete
with respect to the Riemannian metric (\ref{eq:defgr});
\item\label{itm:hp6m} the function $-\iip YY$ is {\em bounded away from $0$\/}
in $U_k$, i.e., there exists a positive constant $\nu$ such that 
$-\iip YY\ge\nu>0$ in $U_k$;
\item\label{itm:hp7m} $p$ and $\gamma$ are not b-conjugate, i.e., for
any brachistochrone $\sigma$ of energy $k$ in $\bpga1$, the points
$\sigma(0)=p$ and $\sigma(1)$ are not b-conjugate along $\sigma$.
\end{enumerate}
\smallskip

We denote by $\bbpg$ the set of brachistochrones in $\bpga1$; moreover,
let $\Cpg$ denote the set of continuous paths joining $p$ and $\gamma$
in $U_k$:
\begin{equation*}
\Cpg=\Big\{z\in C^0([0,1],U_k):z(0)=p,\ z(1)\in\gamma(\R)\Big\},
\end{equation*}
endowed with the topology of the uniform convergence.

The following is the main result of the Section:
\begin{thm8.1}\label{thm:thm8.1}
Under the assumptions \ref{itm:hp1m}---\ref{itm:hp7m} above, given any
coefficient field $\mathcal K$, the following
equality between formal power series in the variable $\lambda\in \mathcal K$ holds
true:
\begin{equation}\label{eq:8.1}
\sum_{\sigma\in\bbpg}\lambda^{\mu(\sigma)}=\sum_{i=1}^\infty
{\rm dim}\left(H_i(\Cpg,{\mathcal K})\right)\,\lambda^i+(1+\lambda)\,Q(\lambda),
\end{equation}
where $\mu(\sigma)$ is the geometric index of the brachistochrone $\sigma$,
$H_i(\Cpg,{\mathcal K})$ is the $i$-th homology vector space of $\Cpg$
with coefficients in $\mathcal K$, and $Q$ is a formal power series
in $\lambda$ with coefficients in $\N\bigcup\{+\infty\}$.
\end{thm8.1}
The Morse relations (\ref{eq:8.1}) can be used to derive a series
of information about the number of brachistochrones joining  $p$ and
$\gamma$ and  with a given
energy value. 
\begin{rem8.2}\label{thm:rem8.2}
If the open set $U_k$ is contractible, then also the space $\Cpg$ is
contractible, and thus, for every field $\mathcal K$, its homology spaces
$H_i(\Cpg,{\mathcal K})$ vanish for all $i>0$ and $H_0(\Cpg,{\mathcal K})\simeq
{\mathcal K}$. In this case, 
under the assumptions \ref{itm:hp1m}---\ref{itm:hp7m} above, formula
(\ref{eq:8.1}) becomes:
\begin{equation}\label{eq:8.2}
\sum_{\sigma\in\bbpg}\lambda^{\mu(\sigma)}=1+(1+\lambda)Q(\lambda).
\end{equation}
Setting $\lambda=1$ in (\ref{eq:8.2}), we get immediately that the number
of travel time brachistochrones of energy $k$ between $p$ and $\gamma$
is either infinite (if $Q(1)=+\infty$) or {\em odd\/} (if $Q(1)<+\infty$).
 
On the other hand, if $U_k$ is not contractible, then, since $\gamma$
is contractible in $U_k$ (because of the injectivity of $\gamma$), then
there are infinitely many indices $i$ such that ${\rm dim}(H_i(\Cpg,{\mathcal
K}))>0$ (see \cite{Se}). Hence, if $U_k$ is not contractible, then there
exist infinitely many brachistochrones of energy $k$ between $p$ and
$\gamma$ in $U_k$.
\end{rem8.2}
In order to prove Theorem~\ref{thm:thm8.1}, we will use the functional
$E_{\phi_k}$ (defined in (\ref{eq:functphi})) in the space $\Opga1=\Opga1(U_k)$,
and $\phi_k$ is the function  defined in (\ref{eq:defphik}).

Since $U_k$ is open, we need to use a {\em penalization\/} argument,
as follows. We define the function:
\begin{equation}\label{eq:8.3}
\Psi_k=\iip YY+k^2;
\end{equation}
It is $\partial U_k=\Psi_k^{-1}(0)$, moreover $\Psi_k(q)>0$ if and only
if $q\in U_k$. By assumption \ref{itm:hp4m}, the Riemannian gradient
$\nablar\Psi_k$ does not vanish on $\partial U_k$, where $\nablar$ denotes
the gradient with respect to the Riemannian metric (\ref{eq:defgr}). 

We define a family $\chi_\varepsilon$ of real functions of class $C^2$,
for $\varepsilon>0$:
\begin{equation}\label{eq:8.4}
\chi(s)=e^s-(1+s+\frac{s^2}2),\quad\chi_\varepsilon(s)=\left\{
\begin{array}{ll}\chi(s-\frac1\varepsilon),&\text{if}\ s\ge\frac1\varepsilon;\\ \\
0,&\text{if}\ s<\frac1\varepsilon.
\end{array}\right.
\end{equation}
Finally, for all $\varepsilon\in\,]0,1]$, we define the {\em penalized\/}
functional:
\begin{equation}\label{eq:8.5}
E_\varepsilon(w)=E_{\phi_k}(w)+\int_0^1\chi_\varepsilon\left(
\frac1{\Psi_k(w)^2}\right)\;{\rm d}t.
\end{equation}
For all $\varepsilon>0$, $E_\varepsilon$ is a functional of class
$C^2$ on $\opga1$, which satisfies good {\em compactness properties},
as it will be discussed below.

By the completeness of $\overline U_k$, it is not too difficult
to prove that, for every $c\in\R$, the sublevel $E_\varepsilon^c$:
\[E_\varepsilon^c=\Big\{w\in\opga1(U_k):E_\varepsilon(w)\le c\Big\}\]
is a complete metric subspace of $\opga1(U_k)$, with respect to the metric
induced by the Hilbert structure (\ref{eq:cazzarola}).

Moreover, using the same techniques employed in \cite{GP},
one proves the following two facts:
\begin{itemize}
\item $E_\varepsilon$ satisfies the {\em Palais--Smale condition\/} at every level
$c\in\R$, i.e., every sequence $\{w_n\}$ in $E_\varepsilon^c$ such that%
\footnote{here, by convergence to $0$, we mean that $\Vvert {\rm
d}E_\varepsilon(w_n)\Vvert$ goes to zero, where $\Vvert\cdot\Vvert$ is the 
operator norm in the dual space of $T_{w_n}\opga1$.}
${\rm d}E_\varepsilon(w_n)$ tends to $0$ as $n\to\infty$, has a convergent
subsequence in $E_\varepsilon^c$;
\smallskip

\item for all $c\in\R$ there exists $\delta(c)>0$ and $\varepsilon(c)\in\,]0,1]$
such that, for all $\varepsilon\in\,]0,\varepsilon(c)]$ and for all
critical point $w_\varepsilon$ of $E_\varepsilon$ in $\opga1(U_k)$
with $E_\varepsilon(w_\varepsilon)\le c$, then $w_\varepsilon$ is also
a critical point for $E_{\phi_k}$, and the following inequality holds:
\[\Psi_k(w_\varepsilon(t))\ge\delta(c),\quad\forall\,t\in[0,1].\]
\end{itemize}
In particular, it follows that if $c$ is a regular value for $E_{\phi_k}$,
i.e., if there are no critical point for $E_{\phi_k}$ in $E_{\phi_k}^{-1}(c)$,
using (\ref{eq:8.4}) and (\ref{eq:8.5}) we obtain the existence
of a number $\varepsilon'(c)\in\,]0,\varepsilon(c)]$ such that, for
all $\varepsilon\in\,]0,\varepsilon'(c)]$, $c$ is a regular value also
for the functional $E_\varepsilon$, and a curve $w\in\opga1$ is a critical
point for $E_\varepsilon$ if and only if it is a critical point for
$E_{\phi_k}$ (with $E_\varepsilon(w)=E_{\phi_k}(w)$) and:
\[m(w,E_\varepsilon)=m(w,E_{\phi_k}),\]
where $m(z,G)$ denotes the {\em Morse Index\/} of the functional $G$
at the critical point $z$.

Then, using assumption~\ref{itm:hp7m}, for all
$\varepsilon\in\,]0,\varepsilon'(c)]$, every critical point $w$
of $E_\varepsilon$ in $E_\varepsilon^c$ is {\em nondegenerate}, which
allows to obtain the Morse Relations in $E_\varepsilon^c$
(see Ref.~\cite{MW}):
\begin{prop8.3}\label{thm:prop8.3}
If $c$ is a regular value for $E_{\phi_k}$, then there exists $\varepsilon'(c)
\in\,]0,1]$ such that, for every $\varepsilon\in\,]0,\varepsilon'(c)]$, we have:
\begin{equation}\label{eq:8.6}
\sum_{w\in{\mathcal G}_{p,\gamma}^c}\lambda^{m(w,E_{\phi_k})}=
\sum_{i=0}^\infty{\rm dim}\left(%
H_i(E_\varepsilon^c,{\mathcal K})\right)\,\lambda^i+(1+\lambda)\,Q_c(\lambda),
\end{equation}
where ${\mathcal G}_{p,\gamma}^c$ is the set of horizontal geodesics
between $p$ and $\gamma$ with energy less than or equal to $c$:
\[{\mathcal G}_{p,\gamma}^c=\Big\{w\in\opga1(U_k): {\rm d}E_{\phi_k}(w)=0,
\ E_{\phi_k}(w)\le c\Big\},\]
and $Q_c(\lambda)$ is a polynomial in the variable $\lambda$ with coefficients
in $\N$.\qed
\end{prop8.3}

We recall that, given a topological pair $(A,B)$,
i.e., a topological space $A$ and a subspace $B\subset
A$ with the induced topology, we say that $B$ is a 
{\em weak deformation retract\/}
of $A$ if there exists a continuous map $H:A\times[0,1]\longmapsto A$ such that:
\begin{enumerate}
\item $H(\cdot,0)$ is the identity map of $A$;
\item $H(B,s)\subset B$ for all $s\in[0,1]$;
\item $H(A,1)\subset B$.
\end{enumerate}
Given a topological pair $(A,B)$, we denote by $P_\lambda(A,B)$ the Poincar\'e
series of $(A,B)$ in the variable $\lambda$, which is given by:
\[P_\lambda(A,B;{\mathcal K})=\sum_{i=0}^\infty{\rm
dim}\left(H_i(A,B;{\mathcal K})\right)\,\lambda^i,\] where $H_i(A,B;{\mathcal K})$
is the
$i$-th relative homology space of $(A,B)$ with coefficients in the field
${\mathcal K}$.
\smallskip

Now, for $\delta>0$, we denote by $\opga1(\delta)$ the set of curves
in $\opga1$ whose image stays at distance greater or equal to $\delta$
from $\partial U_k$:
\[\opga1(\delta)=\Big\{w\in\opga1:\Psi_k(w(t))\ge\delta\
\forall\,t\in[0,1]\Big\}.\] Using the results of Ref.~\cite{GM}, we can prove that
if
$c$ is a regular value of $E_{\phi_k}$, there exists $\delta_0=\delta_0(c)>0$
and $\varepsilon_0=\varepsilon_0(c)$ such that, for all $\delta\in\,]0,\delta_0]$
and for all $\varepsilon\in\,]0,\varepsilon_0]$, the following two statements
hold:
\begin{eqnarray}
\label{eq:8.7}
&& \opga1(\delta)\cap E_{\phi_k}^c\ \text{is a weak deformation retract of}\ 
E_{\phi_k}^c,
\\ \nonumber\\
\label{eq:8.8}
&& \opga1(\delta)\cap E_\varepsilon^c\ \text{is a weak deformation retract of}\ 
E_\varepsilon^c.
\end{eqnarray}
Observe that, if $\varepsilon$ is sufficiently small,
we have 
\[\opga1(\delta)\cap E_{\phi_k}^c=\opga1(\delta)\cap E_{\varepsilon}^c.\]
Then, 
using standard techniques in Algebraic Topology, from (\ref{eq:8.7})
and (\ref{eq:8.8}) we deduce easily that, if $c_1$ and $c_2$ are critical
values of $E_{\phi_k}$, with $c_1<c_2$, then there exists $\varepsilon_0
\in\,]0,1]$ such that, for all $\varepsilon\in\,]0,\varepsilon_0]$, the following
identities between Poincar\'e series hold:
\begin{itemize}
\item $\displaystyle P_\lambda(E_\varepsilon^{c_2};{\mathcal
K})=P_\lambda(E_{\phi_k}^{c_2};{\mathcal K})$;
\item $\displaystyle P_\lambda(E_\varepsilon^{c_2}, E_\varepsilon^{c_1};
{\mathcal K})=
P_\lambda(E_{\phi_k}^{c_2},E_{\phi_k}^{c_1};{\mathcal K})$.
\end{itemize}
Using the above identities and the same technique of
\cite[Theorem~1.6]{GM}, one passes to the limit as $c\to+\infty$ in
(\ref{eq:8.6}), obtaining the Morse relations for the functional
$E_{\phi_k}$ in $\opga1(U_k)$:
\begin{thm8.4}
Under assumptions \ref{itm:hp1m}---\ref{itm:hp7m}, for all
coefficient field ${\mathcal K}$, we have:
\begin{equation}\label{eq:8.9}
\sum_{w\in{\mathcal G}_{p,\gamma}}\lambda^{m(w,E_{\phi_k})}=
\sum_{i=0}^\infty {\rm
dim}\left(H_i(\opga1(U_k);{\mathcal K})\right)\,\lambda^i+(1+\lambda)\,Q(\lambda),
\end{equation}
where ${\mathcal G}_{p,\gamma}$ is the set of all horizontal
geodesics between $p$ and $\gamma$:
\[{\mathcal G}_{p,\gamma}=\Big\{w\in\opga1:{\rm d}E_{\phi_k}(w)=0\Big\},\]
and $Q(\lambda)$ is a formal power series in $\lambda$ (depending on the choice of
${\mathcal K}$) with   coefficients in $\N\cup\{+\infty\}$.\qed 
\end{thm8.4}
We are finally ready to prove Theorem~\ref{thm:thm8.1}:
\begin{proof}[Proof of Theorem~\ref{thm:thm8.1}]
By Proposition~\ref{thm:first} and Lemma~\ref{thm:stessipunti} we see
that $w\in {\mathcal G}_{p,\gamma}$ if and only if $w={\mathcal D}(\sigma)$,
where $\mathcal D$ is the deformation map of (\ref{eq:defdeformation})
and $\sigma$ is a travel time brachistochrone of energy $k$ between 
$p$ and $\gamma$. 
By the first part of Theorem~\ref{thm:finaleMorse}, the hypothesis
\ref{itm:hp7m} implies that every $w\in{\mathcal G}_{p,\gamma}$
is a nondegenerate critical point of $E_{\phi_k}$ in $\opga1(U_k)$.
Moreover, by Theorem~\ref{thm:MorseRiem}, we
have $m(w,E_{\phi_k})=\mu^{\{k\}}$, while, by Theorem~\ref{thm:finaleMorse},
it is $\mu^{\{k\}}(w)=\mu(\sigma)$. Then, formula (\ref{eq:8.9})
can be written as:
\[\sum_{\sigma\in\bbpg }\lambda^{\mu(\sigma)}=\sum_{i=0}^\infty
{\rm
dim}\left(H_i(\opga1(U_k);{\mathcal
K})\right)\,\lambda^i+(1+\lambda)\,Q(\lambda).\] Finally, it is well known
(\cite[Theorem~17.1]{M}) that $\opga1(U_k)$ has the same homotopy type of
${\mathcal C}_{p,\gamma}^0(U_k)$, which concludes the proof.
\end{proof}
\end{section}

\newpage
\appendix

\begin{section}{An explicit calculation of the Hessian of $T$ on $\bpga1$}
\label{sec:explicit}
In this appendix we show how to compute 
explicitly the second variation of the functional
$T$, or, recalling (\ref{eq:relhessiani}),
equivalently, of the functional $F$.

To this aim, we fix a brachistochrone $\sigma$
of energy $k$ between $p$ and $\gamma$, and we consider the corresponding
Lagrange multipliers $\lambda$ and $\mu$, given by (\ref{eq:l0}) and
(\ref{eq:esprmu}). \smallskip

In the next Lemma it is shown how to compute the second variation on
constrained critical points with the method of Lagrange multipliers.
For simplicity, the result will be stated and proved only for Hilbertian
manifolds.
\begin{abstract2}\label{thm:abstract2}
Let $M$ be a Hilbert manifold and $E$ be a Hilbert space.
Let $f:M\longmapsto \R$ and $g:M\longmapsto E$ be smooth maps.
Suppose that $0\in E$ is a regular value for $g$, i.e., the differential
${\rm d}g(x)$ is surjective for all $x\in g^{-1}(0)$, in such a way that
$N=g^{-1}(0)$ is a smooth submanifold of $M$. Let $x_0\in N$ be a critical
point for the restriction $f\big\vert_N$ and let $\Lambda\in E^*$ be the 
(unique\footnote{%
the Lagrange multiplier $\Lambda$ is unique, because ${\rm d}g(x_0)$
is surjective. The relation ${\rm d}(f-\Lambda\circ g)(x_0)=0$
defines $\Lambda$ uniquely.})
associated Lagrange multiplier, i.e., ${\rm d}(f-\Lambda\circ g)(x_0)=0$.
Then, the Hessian of $f\big\vert_N$ at $x_0$ in $T_{x_0}N$ is given by the
restriction of the Hessian of $(f-\Lambda\circ g)$ to $T_{x_0}N$:
\begin{equation}\label{eq:relhessdan}
H^{f-\Lambda\circ g}(x_0)\big\vert_{T_{x_0}N\times T_{x_0}N}=H^{f \vert_N}(x_0).
\end{equation}
\end{abstract2}
\begin{proof}
Let $v\in T_{x_0}N$ and $y:]-\varepsilon,\varepsilon\,[\longmapsto N$
be a smooth curve such that $y(0)=x_0$ and $y'(0)=v$. By (\ref{eq:forhess}),
we have:
\begin{equation}\label{eq:primabis}
H^{f-\Lambda\circ g}(x_0)[v,v]=\frac{{\rm d}^2\big( (f-\Lambda\circ g)\circ
y\big)}{{\rm d}s^2}\,\big\vert_{s=0};
\end{equation}
since $(g\circ y)\equiv0$, then
\begin{equation}\label{eq:secondabis}
\frac{{\rm d}^2\big( (f-\Lambda\circ g)\circ
y\big)}{{\rm d}s^2}\,\big\vert_{s=0}=\frac{{\rm d}^2(f\circ y)}{{\rm d}s^2}\,\big\vert_{s=0}=
H^{f \vert_N}(x_0)[v,v],
\end{equation}
which concludes the proof.
\end{proof}

By Lemma~\ref{thm:abstract2}, the Hessian $H^F(\sigma)$ is given by the restriction of the
Hessian
$H^{F_{\lambda,\mu}}(\sigma)$ to the space $T_\sigma\bpga1$, where $F_{\lambda,\mu}:\opga1
\longmapsto\R$ is the functional given by:
\begin{eqnarray}\label{eq:defFlm}
F_{\lambda,\mu}(\sigma)&=&\int_0^1\Big[\frac12\iip{\dot\sigma}{\dot\sigma}-
\lambda\iip{\dot\sigma}Y-\mu\left(\iip{\dot\sigma}Y^2+k^2\iip{\dot\sigma}{%
\dot\sigma}\right)\Big]\;{\rm
d}t=\nonumber
\\&=&\int_0^1\Big[\frac12\iip{\dot\sigma}{\dot\sigma}
-\mu\left(\iip{\dot\sigma}Y^2+k^2\iip{\dot\sigma}{%
\dot\sigma}\right)\Big]\;{\rm
d}t.
\end{eqnarray}
In order to compute the second variation of $F_{\lambda,\mu}$, we
will consider smooth variations in $\opga1$ of a brachistochrone  $\sigma$,
defined as follows.
\begin{defvariazione}\label{thm:defvariazione}
Given a curve $z\in\opga1$, by a {\em variation\/} of $z$ we will
mean a map $\eta:]-\varepsilon,\varepsilon\,[\times[0,1]\longmapsto\M$
such that:
\begin{enumerate}
\item $\eta(s,\cdot)\in\opga1$ for all $s\in\,]-\varepsilon,\varepsilon\,[$;
\item $\eta(0,\cdot)=z$;
\item the map $s\longmapsto\eta(s,\cdot)$ is smooth from $]-\varepsilon,\varepsilon\,[$
to $\opga1$.
\end{enumerate}
Given a variation $\eta$ of $z\in\opga1$, for all $s$ and $t$ there exists
the derivative $\frac{\partial\eta}{\partial s}(s,t)\in T_{\eta(s,t)}\M$; the vector field
along
$z$ given by $V(t)=\frac{\partial\eta}{\partial s}(0,t)$ is called the {\em variational
vector field \/} corresponding to the variation $\eta$.
\end{defvariazione}
In the rest of this section, given a  variation 
$\eta(s,t)=\sigma_s(t)$ in $\opga1$ of a curve $\sigma\in\bpga1$, 
we will denote by $\Dds$ and $\Ddt$,
the operations of covariant derivative of the Levi--Civita connection of
$g$ in the directions of $\frac{\partial\eta}{\partial s}$ and
$\frac{\partial\eta}{\partial t}$ for vector fields along $\eta$;
the usual symbols
$\dds$ and
$\ddt$ will denote the differentials with respect to  $s$ and $t$ of functions along $\eta$.

Since the Lie bracket $[\Dds,\Ddt]$ vanish and the Levi--Civita connection is
torsion free, 
we have the following commutation relations
involving the operators
$\Dds$,
$\Ddt$, $\dds$ and $\ddt$:
\begin{equation}\label{eq:commutation}
\Dds\,\ddt=\Ddt\,\dds,\quad \Dds\,\Ddt=\Ddt\,\Dds+R(\Dds,\Ddt), 
\quad \Ddt\,\Dds=\Dds\,\Ddt+R(\Ddt,\Dds);
\end{equation}
where $R$ is the curvature tensor of the Lorentzian metric $g$, defined 
in (\ref{eq:convention}).

We have the following:
\begin{hessianoF}\label{thm:hessianoF}
Let $\sigma\in\bpga1$ be a brachistochrone of energy $k$ between $p$ and $\gamma$
of travel time ${\Tsigma}$.
Then, the Hessian $H^F(\sigma)$ of the action functional $F$ (see~(\ref{eq:F})) at
$\sigma$ is given by the following formula:
\begin{eqnarray}\label{eq:hessianoF}
H^F(\sigma)[\,\zeta,\zeta\,]&=&\int_0^1\frac{\iip YY}{k^2+\iip YY}\Big[%
\iip{\nabla_{\dot\sigma}\zeta}{\nabla_{\dot\sigma}\zeta}+
\iip{R(\zeta,\dot\sigma)\,\zeta}{\dot\sigma}\Big]\;{\rm d}t+\nonumber\\
&+&2k\,{\Tsigma}\int_0^1\frac{\iip{\nabla_{\dot\sigma}\zeta}{%
\nabla_\zeta Y}+\iip{R(\zeta,\dot\sigma)\,\zeta}{Y}}%
{k^2+\iip YY}\;{\rm d}t+\\
&+&\frac{\iip YY}{k^2+\iip YY}\,a_\zeta^2
\,\iip{\nabla_{Y}{Y}}{\dot\sigma} \Big \vert_{t=1},
\nonumber
\end{eqnarray}
for all $\zeta\in T_\sigma\bpg$, where 
 $a_\zeta$ is defined by $\zeta(1)=a_\zeta\cdot Y(\sigma(1))$.
\end{hessianoF}
\begin{proof}
The computation is done by brute force, as follows.

Let $\sigma\in\bpga1$ be a fixed brachistochrone. 
Observe that $\sigma$ is smooth. Hence, by a density argument, it suffices
to restrict our attention to smooth variations.
Let $\zeta\in T_\sigma\bpga1$ be a fixed smooth variational vector
field and let $\sigma_s$, $s\in\,]-\varepsilon,\varepsilon\,[$, be a
smooth variation%
\footnote{the existence of such variations, at least in the smooth case, 
is easily proven using the exponential map  and standard arguments 
in Riemannian manifolds.}  
of $\sigma$ in $\opga1$ corresponding to  $\zeta$.
This means that $\sigma_s\in\opga1$ for all $s$, $\sigma_0=\sigma$,
the map $(s,t)\longmapsto\sigma_s(t)\in\M$ is smooth, and
$\zeta=\dds\,\big\vert_{s=0}\sigma_s$.

We differentiate the expression $F_{\lambda,\mu}(\sigma_s)$ with respect to $s$ twice,
and we evaluate at $s=0$.  From
(\ref{eq:defFlm}), we have:
\begin{equation}\label{eq:a1}
\begin{split}
\frac{{\rm d}^2}{{\rm d}s^2}\Big\vert_{s=0}F_{\lambda,\mu}(\sigma_s)= &
\int_0^1\left(1-2\mu\,k^2\right)\Big(\iip{\Dds\,\Ddt\,\dds\,\sigma_s}{\dot\sigma}+
\iip{\nabla_{\dot\sigma}\zeta}{\nabla_{\dot\sigma}\zeta}\Big)\;{\rm d}t\\
&-2\int_0^1\mu\big(\iip{\nabla_{\dot\sigma}\zeta}Y-\iip\zeta{\nabla_{\dot\sigma}Y}\big)^2
\;{\rm d}t\allowbreak\\
&+2\int_0^1\mu\,k\,{\Tsigma}\Big(
\iip{\Dds\,\Ddt\,\dds\,\sigma_s}{Y}+
\iip{\nabla_{\dot\sigma}\zeta}{\nabla_\zeta Y}\Big)\;{\rm d}t \\
&-2\int_0^1\mu\,k\,{\Tsigma}\Big(
\iip{\Dds\,\dds\sigma_s}{\nabla_{\dot\sigma}Y}+\iip{\zeta}{\Dds\,\Ddt\,Y}\Big)\;{\rm d}t.
\end{split}
\end{equation}
Since $\sigma$ is a brachistochrone,
by Corollary~\ref{thm:carbrach}  the second integral in (\ref{eq:a1})
vanishes:
\begin{equation}\label{eq:a0}
\int_0^1\mu\big(\iip{\nabla_{\dot\sigma}\zeta}Y-\iip\zeta{\nabla_{\dot\sigma}Y}\big)^2
\;{\rm d}t=0.
\end{equation}
By (\ref{eq:commutation}), the last term in (\ref{eq:a1}) can be written as:
\begin{equation}\label{eq:a2}
\iip{\zeta}{\Dds\,\Ddt\,Y}\,\Big\vert_{s=0}=\iip\zeta{\nabla_{\dot\sigma}\nabla_\zeta Y}+
\iip{R(\zeta,\dot\sigma)\,Y}{\zeta}.
\end{equation}
We now consider the three terms in (\ref{eq:a1}) that contain two derivatives with
respect to $s$, and, using (\ref{eq:commutation}), we write them as follows:
\begin{equation}\label{eq:a3}
\begin{split}
\int_0^1&(1-2\mu k^2)\iip{\Dds\,\Ddt\,\dds\,\sigma_s}{\dot\sigma}\;{\rm d}t+\\
&+2\int_0^1\mu k {\Tsigma}\left(\iip{\Dds\,\Ddt\,\dds\,\sigma_s}{Y}-
\iip{\Dds\,\dds\sigma_s}{\nabla_{\dot\sigma}Y}\right)\;{\rm d}t=\allowdisplaybreaks \\
&=\int_0^1\Big[(1-2\mu k^2)\iip{R(\zeta,\dot\sigma)\,\zeta}{\dot\sigma}
+2\mu k {\Tsigma}\,\iip{R(\zeta,\dot\sigma)\,\zeta}{Y}\Big]\;{\rm d}t+\\
&\quad +\int_0^1(1-2\mu k^2)\iip{\Ddt\,\Dds\,\dds\,\sigma_s}{\dot\sigma}\;{\rm d}t+
\\
&+2 k {\Tsigma}\int_0^1\mu\left(\iip{\Ddt\,\Dds\,\dds\,\sigma_s}{%
Y}-\iip{\Dds\,\dds\,\sigma_s}{\nabla_{\dot\sigma}Y}\right)\;{\rm d}t.
\end{split}
\end{equation}
Integration by parts in the last two integrals of (\ref{eq:a3}) gives:
\begin{equation}\label{eq:b1}
\begin{split}
&\int_0^1(1-2\mu k^2)\iip{\Ddt\,\Dds\,\dds\,\sigma_s}{\dot\sigma}\;{\rm d}t\\
&+2 k {\Tsigma}\int_0^1\mu\left(\iip{\Ddt\,\Dds\,\dds\,\sigma_s}{%
Y}-\iip{\Dds\,\dds\,\sigma_s}{\nabla_{\dot\sigma}Y}\right)\;{\rm d}t=\allowbreak\\
&\qquad=\int_0^1(1-2\mu k^2)\iip{\Ddt\,\Dds\,\dds\,\sigma_s}{\dot\sigma}\;{\rm d}t\\
&\qquad+2k{\Tsigma}\int_0^1\mu\Big(\ddt\,\iip{\Dds\,\dds\,\sigma_s}Y-2\iip{%
\Dds\,\dds\,\sigma_s}{\nabla_{\dot\sigma}Y}\Big)\;{\rm d}t=
\allowbreak\\
&\qquad=\Big((1-2\mu k^2)\iip{\Dds\,\dds\,\sigma_s}{\dot\sigma}+
2\mu k{\Tsigma}\iip{\Dds\,\dds\,\sigma_s}Y\Big)\Bigg\vert_{t=0}^{t=1}\\
&\qquad+\int_0^1\iip{\Dds\,\dds\,\sigma_s}{2\mu'k^2\dot\sigma-(1-2\mu
k^2)\nabla_{\dot\sigma}\dot\sigma-2k{\Tsigma}\mu'Y-4\mu k{\Tsigma} Y}=
\allowbreak\\
&\qquad=\iip{\Dds\,\dds\,\sigma_s}{(1-2\mu k^2)\,\dot\sigma+2\mu k {\Tsigma}
Y}\,\Big\vert_{t=1}=\allowbreak\\
&=\iip{\Dds\,\dds\,\sigma_s}{\frac{\iip YY}{k^2+\iip YY}\,\dot\sigma+
\frac{k{\Tsigma}}{k^2+\iip YY}\,Y}\,\Big\vert_{t=1},
\end{split}
\end{equation}
because, by (\ref{eq:diffeq}), we have:
\[2\mu'k^2\dot\sigma-(1-2\mu
k^2)\nabla_{\dot\sigma}\dot\sigma-2k{\Tsigma}\mu'Y-4\mu k{\Tsigma} Y=0,\]
and, since $\sigma_s(0)\equiv p$, \[\Dds\,\dds\,\sigma_s(0)=0.\]
Here, we have used the equalities:
\[1-2\mu k^2=\frac{\iip YY}{k^2+\iip YY},\quad\text{and}\quad \mu'=-\frac{%
\iip{\nabla_{\dot\sigma}Y}Y}{(k^2+\iip YY)^2}.\]

Let now $\alpha(s)$ be defined by:
\begin{equation}\label{eq:b2}
\sigma_s(1)=\gamma(\alpha(s)).
\end{equation}
We have:
\[\zeta(1)=\alpha'(0)\cdot Y(\sigma(1)),\]
hence, multiplying by $Y(\sigma(1))$, we obtain
\begin{equation}\label{eq:b3}
\alpha'(0)=\frac{\iip\zeta Y}{\iip YY}\Big\vert_{t=1}.
\end{equation}
Moreover, from (\ref{eq:b2}) we easily get:
\begin{equation}\label{eq:b4}
\Dds\Big\vert_{s=0}\dds\,\Big[\sigma_s(1)\Big]=\alpha'(0)\cdot\nabla_{\zeta(1)}Y+\alpha''(0)\cdot
Y(\sigma(1)).
\end{equation}
Substitution of (\ref{eq:b3}) and (\ref{eq:b4}) into (\ref{eq:b1}) gives:
\begin{eqnarray}\label{eq:b5}
&&\iip{\Dds\,\dds\,\sigma_s}{\frac{\iip YY}{k^2+\iip YY}\,\dot\sigma+
\frac{k{\Tsigma}}{k^2+\iip YY}\,Y}\,\Big\vert_{t=1}=\\
&&\iip{\alpha'(0)\cdot\nabla_{\zeta(1)}Y+\alpha''(0)\cdot
Y(\sigma(1))}{\frac{\iip YY}{k^2+\iip YY}\,\dot\sigma+
\frac{k{\Tsigma}}{k^2+\iip YY}\,Y}=\nonumber\\
&&=\iip{\frac{\iip\zeta Y}{\iip YY}\cdot\nabla_{\zeta(1)}Y}{\frac{\iip YY}{k^2+\iip
YY}\,\dot\sigma+
\frac{k{\Tsigma}}{k^2+\iip YY}\,Y}\,\Big\vert_{t=1}.
\nonumber
\end{eqnarray}
In conclusion, we have proven the equality:
\begin{eqnarray}\label{eq:bordo}
&&\int_0^1(1-2\mu k^2)\iip{\Dds\,\Ddt\,\dds\,\sigma_s}{\dot\sigma}\;{\rm d}t+\nonumber\\
&&+2\int_0^1\mu k {\Tsigma}\left(\iip{\Dds\,\Ddt\,\dds\,\sigma_s}{Y}-
\iip{\Dds\,\dds\sigma_s}{\nabla_{\dot\sigma}Y}\right)\;{\rm d}t=\nonumber \\
&&\quad=\int_0^1\frac{\iip YY}{k^2+\iip%
YY}\,\iip{R(\zeta,\dot\sigma)\,\zeta}{\dot\sigma}\;{\rm d}t+\\
&&\qquad+\int_0^1\frac{k{\Tsigma}}{k^2+\iip
YY}\,\iip{R(\zeta,\dot\sigma)\,\zeta}{Y}\;{\rm d}t+
\nonumber\\
&&\qquad+\frac{\iip\zeta Y}{\iip YY\,(k^2+\iip YY)}\Big(-\iip
YY\iip{\zeta}{\nabla_{\dot\sigma}Y}+k\,{\Tsigma}\,\iip Y{\nabla_\zeta
Y}\Big)\Big\vert_{t=1}.
\nonumber
\end{eqnarray}
Observe that, since $\zeta(1)=a_\zeta\cdot Y(\sigma(1))$, with
\begin{equation}\label{eq:defazeta}
a_\zeta=\frac{\iip\zeta Y}{\iip YY}\,\Big\vert_{t=1},
\end{equation}
then the boundary term in (\ref{eq:bordo}) vanishes:
\begin{equation}\label{eq:correzione}
\iip Y{\nabla_\zeta Y}\,\Big\vert_{t=1}=0.
\end{equation}
Since $\iip\zeta{\nabla_\zeta Y}\equiv0$, then:
\[0=\ddt\,\iip\zeta{\nabla_\zeta Y}=\iip{\nabla_{\dot\sigma}\zeta}{\nabla_\zeta Y}+
\iip\zeta{\nabla_{\dot\sigma}\nabla_\zeta Y},\]
hence
\begin{equation}\label{eq:c1}
-\iip\zeta{\nabla_{\dot\sigma}\nabla_\zeta Y}=\iip{\nabla_{\dot\sigma}\zeta}{\nabla_\zeta Y}.
\end{equation}
Finally, by the anti-symmetry of the curvature tensor $R$, we have:
\begin{equation}\label{eq:c2}
-\iip{R(\zeta,\dot\sigma)\,Y}\zeta=\iip{R(\zeta,\dot\sigma)\,\zeta}Y.
\end{equation}
Formula (\ref{eq:hessianoF}) now follows from (\ref{eq:a1}), (\ref{eq:a0}),
(\ref{eq:a2}), (\ref{eq:bordo}), (\ref{eq:defazeta}),
(\ref{eq:correzione}), (\ref{eq:c1}) and (\ref{eq:c2}).
\end{proof}

Let's assume now that $\gamma$ has no self intersection, which in particular
implies that $\gamma(\R)$ is an embedded submanifold of $\M$.

\begin{rembordo}\label{thm:rembordo}
If we consider the submanifold $\Sigma=\gamma(\R)$, then the second fundamental form
$S^\gamma$ takes the following form. For $q=\gamma(s_0)$ and $v_i=\nu_i\cdot Y(q)$,
$i=1,2$, given a vector $n\in T_q\M$ which is orthogonal to $Y(q)$, we have:
\[S^\gamma_n(v_1,v_2)=\nu_1\nu_2\cdot\iip {\nabla_Y Y\,\big\vert_q}n.\]
This formula resembles the factor $a_\zeta^2\cdot\iip{\nabla_Y
Y}{\dot\sigma}\,\big\vert_{t=1}$ that appears in the boundary term of
$H^F(\sigma)[\zeta,\zeta]$ in formula (\ref{eq:hessianoF}). The reader should observe,
though, that the vector
$\dot\sigma(1)$ is not orthogonal to $Y(\sigma(1))$, because
$\iip{\dot\sigma}Y\equiv-k{\Tsigma}\ne0$.
\end{rembordo}
\end{section}

\begin{section}{$F$ does not satisfy the Palais--Smale condition
in $\bpg$}\label{sec:example}
We discuss a very simple example to prove that, in general,
the travel time functional $T$, or the action functional
$F$ {\em do not\/} satisfy the Palais--Smale compactness condition
in $\bpg$.\smallskip

Let's consider $\M=\R^3$ to be the flat 3-dimensional Minkowski spacetime,
with metric $\iip\cdot\cdot$ given by
${\rm d}x^2+{\rm d}y^2-{\rm d}z^2$  and
$Y=\frac\partial{\partial z}$ the timelike Killing vector field on $\M$. Let
$\iip\cdot\cdot_o$ denote the Euclidean metric ${\rm d}x^2+{\rm d}y^2$ in $\R^2$.

We fix a point $p=(p_0,0)$ in $\M$ and a curve $\gamma(r)=(p_1,r)$,
where $x_0,x_1\in\R^2$, and a real constant $k>-\iip YY\equiv1$. 

In this case, the set $\bpg$ consists of curves $\sigma(t)=(x(t),y(t),z(t))$
where ${\mathbf x}(t)=(x(t),y(t))$ is in $H^2([0,1],\R^2)$ is a curve in $\R^2$
that joins $p_0$ and $p_1$, $z\in H^2([0,1],\R)$, $z(0)=0$,   and there exists a
positive constant ${\Tsigma}$ such that:
\[\dot z=k\,{\Tsigma},\quad\iip{\dot{\mathbf x}}{\dot{\mathbf x}}_o
-\dot z^2=-{\Tsigma}^2,\]
and so:
\[\iip{\dot{\mathbf x}}{\dot{\mathbf x}}_o=(k^2-1)\,{\Tsigma}^2>0\ 
\text{(constant)}.\]
It is easy to see that the map $({\mathbf x},z)\longmapsto{\mathbf x}$ gives a diffeomorphism
of $\bpg$ and the Hilbert manifold:
\begin{equation}\label{eq:nuovaman}
\begin{split}
\Omega^{(2)}_c(p_0,p_1)&=\\&\Big\{{\mathbf x}\in H^2([0,1],\R^2:{\mathbf
x}(0)=p_0,\  {\mathbf x}(1)=p_1,\ \iip{\dot{\mathbf x}}{\dot{\mathbf x}}_o\equiv
C_{\mathbf x}={\rm const.}  >0 \Big\};
\end{split}
\end{equation}
moreover, the travel time functional $T$ and the action functional
$F$ on $\bpg$ are transformed respectively into (constant multiples of)
the Euclidean length functional $L$ and the Euclidean
energy functional  $E$ on $\Omega^{(2)}_c(p_0,p_1)$:
\[L({\mathbf x})=\int_0^1\iip{\dot{\mathbf x}}{\dot{\mathbf x}}_o\;{\rm d}t,
\quad
E({\mathbf x})=\frac12\int_0^1\iip{\dot{\mathbf x}}{\dot{\mathbf x}}_o^2\;{\rm d}t.\]
It is not hard to prove that 
the only  critical point  of $L$ and $E$ on
$\Omega^{(2)}_c(p_0,p_1)$ is the Euclidean geodesic, i.e., the straight segment, between
$p_0$ and
$p_1$ in $\R^2$. 

On the other hand, if $p_0\ne p_1$, the manifold $\Omega^{(2)}_c(p_0,p_1)$ is
complete, and it is easy to see that its first homotopy group is infinite. 
Thus, if either $L$ or $E$
satisfied the Palais--Smale condition on $\Omega^{(2)}_c(p_0,p_1)$, 
by standard techniques of Critical Point Theory one could
prove the existence of infinitely many 
distinct geodesics between $p_0$ and $p_1$ in $\R^2$, which is clearly absurd.

It follows that neither $T$ nor $F$ satisfies the Palais--Smale condition
on $\bpg$. The same argument shows that neither $T$ nor $F$ satisfies
the Palais--Smale condition in any set of curves satisfying
a regularity that implies the $C^1$-regularity.
\end{section}


\end{document}